\documentclass[12pt]{article}
\usepackage[margin=1in]{geometry} 
\RequirePackage{multicol}
\RequirePackage{setspace}
\RequirePackage[leqno]{amsmath}

\makeatletter
\renewcommand\section{\@startsection{section}{1}{\z@}%
  {-3.5ex \@plus -1ex \@minus -.2ex}%
  {2.3ex \@plus.2ex}%
  {\bfseries\large\center}} % Adjust the size (\large) as per requirement

\renewcommand\subsection{\@startsection{subsection}{2}{\z@}%
  {-3.25ex\@plus -1ex \@minus -.2ex}%
  {1.5ex \@plus .2ex}%
  {\itshape\large\center}} % Adjust the size (\large) as per requirement

\renewcommand\refname{REFERENCES}
\renewenvironment{thebibliography}[1]{
  \if@JEP
    \subsection*{\MakeUppercase{\refname}}
  \else
    \section*{\MakeUppercase{\refname}}
  \fi
  \@mkboth{\MakeUppercase\refname}{\MakeUppercase\refname}
  \list{\@biblabel{\@arabic\c@enumiv}}%
       {\settowidth\labelwidth{\@biblabel{#1}}%
        \leftmargin\labelwidth
        \advance\leftmargin\labelsep
        \@openbib@code
        \usecounter{enumiv}%
        \let\p@enumiv\@empty
        \renewcommand\theenumiv{\@arabic\c@enumiv}}%
  \sloppy
  \clubpenalty4000
  \@clubpenalty \clubpenalty
  \widowpenalty4000%
  \sfcode`\.\@m}
  {\def\@noitemerr
    {\@latex@warning{Empty `thebibliography' environment}}
  \endlist}

\newcommand\TenPtJEL{\@setfontsize\TenPtJEL{10.5}{12}}
\long\def\JEL#1{\long\gdef\@JEL{#1}}

\makeatother
	\usepackage[english]{babel} 
	\usepackage{authblk}
	\usepackage{amsmath, amssymb, textcomp, amsthm, mathtools,mathrsfs} 
	\usepackage{wasysym} 
	\usepackage{stmaryrd}
	\usepackage{dsfont}
 \usepackage{setspace}
 \usepackage{adjustbox}
 \usepackage{fancyhdr}
   	\usepackage{booktabs}
    \usepackage{enumerate}
   	\usepackage{inputenc}
	\usepackage{csquotes}
	\usepackage{floatrow}
    \usepackage{nicefrac}
	\usepackage{footnote,footmisc} 
	\usepackage{setspace} 
	\usepackage{threeparttable}
	\usepackage{tikz}
 \usetikzlibrary{calc, decorations.pathreplacing}
\usepackage{comment}
\usepackage{graphicx}
 \usepackage{subcaption}
 \usepackage{pgfplots}
\usepgfplotslibrary{statistics}
\pgfplotsset{compat=1.16}

	\usepackage{pgfplots}
	\pgfplotsset{compat=1.17}
	\usetikzlibrary{shapes.misc}
	\usetikzlibrary{patterns}
	\usepackage{xcolor}
    \usepackage{multibib}
\newcites{software}{Software References}
	\usetikzlibrary{decorations.pathreplacing}
	\usetikzlibrary{arrows}
	\usepackage{float} 
	\usepackage{pifont}% http://ctan.org/pkg/pifont

\usepackage{floatrow}

	\usepackage{hyperref}
	\hypersetup{
    	colorlinks,
    	linkcolor={red!50!black},
    	citecolor={blue!50!black},
    	urlcolor={blue!80!black}
	}

\DeclareFontFamily{U}{mathx}{}
\DeclareFontShape{U}{mathx}{m}{n}{<-> mathx10}{}
\DeclareSymbolFont{mathx}{U}{mathx}{m}{n}
	\DeclareMathAccent{\widecheck}{0}{mathx}{"71}
\newcommand*\dd{\mathop{}\!\mathrm{d}}

	\usepackage[toc,page]{appendix}
	
	\newtheorem{theorem}{Theorem}
	\newtheorem{lemma}{Lemma} 
	\newtheorem{proposition}{Proposition} 
	\newtheorem{definition}{Definition}

    \newtheorem{corollary}{Corollary}
	\newtheorem{assumption}{Assumption}
	\newtheorem{example}{Example}
	
	\newtheorem*{example*}{Example}

\newtheorem{assumpL}{}
\newtheorem{assumpM}{}

\newtheorem{assumpK}{}
\newtheorem{assumpI}{}
\newtheorem{assumpE}{}

\newtheorem*{assumption*}{Assumption}

	\usepackage{graphicx}  
	\usepackage{marvosym}

    \DeclareMathOperator*{\argmin}{\arg\!\min}

    \newtheorem*{assumptions*}{\assumptionnumber}
    \usepackage{stackrel}

\providecommand{\assumptionnumber}{}

\usepackage[position=top]{caption}
\usepackage{mwe}

\definecolor{cyan}{cmyk}{1, 0.4, 0, 0}

\newcommand{\nddvd}[1]{{\color{blue}[David: #1]}}
\newcommand{\sumn}{\sum_{i=1}^n}
\newcommand{\var}{\operatorname{var}}
\newcommand{\dpar}[2]{\frac{\partial #1}{\partial #2}}

\definecolor{mypink}{RGB}{219, 48, 122}

\def\XXint#1#2#3{{\setbox0=\hbox{$#1{#2#3}{\int}$ }
\vcenter{\hbox{$#2#3$ }}\kern-.6\wd0}}

\usepackage[round]{natbib}
\makeatletter

\allowdisplaybreaks[4]

\makeatother

\makeatletter
\newcommand*{\lesseqgtrslant}{\mathrel{\mathpalette\@gtr@less@eq{\leqslant>}}}
\newcommand*{\gtreqlessslant}{\mathrel{\mathpalette\@gtr@less@eq{\geqslant<}}}
\newcommand*{\@gtr@less@eq}[2]{%
   \vcenter{%
      \offinterlineskip
      \m@th
      \setbox0=\hbox{$#1\@secondoftwo#2$}%
      \hbox{$#1\@firstoftwo#2$}%
      \kern-.2\ht0  % <--- had to guess this...
      \box0
   }%
}
\makeatother
\title{\vspace{-1.5em}Regression Discontinuity Design with Distribution-Valued Outcomes \vspace{-0.5em}}
\author{David Van Dijcke\thanks{\url{dvdijcke@umich.edu}. All errors are mine. I thank Zach Brown, Matias Cattaneo, Jacob Dorn, Ying Fan, Florian Gunsilius, Eva Janssens, Harry Kleyer, Benjamin Scuderi, Kaspar Wuthrich, and participants at the University of Michigan IO-Econometrics workshop, the University of Michigan Labor lunch, and the New Connections in the Study of Political Economy workshop for helpful discussion and comments. An R package accompanying this paper is available at \url{https://davidvandijcke.com/R3D}.} 
}

\affil{\small Department of Economics\\ University of Michigan, Ann Arbor}

\date{\vspace{-0.5em}\small\today}

\begin{document}
\thispagestyle{empty}
\begin{spacing}{0.9} % Slightly reduced spacing for title page
\maketitle

\begin{abstract} 
\small % Reduce font size for abstract
\noindent 

This article introduces Regression Discontinuity Design (RDD) with Distribution-Valued Outcomes (R3D), extending the standard RDD framework to settings where the outcome is a distribution rather than a scalar. Such settings arise when treatment is assigned at a higher level of aggregation than the outcome—for example, when a subsidy is allocated based on a firm-level revenue cutoff while the outcome of interest is the distribution of employee wages within the firm. Since standard RDD methods cannot accommodate such two-level randomness, I propose a novel approach based on random distributions. The target estimand is a ``local average quantile treatment effect'', which averages across random quantiles. To estimate this target, I introduce two related approaches: one that extends local polynomial regression to random quantiles and another based on local Fréchet regression, a form of functional regression. For both estimators, I establish asymptotic normality and develop uniform, debiased confidence bands together with a data-driven bandwidth selection procedure. Simulations validate these theoretical properties and show existing methods to be biased and inconsistent in this setting.  
I then apply the proposed methods to study the effects of gubernatorial party control on
within-state income distributions in the US, using a close-election design. The results suggest a classic equality--efficiency tradeoff under Democratic governorship, driven by reductions in income at the top of the distribution. \\
\vspace{0.5em}

\end{abstract}
\end{spacing}
\clearpage

\onehalfspacing

\section{Introduction}

The regression discontinuity design (RDD) is a popular non-experimental method for causal inference and program evaluation. It exploits a cutoff rule in the assignment variable—often a running variable such as a test score or an income threshold—to identify sharp changes in treatment status among units just above and just below the cutoff. In recent years, it has been widely adopted in economics and political science. 

The conventional RDD setup typically assumes that the running variable and outcome are measured at the same level of aggregation-—each unit has its own running variable and a single outcome measurement. In many policy and program contexts, however, the outcome of interest takes the form of an entire \textit{distribution} within an aggregate unit that receives treatment, rather than a single scalar. For example, when a school district implements an educational policy based on a district-wide threshold (like a cutoff in the district’s poverty rate), one may be interested in the effect on the distribution of student test scores in each district. Similarly, when a minimum wage is implemented along a state border, the outcome of interest could be the distribution of goods prices in each establishment, rather than a single average price. These settings are marked by two layers of randomness: one \textit{across} units (districts, establishments), and one \textit{within} (students within districts, goods sold in an establishment). This motivates the development of a more general framework, where the local average treatment effect is defined over distributions rather than scalars.

\begin{figure}[ht!]
\centering
\begin{subfigure}[b]{0.45\textwidth}
\includegraphics[width=\textwidth]{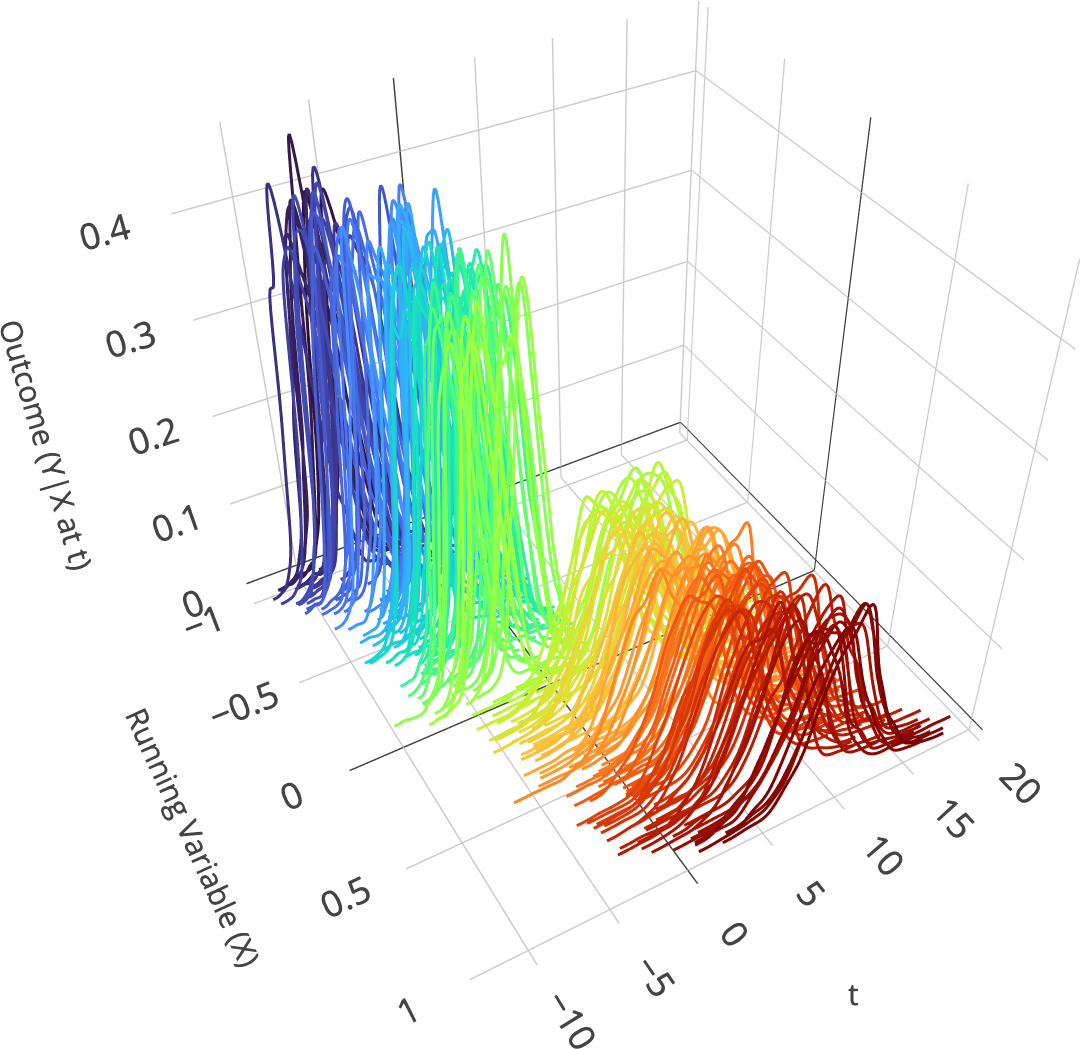}
\caption{R3D: Sample}
\label{fig:gauss_sample}
\end{subfigure} 
\begin{subfigure}[b]{0.45\textwidth}
\includegraphics[width=\textwidth]{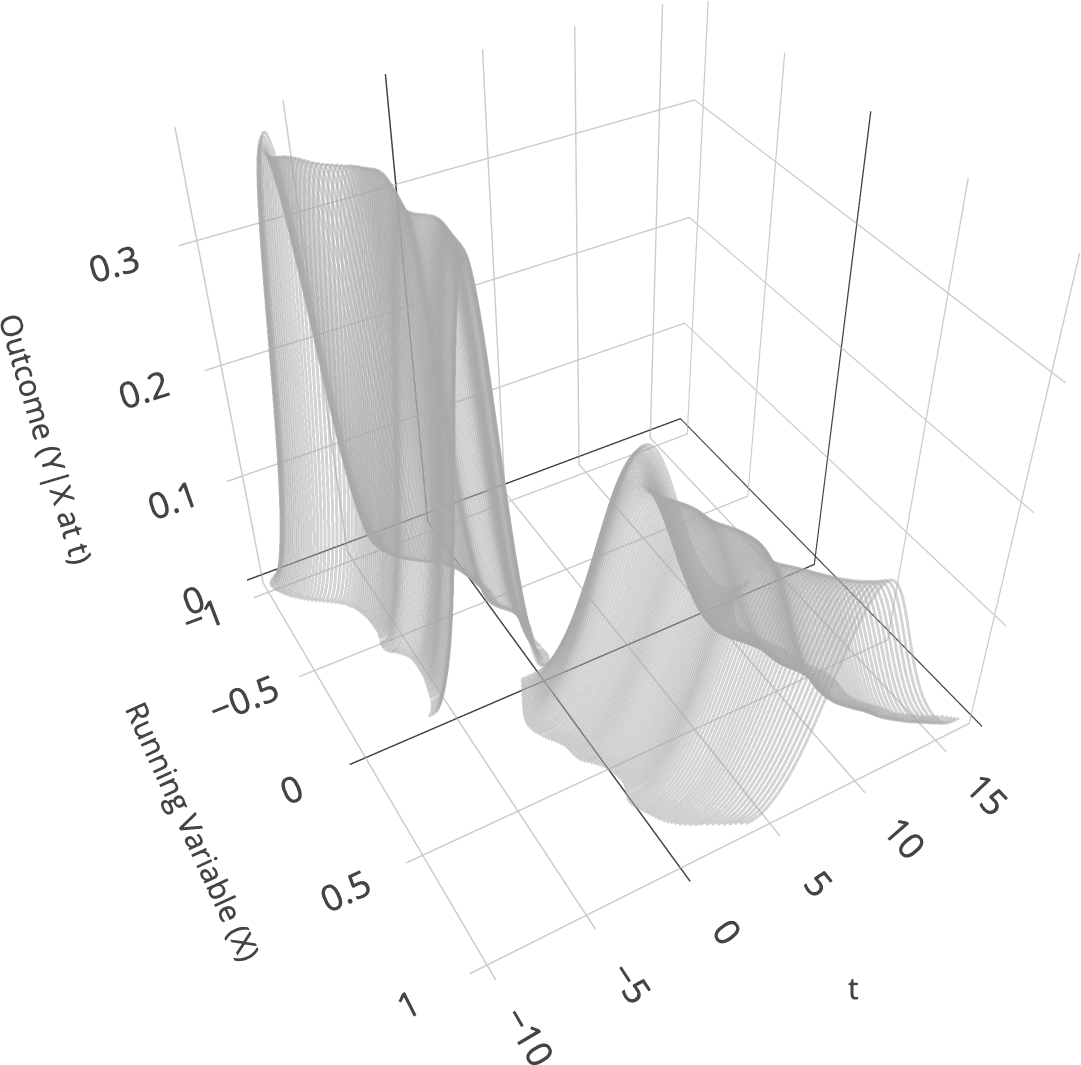}
\caption{R3D: Conditional Expectation}
\label{fig:gauss_frechet}
\end{subfigure}\
\begin{subfigure}[b]{0.5\textwidth}
\includegraphics[width=\textwidth]{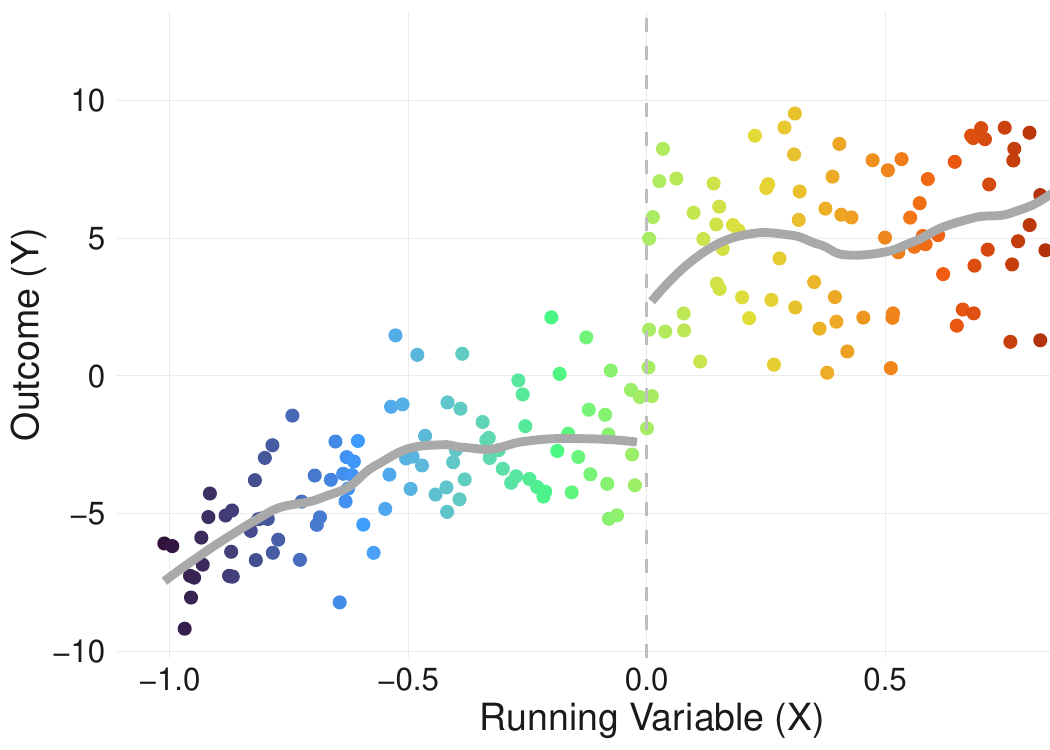}
\caption{RDD: Sample + Conditional Expectation}
\end{subfigure} 
\caption{Example of a Distribution-Valued RDD}
\label{fig:gauss_example}
\end{figure}

In this article, I extend the standard RDD framework to a functional data setting that can accommodate distribution-valued outcomes, allowing one to capture how an intervention shifts entire distributions rather than just their means or fixed quantiles. I call this the \textit{Regression Discontinuity Design with Distributions (R3D)}. Its key distinction from classical settings is that it models the data-generating process as sampling \textit{entire distributions} together with the running variable. Hence, in this setting, distributions themselves are treated as random objects. This naturally leads to a novel concept of distribution-valued treatment effects, the ``local average quantile treatment effect'' (LAQTE), which captures the shift in the underlying \emph{average} quantile function around the cutoff, where the average is with respect to the \textit{distribution of distributions}. Identification is obtained by assuming that this conditional average distribution evolves smoothly. This constitutes an intuitive generalization of the canonical RD smoothness assumption to distribution-valued outcomes. The setting is illustrated in Figure \ref{fig:gauss_sample}. In classical RDD (bottom panel), the data points are a random point cloud (rainbow colors), and their conditional expectation (gray color) is a smooth scalar-valued function. In R3D, the data points are random distributions (rainbow colors), and their conditional expectations (gray color) are a smooth path of distributions.

To estimate these average quantile treatment effects in practice, I propose two closely related estimators. The first estimator extends the canonical local polynomial regression estimator to random quantiles. The idea is to compute the observed outcome quantile function within each aggregate unit, pick a given quantile on these quantile functions, and estimate a local polynomial regression on the resulting ``random quantiles''. This process is then repeated for every point on the quantile function. Such an approach accounts for both the vertical (distribution within unit) and horizontal (distribution across units) sampling that distinguishes the R3D setting from the canonical one. However, while doing this quantile-by-quantile may be intuitive, it is suboptimal in the sense that it does not properly treat the quantile function as a functional object. 

Hence, I propose a second estimator, based on local Fr\'echet regression in 2-Wasserstein space \citep{petersen2019frechet}, that estimates a local polynomial regression for the entire quantile function at once.
Such a functional approach is preferred because it uses all information from the entire quantile function, leading to better finite-sample performance, and only requires picking a single bandwidth, which makes it more computationally efficient. Moreover, the resulting estimate is a conditional Wasserstein barycenter (Fr\'echet mean), which has the important property of being the central tendency of the observed quantile functions in probability space \citep{agueh2011barycenters, fan2024conditional}. Importantly, this Fr\'echet (second) estimator is closely linked to the local polynomial (first) estimator. In particular, the Fr\'echet estimator is equivalent to an $L^2$ projection of the local polynomial (first) estimator onto the space of quantile functions. This close link between both estimators allows me to derive uniform, debiased confidence bands for both, by leveraging general theoretical results for local polynomial estimators developed in \citet{chiang2019robust}. Deriving confidence bands for local Fr\'echet regression is in general not feasible due to the absence of the required algebraic structure on general metric spaces \citep{dubey2019frechet}. A notable exception is \citet{petersen2021wasserstein}, who derived confidence bands for global Fr\'echet regression in 2-Wasserstein space by leveraging that space's optimal transport geometry and the linearity of the global regression model. My results complement theirs by deriving the first confidence bands for the \textit{local} Fr\'echet regression estimator in 2-Wasserstein space. I do so by similarly leveraging that space's optimal transport geometry without requiring a linear response model, but instead exploiting the connection to the pointwise local polynomial estimator.  

Empirically, I first validate the estimators through extensive simulations. These results show that, unlike state-of-the-art quantile RD estimators \citep{qu2019uniform}, the proposed R3D estimators do not suffer from asymptotic bias. Moreover, I show that the uniform confidence bands are asymptotically valid and consistent, quickly converging to the nominal 95\% level and to a power of 1. 

Further, I demonstrate the estimators' use in an empirical application. The question studied is, ``what is the effect of partisan control of the state governor's office on the within-state income distribution?''. To answer this question, I leverage a close-election R3D design, which compares states where the Democratic candidate narrowly won their election to states where they narrowly lost. Because each state has only a single election outcome but an entire distribution of family incomes, this is a prototypical R3D setting. Applying the proposed estimators to this setting, I estimate reductions in income for above-median earners that get stronger with income and become statistically significant for the top 10 percentiles, but no such effects for lower-income families. These results point to a classical equality--efficiency tradeoff \citep{okun1975equality}, where a decrease in income inequality can only be achieved at the cost of an overall loss of income. 

To conclude the introduction, I note that the setting considered here is distinct from that of the quantile RD (Q-RD) setting first developed in \citet{frandsen2012quantile}. That approach estimates quantile treatment effects for \emph{scalar-valued} outcomes, and thus does not apply to the distribution-valued setting considered here. Indeed, in what follows, I show that the quantile RD estimator is biased and inconsistent in the R3D setting, both theoretically and in simulations. This bias results from the fact that its scalar-valued sampling framework is inappropriate for the R3D setting, and its identifying smoothness assumption highly restrictive. This can be seen in Figure \ref{fig:gauss_example}. Because of random sampling, the \emph{observed} distributions (rainbow) exhibit \emph{discontinuous} changes, violating the Q-RD assumption. The \emph{average} distributions  (gray) do evolve smoothly, however. Of course, when treatment and outcome are measured at the same level -- i.e., we find ourselves in the classical RD setting -- the quantile RD estimator is preferred over the R3D estimator.

\subsection*{Literature}

This article contributes to several strands of literature, the primary one being the literature on regression discontinuity design \citep{thistlethwaite1960regression, hahn2001identification}, see \citet{lee2010regression} and \citet{cattaneo2022regression} for an older and more recent overview. I contribute to this large area of research in three ways. 

First, I extend the literature on quantile treatment effects in RDD to allow for distribution-valued outcomes. \citet{frandsen2012quantile} first developed the framework for quantile RD and derived uniform convergence results, though they did not derive uniform confidence bands. These were developed later for different types of quantile RD estimators in \citet{qu2019uniform, qu2024inference, chiang2019robust}. Further variations of the classical quantile RD were studied in \citet{jin2025identification, chiang2019causal, qu2024inference, chen2020quantile}. I build on this literature, in particular the general framework of \citet{chiang2019robust}, to derive uniform confidence bands for distribution-valued RD designs. This also connects to the larger literature on distributional inference \citep{chernozhukov2013inference} and quantile regression and treatment effects \citep{koenker1978regression, firpo2009unconditional, firpo2007efficient, chernozhukov2005iv}. 

Second, I contribute to the strands of literature that have developed robust, debiased confidence bands for local polynomial estimators with mean-squared error (MSE) based bandwidth selection procedures \citep{calonico2014robust, calonico2018effect, calonico2020optimal, calonico2022coverage, armstrong2018optimal, imbens2012optimal}, by extending these tools to distribution-valued settings. That places this paper in a rich literature built on the foundational contributions in local polynomial regression, particularly related to bias reduction and bandwidth selection, made by \citet{fan1992variable, fan1993local, fan1995adaptive, linton1994multiplicative}.

Third, this article relates to several other papers that have considered RD designs with varying levels of aggregation. \citet{borusyak2024regression} considered the opposite design, where the treatment assignment is at a \emph{lower} instead of a higher level of aggregation than the outcome. \citet{cattaneo2016interpreting, cattaneo2021extrapolating, bertanha2020regression} considered aggregation schemes for RD with multiple cutoffs. Relatedly, \citet{gunsilius2023free, papay2011extending, cheng2023estimation} considered RD designs with multi-dimensional or multiple assignment variables. 

The other main strand this paper contributes to is the literature on \citet{frechet1948elements} regression, which was originally developed by \citet{petersen2019frechet} for general metric spaces, with several further contributions for distribution regression in Wasserstein space \citep{chen2023wasserstein, fan2022conditional, chen2023sliced, ghodrati2022distribution, zhou2024wasserstein} and for local Fr\'echet regression \citep{chen2022uniform, iao2024deep, qiu2024random}. As noted above, I contribute to this literature by deriving uniform confidence bands for local Fr\'echet regression in Wasserstein space, complementing related results for global Fr\'echet regression in \citet{petersen2021wasserstein} and for Wasserstein barycenters in \citet{carlier2021entropic, agueh2017vers, kroshnin2021statistical}. My results also hold for general polynomial orders while the literature has mostly focused on local linear regression, with the exception of \citet{schotz2022nonparametric}. More broadly, this article contributes to the large literature on functional data analysis \citep{ramsay2005functional}. 

Relatedly, my results leverage the fact that Fr\'echet regression in 2-Wasserstein space is an $L^2$ projection of the local polynomial estimator onto the space of quantile functions. This relates closely to isotonic regression \citep{barlow1972statistical} and  monotone rearrangement methods, \citep{chernozhukov2010quantile}, as well as shape-constrained inference with convex projection operators \citep{chetverikov2018econometrics, groeneboom2014nonparametric, fang2021projection, dumbgen2024shape}.

This article also contributes to the literature applying optimal transport tools to causal inference -- see \citet{gunsilius2025primer} for a recent overview. In particular, \citet{gunsilius2023distributional} considered a similar setting to mine, where treatment is at a higher level than the outcome, in the context of synthetic controls (see also \citet{van2024return} for an application to firm tenure distributions). \citet{kurisu2024geodesic} introduced causal inference for objects in general metric spaces using geodesics, while \citet{zhou2025geodesic} applied this to the well-known difference-in-differences estimator, thus complementing the distributional estimators of \citet{athey2006identification, torous2024optimal, callaway2018quantile}. 

Finally, this paper’s empirical application-—to gubernatorial party control and income distributions—fits into a rich literature linking partisan control of US state governments to inequality and other economic outcomes. Building on Hibbs' partisan theory \citep{hibbs1977political} and Kelly’s \emph{market conditioning} \citep{kelly2009politics}, research has generally argued that Democrats, allied with lower income groups, adopt policies that narrow income gaps, whereas Republicans, favoring upper and business income constituencies, may widen them. Panel studies show that Democratic legislatures raise taxes and spending \citep{reed2006democrat}, implying stronger redistribution. Though recent evidence from difference-in-difference designs and close-election RDDs found no evidence that party control significantly affects most state-level economic outcomes within a governor's tenure \citep{dynes2020noisy}, other close-election RDs have shown that Democratic state control often increases minimum wages and welfare caseloads, compressing the post‐tax income distribution \citep{leigh2008estimating}, 
and leads to more liberal policies \citep{caughey2017incremental}. I contribute to this literature by providing credible causal estimates of the effect of gubernatorial party control on the income distribution, using rich individual-level data within each state instead of just state-level aggregates. That way, I estimate significant declines in pre-tax income for upper-income families, which compress the income distribution.

\section{Regression Discontinuity with Distribution-Valued Outcomes}

In this section, I present the distribution-valued version of the canonical regression discontinuity design. First, I formally introduce the setting, before providing several concrete examples from the literature. Then, I introduce a new definition of ``local average quantile treatment effects'' (LAQTE) appropriate for this setting, where the average is over \emph{random} quantile functions. Before presenting two consistent estimators for these LAQTEs, I briefly discuss the distinction between my R3D setting and the classical quantile RD setting of \citet{frandsen2012quantile}. I conclude providing an overview of the statistical inference tools developed in Section \ref{sec:statistics}, including extensions to fuzzy RDD and empirical quantile functions.

\subsection{Setting}

First, I define and discuss the R3D setting. Let $\mathcal{Y}$ be the space of cumulative distribution functions (cdfs) $G$ on $\mathbb{R}$ with finite variance, $\int_{\mathbb{R}} x^2 \dd G(x) < \infty$. Let $(X,Y) \sim F$ be a random element with joint distribution $F$ on $\mathbb{R} \times \mathcal{Y}$. I call $X$ the running variable and $Y$ the outcome variable. Unlike the canonical RD design,  here $Y$ is a random \textit{distribution} rather than a random variable. Hence, each draw $(X_i, Y_i)$ from $(X,Y)$ provides a full distribution $Y_i$ at the running variable value $X_i$, rather than a single real number. Then, denote $T \in \{0,1\}$ the treatment status. I assume that $T$ is some monotonic function of $X$ such that,
\[T=\begin{cases} 0 & \text{if } X < c \\ 1 & \text{if } X \geq c \end{cases}\] 
for some threshold $c$. That is, treatment is assigned deterministically when the running variable $X$ crosses the threshold $c$, where I assume without loss of generality that $c=0$. This is the so-called ``sharp'' RD design, on which I focus in the main text for expositional clarity, though I derive statistical results for the fuzzy RDD case as well (see Section \ref{sec:fuzzy}). 

In addition, denote the marginal distributions of $X$ and $Y$ as $F_X, F_Y$. I assume that $\mu = E[X], \Sigma = \text{var}(X)$ and the conditional distributions $F_{X|Y}, F_{Y|X}$ exist with $\Sigma$ positive definite. Here,  $F_{Y|X}$ is a probability measure supported on the set of cdfs $\mathcal{Y}$, $F_{Y \mid X=x}(A) \coloneqq P(Y \in A \mid X=x)$, $A \subseteq \mathcal{Y}$ with $A$ measurable. 

\subsection{Motivating Examples} \label{sec:examples}

To make the setting more concrete, I now provide several prominent examples from the literature that can be viewed as R3D designs. They are instances of broader classes of settings where treatment is assigned to units at a higher level of aggregation than the outcomes.

\begin{example}[Administrative units] 
\label{ex:admin}
In an influential article, \citet{ludwig2007does} study the impact of Head Start, an early childhood education and development program, on child mortality and educational attainment. Counties above a threshold poverty rate received grant writing assistance from the federal government to develop Head Start proposals, causing a discontinuity in Head Start funding rates at the cut-off point. In an R3D setting, this discontinuity could be exploited to estimate the effect of the program on the life expectancy and test score distributions of children growing up in counties just above the cut-off point. At what ages did child mortality drop the most? Did the program's positive impact on years of schooling help all students equally, or mostly those with less years of schooling? More broadly, the R3D design applies whenever a treatment is jointly assigned to all members of an administrative unit, such as counties, school districts, or government agencies. 
\end{example}

\begin{example}[Institutions] \label{ex:institution}
 \citet{clark2009performance} considers a British reform allowing public high schools to become autonomous (directly funded by the central instead of the local government) if a majority of parents vote in favor. The paper finds large increases in examination pass rates at schools that narrowly won the vote, compared to those that narrowly lost. This can be cast as an R3D design, by considering the effect of school autonomy on the entire distribution of student test scores within a school. Does school autonomy lead to a broad-based increase in test scores, or do only the lowest-scoring students benefit? More generally, the R3D design comprises any setting where an entire institution is exposed to a treatment, but the outcome of interest affects its members. Furthermore, since vote-based allocation systems typically aggregate decisions of many individuals into higher-level outcomes, nearly all instances of the ubiquitous ``close-election'' RD design fall under the R3D framework.
\end{example}

\begin{comment}
\begin{example} \textit{Institutions}
A classical paper in the RDD literature is \citet{angrist1999using}, which exploits a rule in Israeli schools that caps class sizes at 40 students. As a result of this rule, the average class size in schools drops sharply whenever total school enrollment exceeds a multiple of 40. The paper exploits this variation to estimate the effect of decreasing class size on average student test scores. This is, in fact, a distributional RDD setting, as a decrease in class size affects the entire distribution of student test scores within a school with a given enrollment count. 
\end{example}
\end{comment}

\begin{example}[Establishments] \label{ex:establishment}
In another seminal article, \citet{card2000minimum} studied the effect of a minimum wage increase in New Jersey on wages, employment, and prices in fast food restaurants, comparing establishments on either side of the border with Pennsylvania. Since establishments typically sell many items and employ tens to hundreds of employees, one could, with the right data, observe entire distributions for each establishment. Wages and tenure (length of employment) could be measured at the employee level, and prices at the product level. Then one could answer questions such as: did the minimum wage increase mainly spur new hires, or did employment increase across the tenure distribution? Did the pass-through of the wage increase to consumers affect all products equally or mostly premium ones? More generally, the R3D design applies to any setting where the establishments are treated as a whole, but one wants to study changes to transactions within the establishment.  
\end{example}

Common to all these examples is that for any value of the running variable (distance to the border, vote share, poverty level), I observe an entire distribution of the outcome (store prices, test scores, child mortality), and these outcome distributions vary across any two units (across restaurants, schools, or counties). This implies that one needs to model the outcome as a \textit{random distribution} instead of a random variable, as discussed above. Consequently, new concepts of average treatment effects and discontinuities that are appropriate for random distributions are required, which I introduce in the next section.

\subsection{Local Average Quantile Treatment Effects}

\subsubsection{Definition} 
To begin, I define a new  treatment effects concept for distribution-valued outcomes appropriate for the setting introduced above. Following Neyman-Fisher-Rubin notation, denote $Y^0 \in \mathcal{Y}$ the counterfactual outcome distribution in the absence of treatment and $Y^1 \in \mathcal{Y}$ the outcome distribution under treatment. Define the observed outcome \[ Y = \begin{cases} Y^0 & \text{if } T = 0 \\
Y^1 & \text{if } T = 1\end{cases},\]
noting that $Y$ is a cdf so I can write $Y(y)$, $y\in \mathbb{R}$ to evaluate the function at a given point $y$. 

Consider, for a moment, the canonical RD setting, such that $Z^T \in \mathbb{R}$ the classical scalar-valued counterfactual outcome. Then, assuming that the treatment effects vary between units, the classical local treatment effect is \citep{hahn2001identification}
\[
E[Z^1 - Z^0 \mid X = 0],
\]
the conditional expectation of the jump in the outcome variable at the threshold. 

In the R3D setting, $Y^T$ is a full distribution function. An intuitive generalization of the classical average treatment effect to settings with distribution-valued outcomes is given in the below definition. Write $Q_{Y}(q)$ for the function mapping the cdf $Y$ to quantiles,
\[
Q_Y(q) \coloneqq \inf \{y \in \mathbb{R}: q \leq Y(y)\}.
\]
Then I get,
\begin{definition}[Local Average Quantile Treatment Effects (LAQTE)] \label{def:aqte}
The local average quantile treatment effects for the R3D design are,
\begin{equation} \label{eq:treatment_effects_quantile}
\begin{aligned}
\tau^{R3D}(q) & \coloneqq E\left[Q_{Y^1}(q)-Q_{Y^0}(q) \mid X=0\right] \\ 
& \coloneqq m_1(q) - m_0(q), \quad q \in [0,1].
\end{aligned}
\end{equation}
\end{definition}
Observe that the expectation is taken with respect to the conditional \textit{distribution of distributions}, $F_{Y^T\mid X=0}$,
\[
m_{T}(q) = E[Q_{Y^T}(q) \mid X=0] = \int_{\mathcal{Y}} Q_{y}(q) \dd F_{Y^T \mid X=0}(0,y), \quad T=0,1.
\]
These average quantile treatment effects (AQTE) are a compelling way to summarize random distributional treatment effects. First, they offer an intuitive generalization of average treatment effects in the Euclidean setting. In particular, they allow one to study what happens to the outcome distribution of the ``average'' unit when it crosses the cutoff and receives treatment. Moreover, as I discuss in more detail below, they are equivalent to a difference of conditional Wasserstein barycenters, which respect the intrinsic geometry of the underlying probability measures being averaged over. In particular, the distribution defined by the LAQTEs has the intuitive interpretation of being the unique distribution with the lowest possible cumulative ``least-squares'' cost of transporting its probability mass into each of the underlying distributions of the individual units. This is exactly analogous to the interpretation of the mean as the ``central tendency'' in the standard Euclidean setting, i.e. the unique quantity that has the lowest expected least-squares distance to all points.  

Next, I show that these unobserved LAQTEs can be identified from observed $(X,Y)$. 

\subsubsection{Identification}

To identify $\tau^{R3D}$ from the data, I impose two assumptions that generalize the canonical RDD requirements. 
First, I assume that the average quantile function is continuous in the running variable around the threshold.
\begin{assumpI}[Continuity] \label{asspt:i_cont}
$E[Q_{Y^T}(q) \mid X=x]$ is continuous in x for all $q \in [0,1]$, for $x \in ]\!-\!\varepsilon, \varepsilon[$, $T \in \{0,1\}$ and $\varepsilon > 0$.
\end{assumpI}
Importantly, this assumption allows for the \textit{observed} random distributions $Y$ to evolve discontinuously with $x$, like in the top left panel of Figure \ref{fig:gauss_example}.
 
The following example may help to clarify this point. Suppose $F_{Y^T|X=x} \sim N(N(g(x) + \tau T,1),1)$ for $T=0,1$, $\tau > 0$ and $supp(X)=[-1,1]$. In words, the counterfactual distribution functions $Y^T$ are drawn from a class of normal distributions with normally distributed means that depend on $X$ and shift with treatment $T$. The distributions in Figure \ref{fig:gauss_example} are an instance of this class. The figure clearly demonstrates what it means for distributions to be drawn randomly: the densities at a given value of the running variable fluctuate, leading to a lack of pointwise continuity with respect to $X$. This directly generalizes the Euclidean setting, where samples form a random point cloud that generally also lacks continuity. By contrast, \ref{fig:gauss_frechet} shows the conditional average distributions estimated on either side of the cutoff using the local polynomial approach set out in Section \ref{sec:estimator}. These average distributions are clearly continuous in the running variable. This demonstrates that even this simple collection of random Gaussian distributions satisfies the weaker continuity assumption in \ref{asspt:i_cont} but still fails continuity in quantiles. The following example establishes this formally. I include the proof here for intuition.
\begin{example} \label{prop:normal_continuity}
Suppose $(X,Y) \in (\mathbb{R}, \mathcal{Y})$ and $Y|X=x \sim N(N(g(x),1),1)$ for some continuous function $g(x)$. Then the conditional distribution of distributions $F_{Y|X}(x,y)$ satisfies assumption \ref{asspt:i_cont}, but the conditional distribution functions $Y_{X}(t,x)$ themselves are not continuous in $x$. 
\begin{proof}
For a given $t$, define a new random variable $Z=\Phi(t-W)$ where $\Phi$ the cdf of the standard normal and $W \sim N(g(x), 1)$. Then $P(Z \leq z) = P(\Phi(t-W) \leq z) = P(t-W \leq \Phi^{-1}(z))$. Since $t-W \sim N(t-g(x),1)$, we have $P(Z \leq z) = \Phi(\Phi^{-1}(z) - (t-g(x)))$ which is continuous in $x$ since $\Phi^{-1}(z)$ is constant for fixed $z \in(0,1), t-g(x)$ is continuous by continuity of $g(x)$, and $\Phi$ is continuous everywhere. Moreover, since $0 < \Phi(x) < 1$, $Y|X=x$ is almost surely not continuous in $x$. 
\end{proof}
\end{example}

The example solidifies the intuition behind Figure \ref{fig:gauss_example}. While the \textit{probability} of drawing a certain distribution varies smoothly in $X$ the actual distributions at any two points $x, x'$ close to each other will always be different with probability 1. This follows from the distributions being \textit{random objects} themselves. In Section \ref{sec:quantile_comparison} how this setting precludes the smoothness assumption used in  the classical quantile RD \citep{frandsen2012quantile}. 

The second assumption I need for identification is a standard RDD assumption which posits no manipulation and a non-zero mass of observations around the threshold.

\begin{assumpI}[Density at threshold] \label{asspt:i_dens}
$F_X(x)$ is differentiable at $c$ and $0 < \lim_{x\to c} f_X(x) < \infty$. 
\end{assumpI}

Then, I obtain the following identification result. 

\begin{lemma}[Identification] \label{lemma:ident}
Under Assumptions \ref{asspt:i_cont} and \ref{asspt:i_dens}, the unobserved $\tau^{\mathrm{R3D}}$ is identified from the joint distribution of the observed $(X,Y)$ as,
\begin{align}
\tau^{\mathrm{R3D}}(q) & = \lim_{x \to 0^+} E[Q_Y(q) \mid X=x] -  \lim_{x \to 0^-} E[Q_Y(q) \mid X=x] \\
& \coloneqq \lim_{x \to 0^+} m(q) -  \lim_{x \to 0^-} m(q) \nonumber \\
& \coloneqq m_{+}(q) - m_{-}(q), \nonumber
\end{align}
\end{lemma}
\noindent where the lemma defines $m_{\pm}(q), m(q)$.

\subsubsection{Discontinuities in Average Distributions}

The weak distributional continuity assumption \ref{asspt:i_cont} introduced above implies that the treatment has an effect when there is a discontinuity in the observed \textit{average} distribution $E[Q_Y(q) \mid X =c]$ at the threshold $X=0$. Thus, I can define a discontinuity in our setting to occur when, for some $q \in [0,1]$
\[
\lim_{x\to 0^+} E[Q_Y(q) \mid X =x] \neq \lim_{x\to 0^-} E[Q_Y(q) \mid X =x].
\]
The uniform confidence bands I derive below allow one to test for the presence of such discontinuities for a given quantile $q$. Alternatively, one can conduct inference on entire segments of the distribution at once. An overview of inference is given in Section \ref{sec:inference_overview}.

\subsection{Comparison to Quantile RDD} \label{sec:quantile_comparison}

Before developing the estimators for the LAQTEs, I briefly discuss the difference between the R3D setting and the quantile RD estimator of \citet{frandsen2012quantile}. The key insight is that quantile RDs are appropriate for estimating quantile treatment effects (QTE) for scalar-valued outcomes, while the R3D estimator can estimate (average) QTEs for distribution-valued outcomes, and there is no overlap in use cases. 

A comparison of the population quantities targeted by each estimator makes this point clearer. Let $Y \in \mathcal{Y}$ and $Z \in \mathbb{R}$ as before. The two population objects targeted are,
\[
\text{R3D}: \lim_{x \to 0^{\pm}}E[Q_Y(q) \mid X=x] \qquad \qquad \text{Q-RDD}: \lim_{x \to 0^{\pm}} E[\mathrm{1}(Z \leq z) \mid X=x].
\]
Thus, the R3D aims to estimate a conditional average quantile. The Q-RDD on the other hand, aims to estimate a fixed distribution function. Practically, they do so with the following local linear estimators, 
\[
\text{R3D}: \frac1n \sumn s_{\pm,i}(h) Q_{Y_i}(q) \qquad \qquad \text{Q-RDD}: \frac1n \sumn s_{\pm,i}(h) \mathrm{1}(Z_i \leq z). 
\]
As can be seen, the R3D approach \textit{first} estimates quantiles and only then runs a local linear regression. This properly accounts for the two-level randomness intrinsic to the R3D setting. Distribution estimation at a given $X=x$ precedes smoothing. By contrast, the Q-RDD estimator intrinsically estimates the distribution \textit{by smoothing}, ignoring the randomness \textit{within} units. In the presence of such randomness, the observed distributions will almost surely not very smoothly, and the Q-RD approach will be biased and inconsistent. 
 
Underlying these arguments are three distinct differences between the R3D and the Q-RDD setting. First, as mentioned, the sampling model imposed by the Q-RDD setting does not correctly represent the underlying data-generating process. In particular, it assumes i.i.d. sampling of scalar-valued outcomes instead of distribution-valued ones, which ignores the within-unit sampling that characterizes the R3D setting. As such, the sampling framework of the Q-RD design could never result in multiple data points having the same value of the (continuous) running variable. Second, as mentioned, the quantile continuity assumption required for the identification of the estimator in \citet{frandsen2012quantile} is highly restrictive in the R3D setting, requiring that two units that are both close to the threshold have essentially identical distributions. In the examples in Section \ref{sec:examples}, this would imply that, conditional on having the same value of the running variable, two different restaurants would have the exact same distributions of product prices, two different schools the same distribution of tests, and two different counties the same distribution of child mortality. Of course, there is no reason why the cheapest product in one restaurant should have the same price as in another, or the best student in one school the same score as in another, even if their running variables did happen to take on the same value. The estimator I propose requires a much weaker continuity assumption in \ref{asspt:i_cont}. In particular, it only demands, for example, that the test score distributions of schools near the cutoff \textit{on average} look the same, while allowing the distributions of \textit{specific} schools to differ. In this way, \ref{asspt:i_cont} is the direct distribution-valued analogue of the conditional mean continuity assumption originally imposed in \citet[A2]{hahn2001identification}, which only requires the expectation of the random outcome variable to be continuous but leaves its distribution otherwise unrestricted. Indeed, while  \ref{asspt:i_cont} is consistent with the common approach of averaging the outcome variable at the level of the aggregate  unit and then estimating a standard RD, the continuity assumption in \citet{frandsen2012quantile} is not, because there would be no random variation left in the averages, which are assumed to evolve smoothly.
Third, and similarly, the standard assumption that treatment effects vary across units automatically implies that the counterfactual distributions must be \textit{random} objects themselves: the outcome is a distribution, and receiving treatment affects this distribution differently for different units. More concretely: if a policy affects the entire workforce of a company, but does so differently at Company A compared to Company B, then \textit{even if} all untreated companies have identical distributions in the absence of treatment (an unrealistically strong assumption), the outcome distributions of those companies under treatment will still differ.

\subsection{Estimators} \label{sec:estimator}

To estimate the local average quantile treatment effects introduced above, I now propose two intuitive estimators that generalize local polynomial regression to the R3D setting with distribution-valued outcomes. The first is based on the simple idea of running local polynomial regressions on the quantile functions, separately at each quantile. The second estimator builds on this by projecting the local polynomial estimator back onto the space of quantile functions. As shown in Proposition \ref{prop:wasserstein}, the resulting estimator coincides with the local Fr\'echet regression estimator of \citet{petersen2019frechet}, restricted to the space of cumulative distribution functions equipped with the 2-Wasserstein distance (see Appendix \ref{app:frechet} for an overview). In Section \ref{sec:statistics}, I derive valid uniform confidence intervals for both approaches, though the Fr\'echet estimator is preferable due to its computational advantages, superior finite-sample performance, and its more meaningful interpretation as the ``average'' distribution.

\subsubsection{Local Polynomial Regression on Quantiles} \label{sec:estimator_simple}

\begin{figure}
\includegraphics[width=0.9\textwidth]{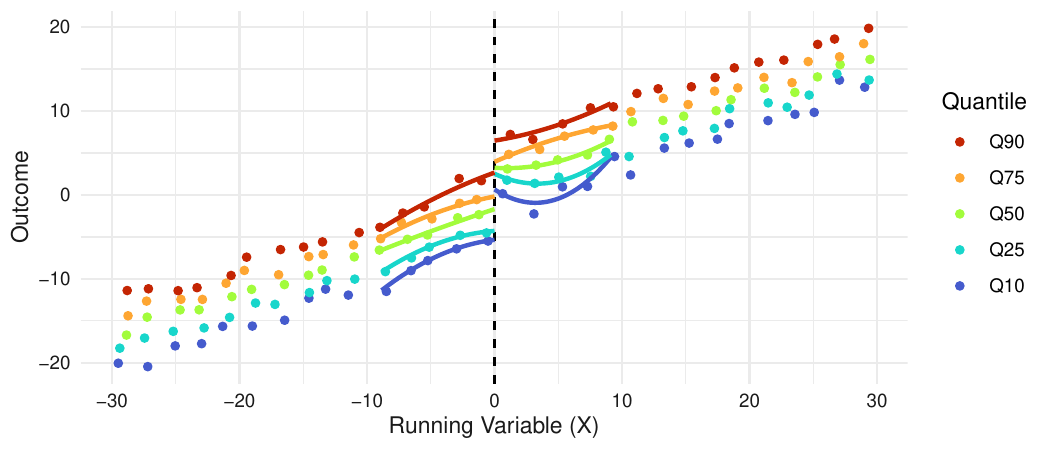}
\caption{Local Polynomial Estimator: Illustration}
\label{fig:naive}
\end{figure}

A simple and intuitive first approach to estimating the average distributions near the threshold is to treat quantiles as the fundamental unit of observation, and estimate their conditional expectations using the local polynomial regression approach that has become canon in RDD \citep{hahn2001identification}. The intuition behind the approach is illustrated in Figure \ref{fig:naive}: regression lines are fitted through data points that represent randomly scattered quantiles. 

The local polynomial R3D estimators $\hat{m}_{\pm, p}(q)$ of order $p$ for each quantile $q$ can then be written in their standard form 
\begin{align} \label{eq:local_linear}
  \hat{m}_{\pm,p}(q) & =
  \Bigl.\Bigl(\text{polynomial fit at }x = 0^{\pm}\Bigr)\Bigr|_{\text{order}=p} \\
  \hat{\boldsymbol{\alpha}}_{\pm,p}
 &  =  \underset{\boldsymbol{\alpha}\,\in\,\mathbb{R}^{p+1}}{\arg\min}
  \;\sum_{i=1}^{n}
    \delta_i^\pm \;\;
    K\!\Bigl(\tfrac{X_{i}}{h}\Bigr)\,\Bigl[
      Q_{Y_i}(q)-\boldsymbol{\alpha}^\top\,r_{p}\Bigl(\tfrac{X_{i}}{h}\Bigr)
    \Bigr]^2, \nonumber
\end{align}
where $K_h(x) \coloneqq \frac1h K(x/h)$, $\delta_i^\pm \coloneqq 1\!\bigl\{X_{i}\,\gtreqlessslant\,c\bigr\}$, and $r_p(x) \coloneqq (1, x, x^2, \ldots, x^p)$. The only difference with the standard local polynomial RDD estimator is that I now have i.i.d. samples $(Q_Y(q), X_i)$ instead of $(Y_i, X_i)$. Standard derivations give the following solution for the conditional mean estimator,
\begin{equation}
\hat{m}_{\pm,p}(q) = \sumn s_{\pm,\,i n}^{(p)}(h) Q_{Y_i}(q)
\end{equation}
where $s_{+,\,i n}^{(p)}(h)$ are the usual empirical weights for a local polynomial regression of order $p$ \citep{fan1996local}, which I derive explicitly in Appendix \ref{sec_app:poly_weights}.

Note that the estimator $\hat{m}_{\pm,p}(q)$ is technically a function of $x$, but I suppress this for all estimators to ease notation, since I only consider the cutoff point $X=0$. Further, observe that since the weights $s_{\pm,in}(h)$ can be negative, $\hat{m}_{\pm,p}$ need not be a quantile function. To resolve this, I use the standard monotone rearrangement from \citet{chernozhukov2010quantile}.

The corresponding R3D estimator then is, for each $q \in [0,1]$,
\begin{equation} \label{eq:estimator_R3D}
\hat{\tau}^{\mathrm{R3D}}_p(q) \coloneqq \hat{m}_{+,p}(q) - \hat{m}_{-,p}(q).
\end{equation}

In Section \ref{sec:statistics} below, I show that, under some assumptions, this estimator converges uniformly to an asymptotic normal distribution centered at the true treatment effect, for $p\geq 1$. Following \citet{chiang2019robust}, I build bias correction into the estimator by leveraging Remark 7 in \citep{calonico2014robust}, which establishes an equivalence between explicitly bias-corrected estimators and estimators where the MSE-optimal bandwidth is chosen based on a pilot estimator of lower order -- which is the approach I will take.

\subsubsection{Local Fr\'echet Regression}  \label{sec:estimator_frechet}

Three intuitive improvements can be made to the local polynomial regression on quantiles introduced above. First, as noted, the resulting function is not guaranteed to be a quantile function because the weights $s^{(p)}_{\pm,in}(h)$ can be negative and thus introduce non-monotonicity (quantile crossing). Second, the pointwise estimation approach ignores global function information, which degrades the estimator's finite-sample performance, as confirmed in the simulations below. Third, the pointwise estimation approach also requires repeated bandwidth selection and estimation for each quantile, leading to computational overhead. To resolve these three issues at once, I consider the following extension of the estimator,
\begin{equation} \label{eq:isotonic_regression}
\hat{m}_{\pm, \oplus, p} \coloneqq \Pi_{\mathcal{Q}}\left( \hat{m}_{\pm,p} \right) \coloneqq \argmin_{m_{\pm} \in Q(\mathcal{Y})} \: \int_{a}^b \left(\hat{m}_{\pm,p}(q) - m_{\pm}(q) \right)^2 \dd q, 
\end{equation}
where $Q(\mathcal{Y})$ is the space of quantile functions of the cdfs in $\mathcal{Y}$, restricted to $[a,b] \subseteq [0,1]$. I define $\Pi_{\mathcal{Q}}$ as the $L^2$ projection onto that space of restricted quantile functions.\footnote{Working on $[a,b]$ instead of $[0,1]$ requires much weaker assumptions on the support of the distributions and is nearly equivalent in practice, see Section \ref{sec:statistics}. } In Proposition \ref{prop:wasserstein}, I show that $\hat{m}_{\pm, \oplus, p}$ is unique and exists under the stated assumptions. 
The estimated treatment effects are then defined as,
\begin{equation} \label{eq:frechet_treatment_effects}
\hat{\tau}^{\mathrm{R3D}}_{\oplus,p}(q) \coloneqq \hat{m}_{+,\oplus,p}(q)-\hat{m}_{-, \oplus,p}(q).
\end{equation}
The augmented estimator in \eqref{eq:isotonic_regression} is an $L^2$ projection of the local polynomial estimator introduced above, with the entire function projected onto the space of quantile functions. As such, it is a form of isotonic regression \citep{Robertson1988}. Indeed, the approach can be viewed as a ``double regression'': a local linear regression on pointwise quantile functions, followed by a functional regression on quantile functions. More importantly, due to the deep connection between $L^2$ space and the 2-Wasserstein space, this extended estimator is equivalent to the local Fr\'echet regression estimator of \citet{petersen2019frechet}, restricted to the space of finite-variance probability distributions $\mathcal{Y}$ equipped with the 2-Wasserstein distance, $d_{W_2}$ (i.e. 2-Wasserstein space). In Appendix \ref{app:frechet}, I define these objects and provide an overview of local Fr\'echet regression. Here, the main thing to note is that $\hat{m}_{\pm, \oplus,p}$ converges to the same population quantile function $m_\pm(q)$ as the local polynomial estimator. This is established in Theorem \ref{thm:bootstrap_frechet} through the insight that the projection of $m_{\pm}(q)$ onto the space of quantile functions is just an identity operator, as $m_{\pm}(q)$ is a valid quantile function. Another way to view this connection is that the local Fr\'echet estimator converges to the conditional Fr\'echet mean on $(\mathcal{Y}, d_{W_2})$, which is the ``conditional Wasserstein barycenter'' \citep{agueh2011barycenters, fan2024conditional} -- the unique distribution that has a quantile function equal to the average of the quantile functions at the cutoff, i.e., $m_\pm$ (see the proof in Proposition \ref{prop:wasserstein}). In short, the Fr\'echet estimator offers a principled functional approach to estimating the LAQTE in Definition \ref{def:aqte}, while converging to the same object in population. 

This connection to local Fr\'echet regression in Wasserstein space explains why the ``double regression'' approach in \eqref{eq:isotonic_regression} is preferred over monotonizing the simple local linear estimator in \eqref{eq:local_linear}. Similar to the population object $m_{\pm}(q)$, the estimator in \eqref{eq:isotonic_regression} can be interpreted as the unique quantile function that minimizes the ``quantile least squares distance'' (the 2-Wasserstein distance) to each of the quantile functions $Q_{Y_i}$ in the space of probability distributions, weighted by the local regression weights $s^{(p)}_{\pm,in}(h)$. In other words, it is the weighted central tendency of the sample quantile functions $\{ Q_{Y_i} \}_{i=1}^n$ in probability space. This interpretation makes the projection approach in \eqref{eq:isotonic_regression} preferable over the mononotization approach of \cite{chernozhukov2010quantile}, as the quantile function resulting from the latter generally does not have this desirable interpretation. Another advantage is that local Fr\'echet regression more naturally leverages global function information by smoothing large deviations across quantiles to minimize the objective function. In comparison, the monotone rearrangement approach just sorts the quantiles but does not otherwise use the global function information to do so in any optimal manner. These superior theoretical qualities of the Fr\'echet regression approach express themselves in better finite-sample performance in the simulations in Section \ref{sec:simulations}. 

\subsection{Overview of Inference} \label{sec:inference_overview}

For these two R3D estimators (one local polynomial, one Fr\'echet),  I derive the asymptotic distribution, uniformly over $q \in [a,b]$, a compact subset of $[0,1]$, in Section \ref{sec:statistics} below. Further, I propose estimated multiplier bootstrap processes $\hat{\mathbb{G}}^{\mathrm{R3D}}$, $\hat{\mathbb{G}}^{\mathrm{F3D}}$ for the sharp and fuzzy design, respectively, that are shown to converge to the uniform limiting law and hence can be used to construct uniform confidence bands. This allows one to determine what quantiles have a statistically significant treatment effect while accounting for multiple testing due to the functional nature of the estimands. 

Moreover, the bootstrapped distributions can also be used to construct critical values for various distributional hypothesis tests. In particular, treatment nullity and homogeneity can be tested in a particular part of the distribution $[\underline{q}, \overline{q}] \subset (0,1)$ through the following tests \citep{chiang2019causal}:
\begin{table}[H]
\centering
\begin{tabular}{ll}
\hline
Test  & Test Statistic \\
\hline
Uniform Treatment Nullity
& 
\(\displaystyle \max_{q \in [\underline{q}, \overline{q}]} \sqrt{nh_n}\,\bigl|\hat{\tau}^{\mathrm{R/F3D}}(q)\bigr|\) 
\\[1em]
Treatment Homogeneity
& 
\(\displaystyle \max_{q \in [\underline{q}, \overline{q}]} \sqrt{nh_n}\,\Bigl|\hat{\tau}^\mathrm{R/F3D}(q) - \frac{1}{\overline{q} - \underline{q}}\int_{[\underline{q}, \overline{q}]} \hat{\tau}^\mathrm{R/F3D}(q')\,dq'\Bigr|\)
\\
\hline
\end{tabular}
\end{table}
\noindent where the critical values can be constructed by taking the $(1-\lambda)$-th quantiles of \\ $ \left\{ \max_{q \in [\underline{q}, \overline{q}]} \left| \hat{\mathbb{G}}^{\mathrm{R/F3D}'}(q) \right| \right\}_{b=1}^B$ and  $\left\{ \max_{q \in [\underline{q}, \overline{q}]} \left| \hat{\mathbb{G}}^{\mathrm{R/F3D}'}(q) - \frac{1}{\overline{q}-\underline{q}}\right.\right.$\linebreak[1]
$\left.\left.\int_{[\underline{q}, \overline{q}]}\hat{\mathbb{G}}^{\mathrm{R/F3D}'}(q')\,dq' \right| \right\}_{b=1}^B$ with $\lambda$ the desired level of statistical significance and $B$ the number of bootstrap repetitions.

\subsection{Extensions}

\subsubsection{Fuzzy R3D} \label{sec:fuzzy}

So far, I have focused on the sharp regression discontinuity design, where treatment assignment is a deterministic function of the cutoff. Now, I show how to define a fuzzy R3D design in which treatment assignment is a random function of the cutoff, so only a fraction of units are treated on either side of it \citep{hahn2001identification}. 

Define $T^0,$ $T^1$, the local potential treatment states as $\lim_{x \to 0^\pm} T(x)$, where $T(x)$ is the potential treatment status as a function of the running variable. Further, define the events,
\begin{itemize}
\item Compliers: $C=\left\{\omega: T^1(\omega)>T^0(\omega)\right\}$.
\item Indefinites: $I=\left\{\omega: T^1(\omega)=T^0(\omega)\right\} \backslash\left\{\omega: T^1(\omega)=T^0(\omega) \in\{0,1\}\right\}$.
\end{itemize}
The treatment effects of interest are,
\begin{definition}[Fuzzy LAQTE]
The local average quantile treatment effects for the fuzzy R3D design are,
\begin{align} \tau^{\mathrm{F3D}}(q) & :=E\left[Q_{Y^1}(q)-Q_{Y^0}(q) \mid X=0, \, C\right] \quad q \in[0,1]. \end{align}
\end{definition}

To identify these, I need the following standard additional assumptions, 
\begin{assumpI}[Fuzzy RD] \label{asspt:i_fuzzyrd}
$\lim_{x \to 0^+} P(T \mid X=x) >\lim_{x \to 0^-} P(T \mid X=x) $.
\end{assumpI}
\begin{assumpI}[Treatment Continuity] \label{asspt:i_treatment_continuity}
$E[T|X=x]$ is continuous in $x$ over $]\!-\!\varepsilon, \varepsilon[$, $\varepsilon > 0$.
\end{assumpI}

\begin{assumpI}[Monotonicity] \label{asspt:i_monotonicity}
$\lim_{x \to 0} P(T^1 > T^0 \mid X=x) = 1$ and $P(\mathrm{Indefinites}) = 0$.
\end{assumpI}
This gives,
\begin{lemma}[Fuzzy Identification] \label{lemma:fuzzy_identification}
Under Assumptions \ref{asspt:i_cont}-- \ref{asspt:i_monotonicity}, the unobserved $\mathrm{\tau}^{F3D}$ is identified from the joint distribution of the observed $(X,Y,T)$ as,
\begin{align} \tau^{\mathrm{F3D}}(q) & =\frac{\lim _{x \rightarrow 0^{+}} E\left[Q_Y(q) \mid X=x\right]-\lim _{x \rightarrow 0^{-}} E\left[Q_Y(q) \mid X=x\right]}{
\lim _{x \rightarrow 0^{+}} E\left[T \mid X=x\right]-\lim _{x \rightarrow 0^{-}} E\left[T \mid X=x\right]
} \\ & \coloneqq \frac{m_{+}(q)-m_{-}(q)}{
m_{+,T}- m_{-,T} \nonumber
}. \end{align}
\end{lemma}
This Wald estimator takes the same form as the standard fuzzy RDD one \citep{hahn2001identification}, except that the outcomes are random quantiles. Note that I can work with this simpler form compared to \citet[Eq. 2-3]{frandsen2012quantile} because I work directly with the random quantiles and hence do not need to invert the CDFs on each side.

The corresponding treatment effect estimator, using local polynomial regressions of order $p$, then becomes,
\begin{equation} \label{eq:estimator_F3D}
\hat{\tau}^{\mathrm{F3D}}_p(q) \coloneqq \frac{\hat{\tau}_p^{R3D}(q)}{\hat{m}_{+,T,p} - \hat{m}_{-,T,p}}
\end{equation}
where 
\begin{equation*}
\hat{m}_{ \pm, T, p} \coloneqq \sum_{i=1}^n s_{ \pm, i n}^{(p)}(h) T_i.
\end{equation*}

The corresponding Fr\'echet estimator is,
\begin{equation}
\hat{\tau}^{\mathrm{F3D}}_{\oplus, p}(q) \coloneqq \frac{\hat{\tau}_{\oplus,p}^{R3D}(q)}{\hat{m}_{+,T,p} - \hat{m}_{-,T,p}}.
\end{equation}

\begin{comment}
TODO
\subsubsection{Discrete Distributions}

\nddvd{Discrete distributions with finite support -- get pointwise convergence of empirical quantile function, and can use standard pointwise convergence of local linear estimator. However for ordinal variables that aren't interval variables, need to do regression on CDFs -- use 1-Wasserstein distance.}
\end{comment}

\begin{comment}
\subsubsection{Local Randomization Framework}
\end{comment}

\subsubsection{Empirical Quantile Functions} \label{sec:empirical_cdfs}

So far, I have assumed that the researcher observes entire quantile functions $Q_{Y_i}$. This is realistic in settings where an entire population of sub-units within a given aggregate unit is observed -- for example, when a researcher has access to the census so that all firms within a US county are in the data. In practice, however, there is often another sampling layer, where one only observes further i.i.d.~\emph{samples} $Z_{ij}, j=1,\ldots,n_i$ from these distribution functions, with $Z_{ij} \in \mathbb{R}$ distributed according to $Y_i$.\footnote{See \citet{chen2023wasserstein} for an analogous setting in the context of distribution-on-distribution regression, and \citet{zhou2024wasserstein} in a similar setting.} 
In Section \ref{sec:statistics}, I show that under standard assumptions, the empirical quantile functions converge to the true quantile functions faster than the R3D estimators and hence do not affect the asymptotic results. The corresponding sharp RD estimator is defined as,
\begin{equation}
\bar{\mu}_{\pm,p}(q) \coloneqq \frac1n \sumn s^{(p)}_{\pm,in}(h) \widehat{Q}_{Y_i}(q),
\end{equation}
where 
\begin{equation} \label{eq:empirical_quantile}
\widehat{Q}_{Y_i}(q) \coloneqq \inf\left\{ x : \widehat{Y}_i(x) \geq q\right\} 
\end{equation}
with 
\[
\widehat{Y}_i(x) \coloneqq \frac{1}{n_i} \sum_{j=1}^{n_i} \mathrm{1}\left( Z_{ij} \leq x \right),
\]
the empirical distribution function. The other estimators are similarly modified by plugging in $\widehat{Q}_{Y_i}$, and denoted with a bar instead of a hat, e.g.\ $\bar{\tau}_p^{\mathrm{R3D}}$. Sampling weights can be incorporated by constructing $\widehat{Q}_{Y_i}$ as weighted quantile functions. The asymptotic results for this setting are established in Section \ref{sec:empirical_cdf_stats}.

\section{Statistical Results} \label{sec:statistics}

In this section, I derive the asymptotic distributions of the local polynomial and Fr\'echet regression estimators. I do so in full generality for $p$-th order local polynomials, accommodating both the sharp and fuzzy RDD setting. The results for the local polynomial estimator follow from an application of the general results in \citet{chiang2019robust}, extended to random distribution-valued outcomes. The corresponding results for the local Fréchet regression follow from the functional delta method and a projection argument that uses a version of Rademacher's theorem for Banach spaces \citep{preiss2014gateaux}. I conclude by extending these asymptotic results to the empirical quantile setting.

\subsection{Assumptions}

Throughout, I work on the restricted set of quantiles $[a,b]$, a compact subset of $(0,1)$, and let $\underline{c} < 0 < \overline{c}$. Also, define $\mathcal{Y}_c \coloneqq \{ Y(\omega): \omega \in \Omega^x, X(\omega) \in[\underline{c}, \bar{c}]\}$ as the set of random cdfs that are realized in a small neighborhood around the cutoff. 

\begin{assumpK}[Kernel] \label{asspt:kernel}
The kernel $K$ is a continuous probability density function, symmetric around zero, and non-negative valued with compact support. 
\end{assumpK}

\begin{assumpK}[Bandwidth] \label{asspt:bandwidth_chiang}
The bandwidths satisfy $h_{1}(q)=c_1(q) h$ and $h_2(q) = c_2(q) h$ for $c_1(q) : [a,b] \to [\underline{c}$ where $\bar{c}] \subset \mathbb{R}$ a bounded Lipschitz function and $c_2(q) = \bar{c}_2 > 0$. The baseline bandwidth $h=h_n$ satisfies $h \to 0$, $n h^2 \to \infty$, $nh^{2p+3} \to 0$.
\end{assumpK}

\begin{assumpL}[Sampling] \label{asspt:sampling} 
\begin{enumerate}
\item[(i)] \label{asspt:sampling_population}
 $\{ (Y_i, T_i, X_i) \}_{i=1}^n$ are i.i.d. copies of a random element $(Y,T,X)$ defined on a probability space $(\Omega^x, \mathcal{F}^x, P^x)$.
 \item[(ii)] $ \{ Z_{ij} \}_{j=1}^{n_i}$ are i.i.d. draws from the random distribution $Y_i$ for each $i=1,\ldots,n$.
\end{enumerate}
\end{assumpL}

\begin{assumpL}[Average Quantile Continuity]\label{asspt:average_quantile_smoothness} 
For each $x\in \mathcal{N} \coloneqq (-\epsilon, \epsilon)$ for $\epsilon > 0$ and $q\in [a,b] \subset (0,1)$, the following conditions hold:
\begin{enumerate}
\item[(i)] 
The maps $(x,q)\to E[Q_{Y}(q) | X=x]$ and $x \to E[T|X=x]$ are $p$-times continuously differentiable in $x$ on $\mathcal{N}$, 
with all partial derivatives (up to order $p$) Lipschitz in $x$ on $\mathcal{N}\times[a,b]$.
\item[(ii)] 
For any fixed $q_1, q_2 \in [a,b]^2$, the map $x \to E\bigl[Q_{Y}(q_{1}) Q_{Y}(q_{2}) \mid X=x\bigr] \in C^1(\mathcal{N} \setminus \{0\})$ with bounded derivatives in $x$ and bounded limits as $x \to 0^{\pm}$.
\end{enumerate}
\end{assumpL}

\begin{assumpL}[Random Quantile Spread] \label{asspt:quantile_spread}
The maps $Y \to \sup_{q \in [a,b]} |Q_Y(q)|$ and  \\
$x \to \sup_{q \in [a,b]} \left| E[Q_Y(q) \mid X=x] \right| $ are in $L^{2+\epsilon}(P^x)$ on $[\underline{c}, \overline{c}] \times \mathcal{Y}_c$.
\end{assumpL}

\begin{assumpM}[Multiplier] \label{asspt:multiplier} $\{ \xi_i \}_{i=1}^n$ is an independent standard normal random sample defined on a probability space $(\Omega^\xi, \mathcal{F}^\xi, P^\xi)$ independent of $(\Omega^x, \mathcal{F}^x, P^x)$.
\end{assumpM}

\ref{asspt:kernel} is a standard kernel assumption and is satisfied by the commonly used triangular and uniform kernels. 
\ref{asspt:bandwidth_chiang} is a standard bandwidth assumption, with the important benefit that for local polynomial order $p > 1$, it accommodates the bandwidth rates implied by common bandwidth selection procedures, which are typically slower than $h=n^{-1/5}$ \citep{calonico2014robust}. Moreover, the assumption accommodates quantile-specific bandwidths. \ref{asspt:average_quantile_smoothness} (i)  
is a stronger version of the standard continuity assumption \ref{asspt:i_cont} that ensures the Taylor expansions required for local polynomial regression of order $p$ are well-defined.  \ref{asspt:average_quantile_smoothness} (ii) further provides some minimal control over the functional objects $E[Q_Y(q) | X=x]$ through the covariance of the quantiles. Note that both (i) and (ii) are implied by the much stronger assumption that the random distribution $F_{Y|X}$ evolves smoothly, which would be the random-distribution equivalent of Assumption E1 in \citet{frandsen2012quantile} and is imposed in \citet{petersen2019frechet}. For Assumption \ref{asspt:quantile_spread}, first note that clearly, for every $Y$, there exists an $M_Y$ such that $\sup_{q \in [a,b]} Q_Y(q) < M_Y$. However, the assumption strengthens this point-wise fact into a statement that these caps cannot `blow up' too often in all possible realizations $Y$. In practice, this means that while each $Y$ can have unbounded support, the family $\mathcal{Y}$ must not produce extremely large quantiles too often around the cutoff. As such, the assumption controls the across-distribution variance, enabling uniform statistical arguments.
 Finally, Assumption \ref{asspt:multiplier} is a standard assumption for multiplier bootstraps that can easily be satisfied in practice.

For the extension with empirical quantile functions, I further impose the following,

\begin{assumpE}[Empirical Quantiles] \label{asspt:E_convergence}
Any distribution function $Y \in \mathcal{Y}_c$ either
\begin{enumerate}[(i)] 
\item has compact support and is $C^1$ with strictly positive density, or
\item has infinite support and for every $0<q_1<q_2<1$ there exists an $\varepsilon>0$ such that $Y$ is continuously differentiable on the interval $\left[Q_{Y}\left(q_1\right)-\varepsilon, Q_{Y}\left(q_2\right)+\varepsilon\right]$ with strictly positive density.
\end{enumerate}
\end{assumpE}

\begin{assumpE} \label{asspt:E_n}
There exists a sequence $m = n^{\gamma}, \gamma \geq 1$ such that $\min\{n_i : i=1,\ldots,n\} \geq m$. Moreover, the sample sizes for each $Z_{ij}$ are asymptotically balanced, i.e. $\frac{n_i}{n_j} \to \eta_{ij}$ with $0 <\eta_{ij} < \infty$ for $i,j \in \{1,\ldots,n\}$.
\end{assumpE}

\ref{asspt:E_convergence} implies that the empirical quantile functions converge uniformly \citep[Corollary 21.5]{van2000asymptotic}. The assumption can be relaxed for the case with discrete distributions and finite support, since then standard pointwise convergence results for local linear regression imply uniform convergence \citep{fan1996local}. Also note that if the entire distribution \emph{is} observed, then these assumptions are not required and the quantile functions are allowed to have discontinuities.
\ref{asspt:E_n} is a weak assumption on the number of measurements per distribution that guarantees the empirical quantile functions will converge faster than the estimators \eqref{eq:local_linear} and \eqref{eq:isotonic_regression}.

\subsection{Asymptotic Distribution}

\subsubsection{Local Polynomial Estimator} \label{sec:asymptotic_local_polynomial}
Under these assumptions, I can derive the asymptotic distribution of the local polynomial estimator. For that, I need a few additional pieces of notation, borrowing from \citet{chiang2019robust}. For the formal results, I assume without loss of generality that the kernel $K$ is supported on $[-1,1]$. Define $g_1: (Y, T, q) \subset (\mathcal{Y}, \{0,1\}, [a,b]) \to Q_Y(q)$, $g_2 : (Y, T, q) \subset  (\mathcal{Y}, \{0,1\}, [a,b]) \to T$. Further, define the population residual $\mathcal{E}_k(y, t, x, q) \coloneqq g_k(y, t, q) - E[g_k(y, t, q) | X_i=x], \: k={1,2}$ and let 
\[ \sigma_{k l}(q, q' \mid x)=E\left[\mathcal{E}_k\left(Y_i, T_i, X_i, q\right) \cdot \mathcal{E}_l\left(Y_i, T_i, X_i, q'\right) \mid X_i=x\right] \] 
with $k,l \in \{1,2\}$, $q,q' \in [a,b]$, and $\sigma_{kl}(q, q' \mid 0^{\pm}) = \lim_{x \to 0^{\pm}} \sigma_{k l}(q, q' \mid x)$.
Moreover, let $e_0$ denote the $0$th standard basis vector of $\mathbb{R}^p$, $(1, 0, \ldots ,0)$, and write $\Gamma_{\pm, p} \coloneqq \int_{\mathbb{R}_{\pm}} K(u) r_p(u) r_p'(u) \dd u$. Also, let $X_n \leadsto X$ denote weak convergence for some sequence of random variables $X_n$ and a random variable $X$, while $X_n \stackrel[\xi]{p}{\leadsto} X$ denotes conditional weak convergence. The latter is defined as $\sup_{h \in BL_1} \left| E_{\xi \mid x}\left[ h(X_n) - E[h(X)] \right] \right| \stackrel[x]{p}{\to} 0$ where $BL_1$ the set of bounded Lipschitz functions with supremum norm bounded by 1 and $\stackrel[x]{p}{\to}$ denotes convergence in probability with respect to probability measure $P^x$ \citep[\S 1.13]{vaart1996weak}. Then, I first get the following preliminary result for the conditional means.
\begin{theorem}[Convergence: Conditional Means] \label{thm:asymptotic_conditional_means}
    Under Assumptions \ref{asspt:i_dens}, \ref{asspt:kernel}, \ref{asspt:bandwidth_chiang}, \ref{asspt:sampling}-(i), \ref{asspt:average_quantile_smoothness}, \ref{asspt:quantile_spread},
    $$
    \sqrt{nh} \left[
    \begin{array}{c} \hat{m}_{\pm,p} - m_\pm  \\  \hat{m}_{\pm, T,p} - m_{\pm, T} \end{array}\right] 
    \leadsto 
     \left[
    \begin{array}{c} 
    c_1(\cdot)^{-1/2} \mathbb{G}_{H\pm}(\cdot,1) \\ 
     c_2(\cdot)^{-1/2} \mathbb{G}_{H\pm}(\cdot,2)
    \end{array}\right] 
    $$
    where $\mathbb{G}_{H \pm} : \Omega^x \to l^{\infty}\left( [a,b] \times \{ 1,2\}\right)$ is a zero-mean Gaussian process with covariance function,
\[
H_{\pm, p}\left((q, k), (q',l)\right) = \frac{
\sigma_{kl}( q, q' | 0^{\pm}) e'_0 \left(\Gamma_{\pm,p}\right)^{-1} \Psi_{\pm,p}\left((q,k),(q',l)\right) \left( \Gamma_{\pm,p}^{-1} \right) e_0
}{
\sqrt{ c_k(q) c_l(q')  } f_{X}(0)
},
\]
where,
\[
\Psi_{\pm,p}\left((q,k),(q',l)\right) \coloneqq \int_{\mathbb{R}} r_p(u  /c_k(q)) r'_p( u / c_l(q')) K(u / c_k(q )) K(u / c_l(q')) \dd u
\]
for each $q, q' \in [a,b]$. 
\end{theorem}

Then, a simple application of the functional delta method yields the following result. 

\begin{theorem}[Convergence: Treatment Effect] 
\label{thm:asymptotic_treatment_effects}
Under the assumptions of Theorem~\ref{thm:asymptotic_conditional_means} it follows that, 
\begin{align*}
& \sqrt{n h}\bigl(\hat{\tau}_p^{\mathrm{R3D}} - \tau^{\mathrm{R3D}}\bigr)
  \leadsto c_1(\cdot)^{-1/2}\,\mathbb{G}_\Delta(\cdot, 1) \coloneqq \mathbb{G}^{\mathrm{R3D}},  \\
\intertext{and under the additional Assumptions \ref{asspt:i_fuzzyrd}--\ref{asspt:i_monotonicity}}
& \sqrt{n h}\bigl(\hat{\tau}_p^{\mathrm{F3D}} - \tau^{\mathrm{F3D}}\bigr)  \\ 
  & \quad \leadsto  \frac{
    c_1(\cdot)^{-1/2}(m_{+,T}(\cdot)-m_{-,T}  (\cdot))\,\mathbb{G}_\Delta(\cdot,1)
    - c_2(\cdot)^{-1/2}(m_{+}(\cdot)-m_{-}(\cdot))\,\mathbb{G}_\Delta(\cdot,2)
  }{(m_{+,T}(\cdot)-m_{-,T}(\cdot))^2},    \\
  & \quad \coloneqq \mathbb{G}^{\mathrm{F3D}} 
\end{align*}
\noindent where, for $k \in \{1,2\}$,
\[
\mathbb{G}_\Delta(\cdot, k) \coloneqq \mathbb{G}_{H+}(\cdot, k) -  \mathbb{G}_{H-}(\cdot, k),
\]
and $\mathbb{G}_{H\pm}(\cdot,k)$ are as defined in Theorem~\ref{thm:asymptotic_conditional_means}.
\end{theorem}

In practice, it is easier to approximate the limiting processes in Theorem \ref{thm:asymptotic_treatment_effects} with a multiplier bootstrap, which preserves the local structure without full resampling. To that end, I use the pseudo-random samples $\{ \xi_i \}_{i=1}^n$ defined in \ref{asspt:multiplier} to define the estimated multiplier process,
\begin{equation} \label{eq:EMP}
\hat{\nu}^\pm_{\xi, n}(q, k) = \sumn 
\xi_i \frac{e_0^{\prime}\left(\Gamma_{\pm, p}\right)^{-1} \hat{\mathcal{E}}_k\left(Y_i, T_i, X_i, q\right) r_p\left(\frac{X_i}{h_{k}\left(q\right)}\right) K\left(\frac{X_i}{h_{k}\left(q\right)}\right) \delta_i^{ \pm}}{\sqrt{n h_{k}\left(q\right)} \hat{f}_X(0)},
\end{equation}
where $\hat{f}_X(0)$ is any uniformly consistent estimator of $f_X(0)$, and $\hat{\mathcal{E}}_k\left(Y_i, T_i, X_i, q \right)$ is any uniformly consistent first-stage estimator of the residual $\mathcal{E}_k$. In practice, I will use the first-stage estimator proposed in \citet[A.6]{chiang2019robust}, described in detail in Appendix \ref{app:first_stage}. The process $\hat{\nu}^\pm_{\xi, n}(q, k)$ is an estimator for the uniform Bahadur representation of the bias-corrected processes $\hat{m}_{\pm,p}(q) - m_{\pm}(q)$, $\hat{m}_{\pm,T,p}(q) - m_{\pm,T}(q)$  \citep{chiang2019robust}, see the proof of Theorem \ref{thm:asymptotic_conditional_means} for more details. Then, I obtain the uniform validity of the multiplier bootstrap,
\begin{theorem}[Bootstrap] \label{thm:bootstrap}
Under the Assumptions of Theorem \ref{thm:asymptotic_conditional_means} and Assumption \ref{asspt:multiplier} it follows that $\hat{\nu}_{\xi,n}^\pm \stackrel[\xi]{p}{\leadsto} \mathbb{G}_{H^\pm}$ and thus,
\begin{align*}
&  \hat{\mathbb{G}}^{\mathrm{R3D}}(\cdot) \coloneqq c_1(\cdot)^{-1/2} \hat{\nu}_{\Delta, n}(\cdot, 1) \stackrel[\xi]{p}{\leadsto} \mathbb{G}^{\mathrm{R3D}} \\ 
\intertext{and under the additional Assumptions 
  \ref{asspt:i_treatment_continuity}, \ref{asspt:i_monotonicity},}
& \hat{\mathbb{G}}^{\mathrm{F3D}}\left(\hat{m}_{+,p}, \hat{m}_{-,p}, \hat{m}_{+,T,p}, \hat{m}_{-,T,p}\right)(\cdot) \coloneqq \\ 
&  \quad \frac{ c_1(\cdot)^{-1/2}\left(\hat{m}_{+,T,p}(\cdot)-\hat{m}_{-,T,p}(\cdot)\right)\hat{\nu}_{\Delta, n}(\cdot, 1)-c_2(\cdot)^{-1/2}\left(\hat{m}_{+,p}(\cdot)-\hat{m}_{-,p}(\cdot)\right)\hat{\nu}_{\Delta, n}(\cdot, 2)
}{
\left(\hat{m}_{+, T,p}(\cdot)-\hat{m}_{-, T,p}(\cdot)\right)^2
}\\[5pt]
&\hspace{5cm} \stackrel[\xi]{p}{\leadsto} \mathbb{G}^{\mathrm{F3D}} 
\end{align*}

where, for $k \in \{1,2\}$,
\[
\hat{\nu}_{\Delta, n}(\cdot, k) \coloneqq \hat{\nu}^+_{\xi, n}(\cdot, k) - \hat{\nu}^-_{\xi, n}(\cdot, k).
\]
\end{theorem}

A practical algorithm for computing the empirical bootstrap is provided in Appendix \ref{app:bootstrap}. The asymptotic validity and consistency of the tests proposed in Section \ref{sec:inference_overview} follow immediately from Theorem \ref{thm:bootstrap}.

\subsubsection{Fr\'echet Estimator}

Turning to the Fr\'echet estimator, I now show that it has the same asymptotic distribution as the local polynomial estimator. I include a proof sketch to explain the intuition behind this striking result. 

\begin{theorem}[Convergence: Conditional Fr\'echet Means] \label{thm:convergence_frechet}
Under the Assumptions of Theorem \ref{thm:asymptotic_conditional_means}, 
\[ \sqrt{n h}
\left( \hat{m}_{ \pm, \oplus, p} -  m_{\pm}\right) \leadsto  \mathbb{G}_{H \pm}(\cdot, 1),
\]
where $\mathbb{G}_{H \pm}$ is the \emph{same} zero-mean Gaussian process as in Theorem \ref{thm:asymptotic_conditional_means}.
\begin{proof}[Proof sketch.]
The result obtains by the fact that the Fr\'echet estimator is the projection of the local polynomial estimator onto the space of quantile functions $\Pi_{\mathcal{Q}}$, but this projection is only active in finite sample. The limit of the local polynomial estimator is the conditional average quantile function, which is a valid quantile function of its own right, and hence the projection simply becomes the identity function in the limit, barring some slight intricay on the boundaries of $\mathcal{Q}$. The projection operator $\Pi_{\mathcal{Q}}$ is a metric projection onto convex sets and hence well-known to be globally  Lipschitz \citep[Prop. 4.16]{bauschke2017}. Then, by a generalization of Rademacher's theorem to Banach spaces, $\Pi_{\mathcal{Q}}$ is Hadamard differentiable except on a special set that is directionally porous, hence negligible in the sense of \citet{preiss2014gateaux}. Since the local polynomial estimator $\hat{m}_{\pm,p}$ converges to the true quantile function $m_{\pm}$ in $L^2$ norm (an implication of Theorem \ref{thm:asymptotic_conditional_means}), it almost surely avoids this exceptional set in large samples. Thus, the Hadamard derivative of $\Pi_{\mathcal{Q}}$ evaluated at the true limit $m_{\pm}$ coincides with the identity map. Consequently, by the functional delta method, the projected estimator inherits the same asymptotic distribution as the original local polynomial estimator.
\end{proof}
\end{theorem}

Note that in this theorem the $c_1(\cdot)$ term does not appear because the Fr\'echet estimator uses a single bandwidth for all quantiles. 
Then, the following result again follows by a simple application of the functional delta method. 

\begin{corollary}[Convergence: Fr\'echet Treatment Effects] \label{corollary:frechet_treatment}
Under the assumptions of Theorem~\ref{thm:asymptotic_conditional_means} it follows that, 
\begin{align*}
 \sqrt{n h}\bigl(\hat{\tau}_{\oplus,p}^{\mathrm{R3D}}- \tau^{\mathrm{R3D}}\bigr)
  & \leadsto \mathbb{G}^{\mathrm{R3D}},  \\
\intertext{and under the additional Assumptions \ref{asspt:i_fuzzyrd}--\ref{asspt:i_monotonicity},}
 \sqrt{n h}\bigl(\hat{\tau}_{\oplus,p}^{\mathrm{F3D}}- \tau^{\mathrm{F3D}}\bigr) &  \leadsto  \mathbb{G}^{\mathrm{F3D}} 
\end{align*}
\end{corollary}

\begin{corollary}[Bootstrap: Fr\'echet] \label{thm:bootstrap_frechet}
Under the assumptions of Theorem \ref{thm:bootstrap}, the estimated bootstrap processes $\hat{\mathbb{G}}^{\mathrm{R3D}}$ and $\hat{\mathbb{G}}^{\mathrm{F3D}}\left(\hat{m}_{+,\oplus,p}, \hat{m}_{-,\oplus,p}, \hat{m}_{+,T,p}, \hat{m}_{-,T,p}\right)$ deliver asymptotically valid confidence intervals for the Fr\'echet estimators $\hat{\tau}_{\oplus,p}^{\mathrm{R3D}}$, $\hat{\tau}_{\oplus,p}^{\mathrm{F3D}}$.
\end{corollary}

\subsection{Extensions}

\subsubsection{Empirical Distribution Functions}
\label{sec:empirical_cdf_stats}

\begin{proposition}[Empirical Quantiles] \label{prop:empirical_quantiles}
Under the same respective Assumptions of Theorems \ref{thm:asymptotic_conditional_means} and \ref{thm:asymptotic_treatment_effects}, as well as Assumptions \ref{asspt:E_convergence}, \ref{asspt:E_n}, the estimators with empirical quantile functions, $\bar{m}_{\pm,p}$, $\bar{m}_{\pm, \oplus,p}$, $\bar{\tau}^{\mathrm{R3D}}_p$, $\bar{\tau}^{\mathrm{F3D}}_p$, $\bar{\tau}^{\mathrm{R3D}}_{\oplus, p}$, $\bar{\tau}_{\oplus,p}^{\mathrm{F3D}}$ converge to the same uniform limiting processes as their respective population analogs.
\end{proposition}

\begin{corollary}[Bootstrap: Empirical Quantiles] Under the assumptions of Theorem \ref{thm:bootstrap}, along with Assumptions \ref{asspt:E_convergence}, \ref{asspt:E_n}, the estimated bootstrap processes $\hat{\mathbb{G}}^{\mathrm{R3D}}$ and $\hat{\mathbb{G}}^{\mathrm{F3D}}(\cdot, \cdot, \cdot, \cdot)$ (with the appropriate conditional mean estimators plugged in) deliver asymptotically valid confidence bands for the treatment effect estimators with empirical quantile functions, $\bar{\tau}_{p}^{\mathrm{R3D}}$, $\bar{\tau}_{p}^{\mathrm{F3D}}$, $\bar{\tau}_{\oplus,p}^{\mathrm{R3D}}$, $\bar{\tau}_{\oplus,p}^{\mathrm{F3D}}$.
\end{corollary}

\begin{comment}
\subsubsection{Discrete Distributions}
\end{comment}

\begin{comment}
\subsubsection{Local Randomization Framework}
\end{comment}

\section{Empirical Applications}

\subsection{Simulations} \label{sec:simulations}

To evaluate the proposed estimators’ performance, I conduct Monte Carlo simulations under several data-generating processes. Throughout this and the next section, I use R3D estimators of quadratic order but with bandwidths that are MSE-optimal for the linear estimators. As argued in Remark 7 \citet{calonico2014robust}, this is equivalent to using explicitly bias-corrected linear estimators. 

In the simulations, I estimate the quantile treatment effects $\tau^{\mathrm{R3D}}$ at 10 quantiles using three estimators: 1) a local polynomial estimator for classical quantile RDDs \citep{qu2019uniform};\footnote{Computed using the \texttt{rd.qte} command in R \citep{qu2024qte}.}  2) the local polynomial R3D estimator in Section \ref{sec:estimator_simple}; 3) the (Fr\'echet) R3D estimator in Section \ref{sec:estimator_frechet}. The Q-RD estimator is corrected for bias using the approach in \citep{qu2024inference}. The reason for using the Q-RD estimator of \citet{qu2019uniform} is to give Q-RD the best possible chance, since this estimator allows for bias-corrected, uniform inference, improving on the estimator in \citet{frandsen2012quantile}.

I consider two data-generating processes, where $X_i \sim \mathrm{Uniform}(-1,1)$. \\
\noindent
\textit{DGP 1: Normal with Normal Means}. For each \(i\), draw
\begin{align}
\mu_i & \sim N\!\Bigl(5+ 5 X_i + \delta^+\,\Delta,\;1\Bigr), \label{eq:dgp1} \\
\sigma_i & \sim \bigl|\,N\!\bigl(1 +  \,X_i , \;1 \bigr)\bigr|, \nonumber
\end{align}
and define \(Y_i = N(\mu_i,\,\sigma_i^2)\).

\medskip

\noindent
\textit{DGP 2: Normal--Exponential Mixture with Normal--Exponential means.} Set \(\mu_i = \mathrm{Uniform}(-5,5) + 2\,X_i\) and \(\lambda_i = \mathrm{Uniform}(0.5,1.5)\). Then, generate
\begin{equation} \label{eq:dgp2}
Y_i = N\!\bigl(\mu_i + \delta^+\,\Delta,\;1\bigr)+ 2\,\mathrm{Exp}\!\bigl(\lambda_i + \delta^+\,\Delta_\lambda\bigr).
\end{equation} In both setups, I let \(\Delta\) vary across different simulations to test different treatment effect magnitudes. For the first DGP, the true treatment effects have the closed-form solution $N(\Delta, 2)$, implying constant treatment effects. The heterogeneous treatment effects in the second DGP are estimated by averaging across a large number of simulated quantile functions.

\begin{figure}[h!]
\centering
\begin{subfigure}[b]{0.95\textwidth}
\includegraphics[width=0.9\textwidth]{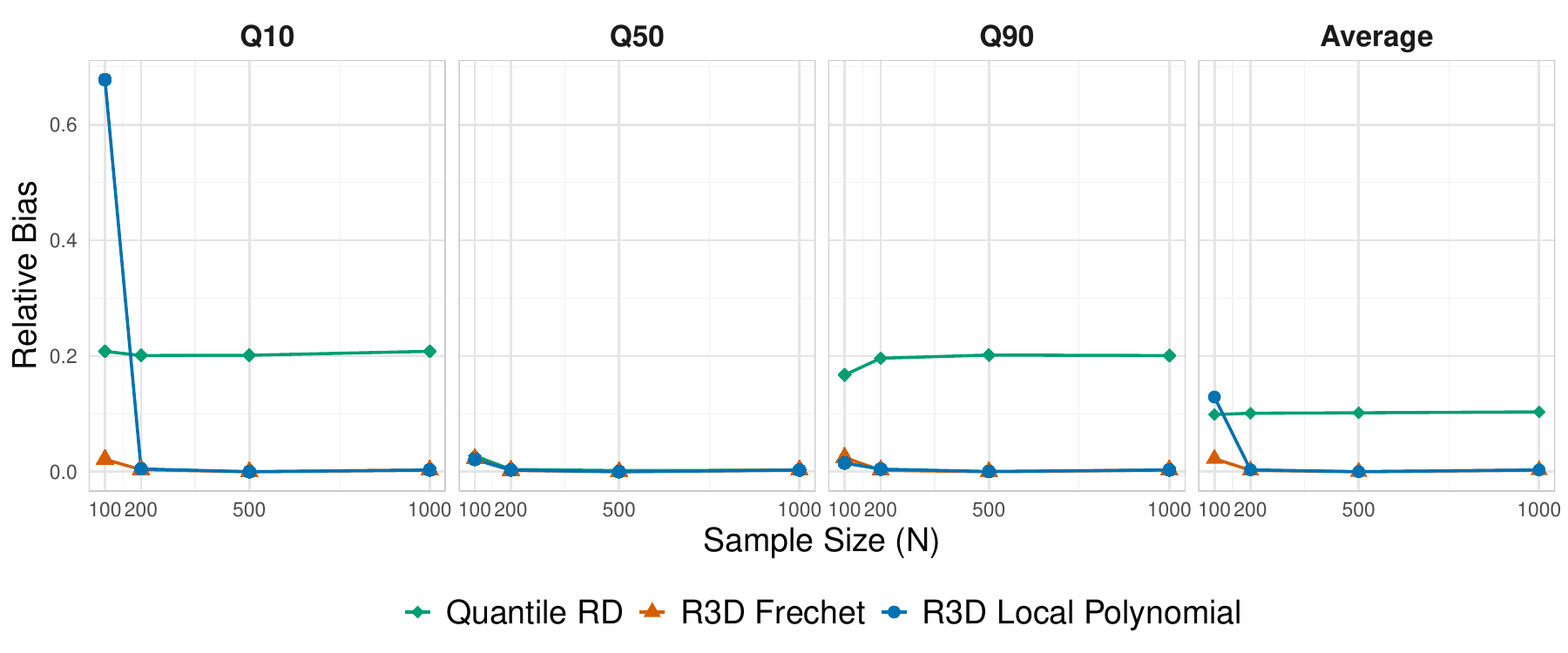}
\caption{Normal Distributions with Normal Means}
\end{subfigure}
\begin{subfigure}[b]{0.95\textwidth}
\includegraphics[width=0.9\textwidth]{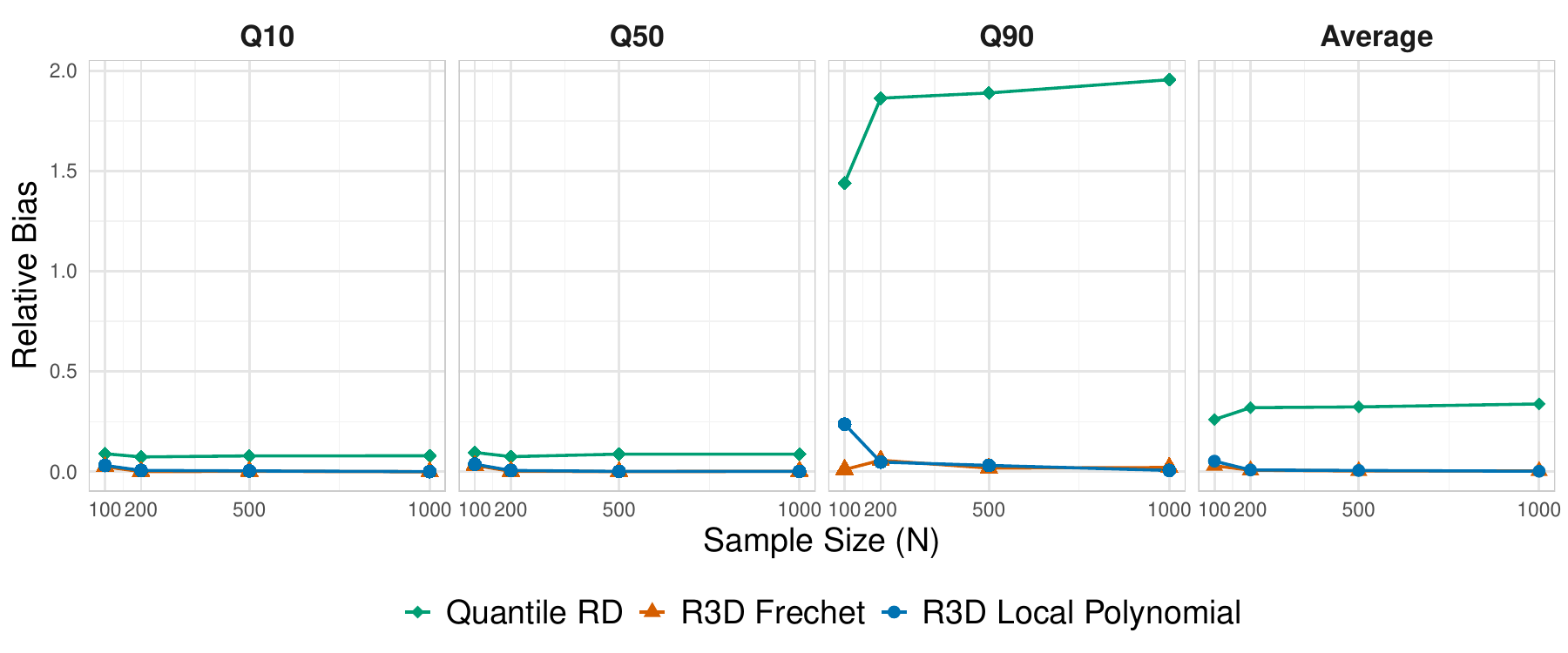}
\caption{Normal--Exponential Distributions with Normal--Exponential Means}
\end{subfigure}
\caption{Simulated Bias of R3D and Q-RD Estimators}
\label{fig:sims_bias}
\floatfoot{\textit{Note}: figure compares relative bias (absolute bias as a percent of the true treatment effect size) of R3D and Q-RD estimators. Each measure is reported for $n=n_i=200, 500, 1000,$ and $2000$ (x axis) at quantiles 10, 50, 90, and the average over all quantiles (facets). Results are averaged over $2,500$ simulations for each sample size. The methods are: 1) a local polynomial estimator for classical quantile RDDs \citep{qu2019uniform} with bias correction \citep{qu2024inference}; 2) the local polynomial R3D estimator in Section \ref{sec:estimator_simple}; 3) the Fr\'echet R3D estimator in Section \ref{sec:estimator_frechet}. Bandwidths are selected using (I)MSE-optimal procedure in Section \ref{sec_app:bandwidth}. Data-generating process: outcome variable $Y$ is a normal distribution with normally distributed means and variances that depend on running variable $X$ and jump across the threshold.}
\end{figure}

Figure \ref{fig:sims_bias} shows the estimators' performance in terms of relative bias, which is the magnitude of the estimated bias at a given quantile as a proportion of the treatment effect at that quantile. I set $\Delta = 2$ but the results are similar for other values. The green line (diamonds) shows the quantile RD estimator, the orange line (triangles) the Fr\'echet estimator, and the blue line (circles) the local polynomial one. In line with theoretical expectations, the quantile RD estimator appears to be inconsistent and suffers from large finite sample bias, with a relative bias that is at least an order of magnitude higher than the R3D estimators', for some quantiles. As expected, the quantile RD estimator performs well at the median in DGP 1, because a mixture of normals approximates the average normal at the median. Similarly, due to the heavy tails of the exponential distribution, it performs much worse at the upper quantiles in DGP 2. Between the two R3D estimators, the Fr\'echet estimator has much lower bias than the local polynomial one for small sample sizes, but both converge quickly to near-zero, supporting the asymptotic theory.

\begin{figure}[ht!]
\centering
\begin{subfigure}[b]{0.95\textwidth}
\includegraphics[width=0.9\textwidth]{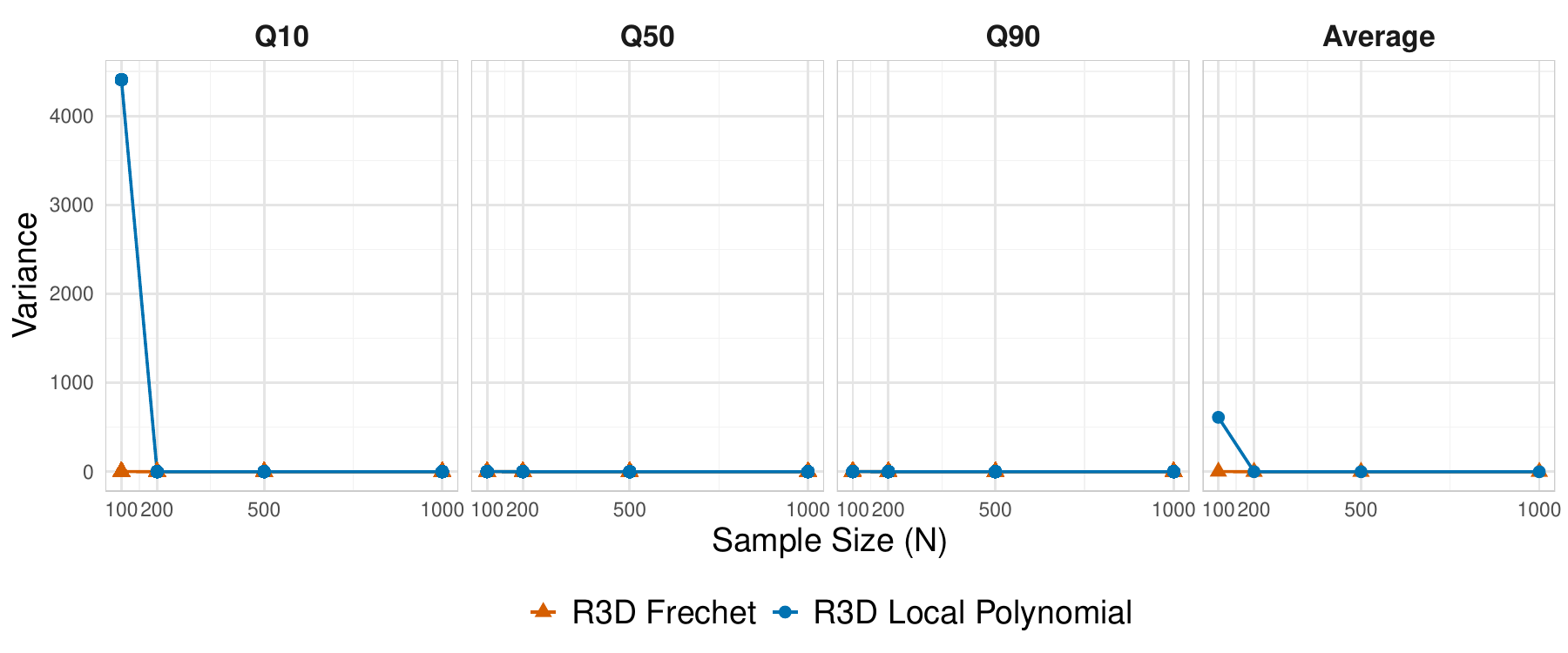}
\caption{Normal Distributions with Normal Means}
\end{subfigure}
\begin{subfigure}[b]{0.95\textwidth}
\includegraphics[width=0.9\textwidth]{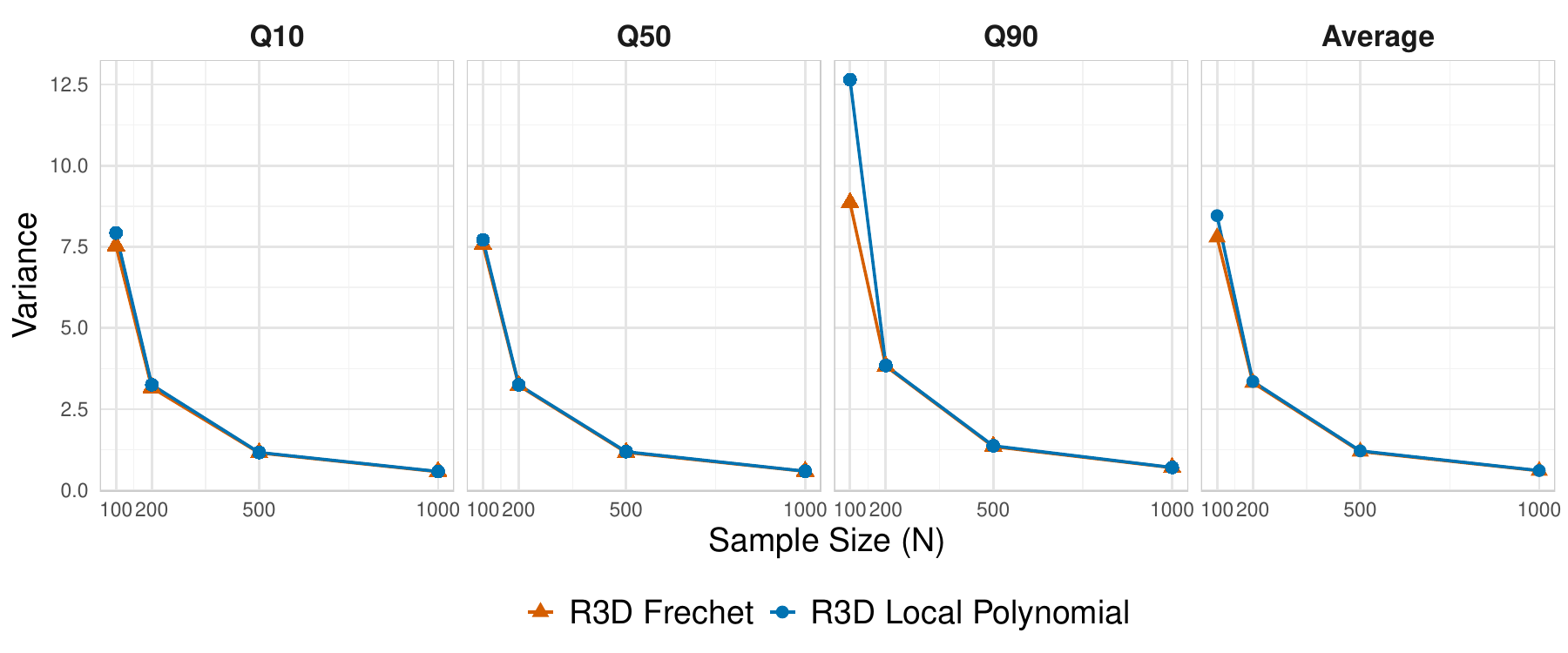}
\caption{Normal--Exponential Distributions with Normal--Exponential Means}
\end{subfigure}
\caption{Simulated Variance of R3D Estimators}
\label{fig:sims_variance}
\floatfoot{\textit{Note}: figure compares performance of the two R3D estimation methods in terms of the variance. Each measure is reported for $N=100, 200, 500, 1000$ (x axis) at quantiles 10, 50, 90, and the average over all quantiles (facets). Results are averaged over $2,500$ simulations for each sample size. The methods are: 1) the local polynomial R3D estimator in Section \ref{sec:estimator_simple}; 2) the Fr\'echet R3D estimator in Section \ref{sec:estimator_frechet}. Bandwidths are selected using (I)MSE-optimal procedure in Section \ref{sec_app:bandwidth}. Data-generating process: outcome variable $Y$ is a normal distribution with normally distributed means and variances that depend on running variable $X$ and jump across the threshold.}
\end{figure}

Figure \ref{fig:sims_variance} further supports the theoretical arguments that the Fr\'echet estimator is preferred over the local polynomial one, as the latter has much larger variance in small samples, though again both estimators quickly converge to a similarly low variance. I do not report results for the quantile RD as its inferential properties are irrelevant due to its inconsistency and bias in the R3D settings.

\begin{table}[hb!]
\centering
\footnotesize
\label{tab:compact}
\renewcommand{\arraystretch}{1}
\begin{tabular}{l cc} Method:
& \textbf{Fréchet} &\textbf{ Local Polynomial} \\
\midrule
\multicolumn{3}{c}{\textbf{ \hspace{4em} DGP 1}} \\
Unif. CIs & 
\begin{tabular}{
cccc
}
\hline
$n$ & $\Delta \!=\! 0$ & $\Delta \!=\! 1.14$ & $\Delta \!=\! 2.27$
\\
\hline
% latex table generated in R 4.4.0 by xtable 1.8-4 package
% Tue Apr  1 15:51:09 2025
 500 & 92.56 & 92.56 & 92.56 \\ 
  1000 & 93.28 & 93.28 & 93.28 \\ 
  2000 & 94.20 & 94.20 & 94.20 \\ 
  \end{tabular}
 &
\begin{tabular}{
cccc
}
\hline
$n$ & $\Delta \!=\! 0$ & $\Delta \!=\! 1.14$ & $\Delta \!=\! 2.27$
\\
\hline
% latex table generated in R 4.4.0 by xtable 1.8-4 package
% Tue Apr  1 15:51:09 2025
 500 & 92.12 & 92.12 & 92.12 \\ 
  1000 & 93.36 & 93.36 & 93.36 \\ 
  2000 & 94.40 & 94.40 & 94.40 \\ 

\end{tabular}
 \\
Homogen. &
\begin{tabular}{
cccc
}
\hline
$n$ & $\Delta \!=\! 0$ & $\Delta \!=\! 1.14$ & $\Delta \!=\! 2.27$
\\
\hline
% latex table generated in R 4.4.0 by xtable 1.8-4 package
% Tue Apr  1 15:51:09 2025
 500 & 92.64 & 92.64 & 92.64 \\ 
  1000 & 93.32 & 93.32 & 93.32 \\ 
  2000 & 94.20 & 94.20 & 94.20 \\ 
  
\end{tabular}
 &
\begin{tabular}{
cccc
}
\hline
$n$ & $\Delta \!=\! 0$ & $\Delta \!=\! 1.14$ & $\Delta \!=\! 2.27$
\\
\hline
% latex table generated in R 4.4.0 by xtable 1.8-4 package
% Tue Apr  1 15:51:09 2025
 500 & 92.52 & 92.52 & 92.52 \\ 
  1000 & 92.60 & 92.60 & 92.60 \\ 
  2000 & 93.64 & 93.64 & 93.64 \\ 

\end{tabular}
 \\
Nullity &
\begin{tabular}{
cccc
}
\hline
$n$ & $\Delta \!=\! 0 $  & $\Delta \!=\! 1.14$ & $\Delta \!=\! 2.27$
\\
\hline
% latex table generated in R 4.4.0 by xtable 1.8-4 package
% Tue Apr  1 15:51:10 2025
500  & -- & 9.36  & 0   \\
1000 & -- & 0.28  & 0   \\
2000 & -- & 0     & 0   \\

\end{tabular}
 &
\begin{tabular}{
cccc
}
\hline
$n$ & $\Delta_\mu \! = \! 0$ & $\Delta \!=\! 1.14$ & $\Delta \!=\! 2.27$
\\
\hline
% latex table generated in R 4.4.0 by xtable 1.8-4 package
% Tue Apr  1 15:51:10 2025
500  & -- & 10.12 & 0   \\
1000 & -- & 0.28  & 0   \\
2000 & -- & 0     & 0   \\

\end{tabular}
 \\[1.5ex] 
\multicolumn{3}{c}{\hspace{4em} \textbf{DGP 2}} \\
Unif. CIs & 
\begin{tabular}{
cccc
}
\hline
$n$ & $\Delta \!=\! 0$ & $\Delta \!=\! 1.86$ & $\Delta \!=\! 3.72$
\\
\hline
% latex table generated in R 4.4.0 by xtable 1.8-4 package
% Tue Apr  1 15:51:10 2025
 500 & 93.20 & 93.20 & 93.20 \\ 
  1000 & 93.12 & 93.12 & 93.12 \\ 
  2000 & 94.36 & 94.36 & 94.36 \\ 

\end{tabular}
 &
\begin{tabular}{
cccc
}
\hline
$n$ & $\Delta \!=\! 0$ & $\Delta \!=\! 1.86$ & $\Delta \!=\! 3.72$
\\
\hline
% latex table generated in R 4.4.0 by xtable 1.8-4 package
% Tue Apr  1 15:51:10 2025
 500 & 92.60 & 92.60 & 92.60 \\ 
  1000 & 93.20 & 93.20 & 93.20 \\ 
  2000 & 94 & 94 & 94 \\ 

\end{tabular}
 \\
Homogen. &
\begin{tabular}{
cccc
}
\hline
$n$ & $\Delta \!=\! 0$ & $\Delta \!=\! 1.86$ & $\Delta \!=\! 3.72$
\\
\hline
% latex table generated in R 4.4.0 by xtable 1.8-4 package
% Tue Apr  1 15:51:10 2025
 500 & 2.40 & 2.40 & 2.40 \\ 
  1000 & 0 & 0 & 0 \\ 
  2000 & 0 & 0 & 0 \\ 
  
\end{tabular}
 &
\begin{tabular}{
cccc
}
\hline
$n$ & $\Delta \!=\! 0$ & $\Delta \!=\! 1.86$ & $\Delta \!=\! 3.72$
\\
\hline
% latex table generated in R 4.4.0 by xtable 1.8-4 package
% Tue Apr  1 15:51:10 2025
 500 & 13.40 & 13.40 & 13.40 \\ 
  1000 & 3.40 & 3.40 & 3.40 \\ 
  2000 & 0.72 & 0.72 & 0.72 \\ 

\end{tabular}
 \\
Nullity &
\begin{tabular}{
cccc
}
\hline
$n$ &   $\Delta \!=\! 0$ & $\Delta \!=\! 1.86$ & $\Delta \!=\! 3.72$
\\
\hline
% latex table generated in R 4.4.0 by xtable 1.8-4 package
% Tue Apr  1 15:51:10 2025
500  & -- & 66.96 & 12.92 \\
1000 & -- & 43.68 & 1.56  \\
2000 & -- & 16.12 & 0.04  \\

\end{tabular}
 &
\begin{tabular}{
cccc
}
\hline
$n$  &  $\Delta \!=\! 0$ & $\Delta \!=\! 1.86$ & $\Delta \!=\! 3.72$
\\
\hline
% latex table generated in R 4.4.0 by xtable 1.8-4 package
% Tue Apr  1 15:51:10 2025
500  & -- & 69.24 & 14.76 \\
1000 & -- & 46.96 & 1.72  \\
2000 & -- & 18.24 & 0     \\

\end{tabular}
 \\
\bottomrule
\end{tabular}
\caption{Acceptance Probabilities of R3D Estimators}
\label{tab:coverage}
\floatfoot{\textit{Note}: table shows simulated acceptance probabilities for the 95\% uniform confidence bands (``Unif. CIs'', probability of coverage), uniform homogeneity test (``Homogen.''), and uniform treatment nullity test (``Nullity'') presented in Section \ref{sec:inference_overview} for various sample sizes, where $n=n_i$ for all simulations, with $n_i$ the sample size for the empirical quantile function. Data-generating processes are described in Equations \eqref{eq:dgp1} and \eqref{eq:dgp2}. All simulations used 2,500 repetitions and 5,000 bootstrap replications and estimated quantile treatment effects at the 9 deciles. Values of $\Delta$ reflects Cohen's d of 0, 0.5, and 1. }
\end{table}

To study the coverage properties of the confidence bands and tests proposed in \ref{sec:inference_overview}, I report their acceptance probabilities for both DGPs with varying values of $\Delta$ in Table \ref{tab:coverage}. The values of $\Delta$ are chosen to reflect an average Cohen's d (treatment effect size relative to standard error) of $0, 0.5,$ and $1$ which correspond roughly to no, medium, and large treatment effects. The coverage and acceptance probabilities of the uniform confidence intervals and the homogeneity test in the first two rows are not affected by the magnitude of the treatment effect. Moreover, both the Fr\'echet and the local polynomial estimator rapidly converge to the correct nominal coverage level, with the Fr\'echet estimator exhibiting slightly better coverage. The slight undercoverage in small samples is expected insofar as the estimators are only asymptotically unbiased, as also illustrated in Figure \ref{fig:sims_bias}. For DGP 2, which has heterogeneous treatment effects, the homogeneity test's coverage rapidly coverges to 0, illustrating the test's consistency and sharp power in finite sample. Finally, for both DGPs, the treatment nullity test also exhibits consistency and significant finite-sample power for rejecting the null hypothesis of no effect.

\subsection{Empirical Illustration: State Governors and the Income Distribution}

To further illustrate the method, I estimate the effect of partisan governorship on the income distribution within US states. To that end, I deploy a classical and widely used RD design in economics and political science: the close-election design \citep{lee2008randomized}. This design compares constituencies where a political party barely won an election to those where it barely lost in order to estimate the effect of that party's win on some outcome of interest. The identification assumption is that the outcome of interest evolves smoothly with the party's vote share in a small window around the 50\% electoral threshold that puts the party in power. Under that assumption, any jump observed in the outcome at the threshold is induced by the party's electoral win, and thus identifies its causal effect locally for states with close election outcomes. Such a close-election design naturally leads to an R3D setting (see also Motivating Example \ref{ex:institution}), since many outcomes of interest are measured at the constituent level, leading to an entire distribution of outcomes within each constituency. 

\subsubsection{Data and Method}

I use data on gubernatorial election outcomes from Congressional Quarterly's Voting and Elections Collection, collating election data from 1984 to 2010. This produces a dataset of 356 state-year combinations where a gubernatorial election took place. Restricting the sample to data before 2010 ensures a stable and clearly defined environment for estimating gubernatorial impacts on state-level income distributions. The year 2010 marked a structural breakpoint in state politics (see e.g.\ the sharp increase in state-level polarization documented in \citet{shor2022two}) due to the significant Republican gains from the Tea Party wave and the subsequent implementation of the Affordable Care Act (ACA). The ACA introduced confounding by influencing state policy choices through federal incentives, while increased partisan polarization changed the nature and meaning of gubernatorial party control itself. Restricting the analysis to pre-2010 thus guarantees a stable treatment definition, ensuring clearer identification of causal effects attributable specifically to Democratic versus Republican gubernatorial control. Indeed, while the magnitude of the effects remains similar when including post-2010 data, their precision and magnitude decrease (see Figure \ref{fig:cps_2018}).

I combine these data with family-level income data from the UNICON extract of the March Current Population Survey (CPS) for the final year of the state governor's tenure, in order to capture the cumulative effect of that tenure on the income distribution. Practically, this means the election data is lagged 3 years, except in New Hampshire and Vermont, which hold gubernatorial elections every 2 years. 

The variables in the sample are defined as follows. The running variable $X_{it}$ is the Democratic candidate's votes in state $i$ in year $t$ as a share of the combined Democratic and Republican votes. When this threshold exceeds 50\%, the Democratic candidate is elected. As such, the treatment $T_{it}$ indicates whether state $i$ elected a Democratic governor in year $t$ compared to a Republican one. 

The outcome variable $Z_{ijt'}$ is real income of family $j$ in state $i$ in year $t'=t+t_j$, where $t_j$ is a state-specific offset to match the income distribution in the final year of a governor's tenure to their electoral results. Real family income is constructed as the ratio of family income in year $t'$ to the federal poverty threshold in that year. Family income is defined in the standard fashion as the combined pre-tax cash income of the family, including earnings and cash transfers, but excluding non-cash benefits or tax credits. The federal poverty threshold is adjusted yearly and depends on family size and the number of children. Normalizing income by the year-specific poverty threshold makes the units of the outcome variable comparable across years, thus accounting for growth in real income levels over time and making the i.i.d.\ assumption required for the R3D estimator more likely to hold. 

The CPS data are a sample of the full census data, thus placing this application in the empirical quantile setting discussed in Section \ref{sec:empirical_cdfs}. In particular, instead of observing the full population income distribution, in each state $i$ in year $t$, I observe a sample of $n_i$ families $j=1,\ldots,n_i$. Based on that, I construct the empirical income quantile functions $\widehat{Q}_{Y_{it}}$, where $Y_{it}$ is the distribution function of family income in state $i$ at time $t$ such that $Z_{ijt} \sim Y_{it}$. I use the family probability weights provided in the CPS to construct these as weighted quantile functions. Further, I winsorize the distribution at the 95th percentile to account for top-coding in the CPS. In practice, I estimate the quantile function on an equally spaced grid of 95 points between $[1 \times 10^{-6}, 0.95+1 \times 10^{-6}]$, where the $1 \times 10^{-6}$ offset ensures I work on a compact subset of $[0,1]$ as required by the theoretical results.

\begin{figure}[h!]
\includegraphics[width=0.8\textwidth]{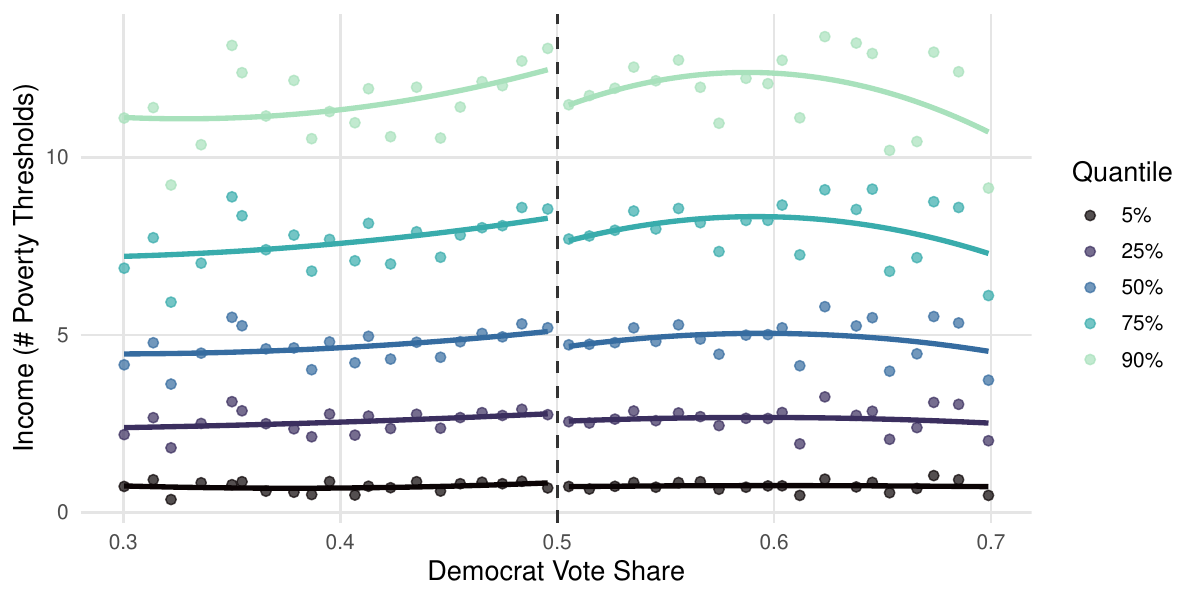}
\caption{R3D Plot: Average Income Quantiles vs. Democrat Vote Share}
\label{fig:quantile_rdplot}
\floatfoot{\textit{Note:} figure shows scatterplot (blue palette) of various average quantiles of within-state income (in multiples of the federal poverty threshold), calculated within bins of width $0.01$. Average quantiles were constructed by computing the weighted quantile functions of family income within each state and year, and then taking the average of the estimated quantile values for a given quantile (0.05, 0.25, etc) within the corresponding bin. These data points were then used to fit separate second-order polynomial regressions for each quantile, shown in the solid lines.  }
\end{figure}

The data are depicted in Figure \ref{fig:quantile_rdplot}, which shows a version of the classical RD plot \citep{calonico2015optimal} appropriate for the R3D setting, similar to Figure \ref{fig:naive}. In particular, it shows a scatterplot of the ``data'', which are the quantile functions at various quantiles $q$, averaged within equal-width bins $B_j $ of the running variable, $\frac{1}{|B_j|} \sum_{j \in B_i} \widehat{Q}_{Y_j}(q),$ with $B_j=\left\{i: X_i \in\left[x_{j, \min }, x_{j, \max }\right)\right\}$ the $j$ bins. For 5 illustrative quantiles $q$, I then fit a second-order polynomial regression line to these data. This simple descriptive plot already suggests that there is a drop in income at the higher (average) quantiles that becomes stronger as it moves up the income distribution. 

Based on these data, I use the Fréchet estimator (Section \ref{sec:estimator_frechet}) to estimate the local average quantile treatment effects in Definition \ref{def:aqte}, plugging in the estimated empirical quantile functions $\widehat{Q}_{Y_{it}}$. For these, 90\% uniform confidence bands are constructed using the bootstrap algorithm described in \ref{app:bootstrap}, where I use the 90\% nominal level to follow the standard in the literature \citep{frandsen2012quantile, qu2019uniform, chiang2019causal}. To address some of the small-sample undercoverage reported in the simulations above, I apply the rule-of-thumb coverage correction of \citet{calonico2018effect} to the IMSE-optimal bandwidth (see Appendix \ref{sec_app:bandwidth}). In addition, I formally test for uniform treatment nullity and homogeneity using the tests described in Section \ref{sec:inference_overview}. 

\subsubsection{Results}

\begin{figure}[ht!]
\includegraphics[width=0.9\textwidth]{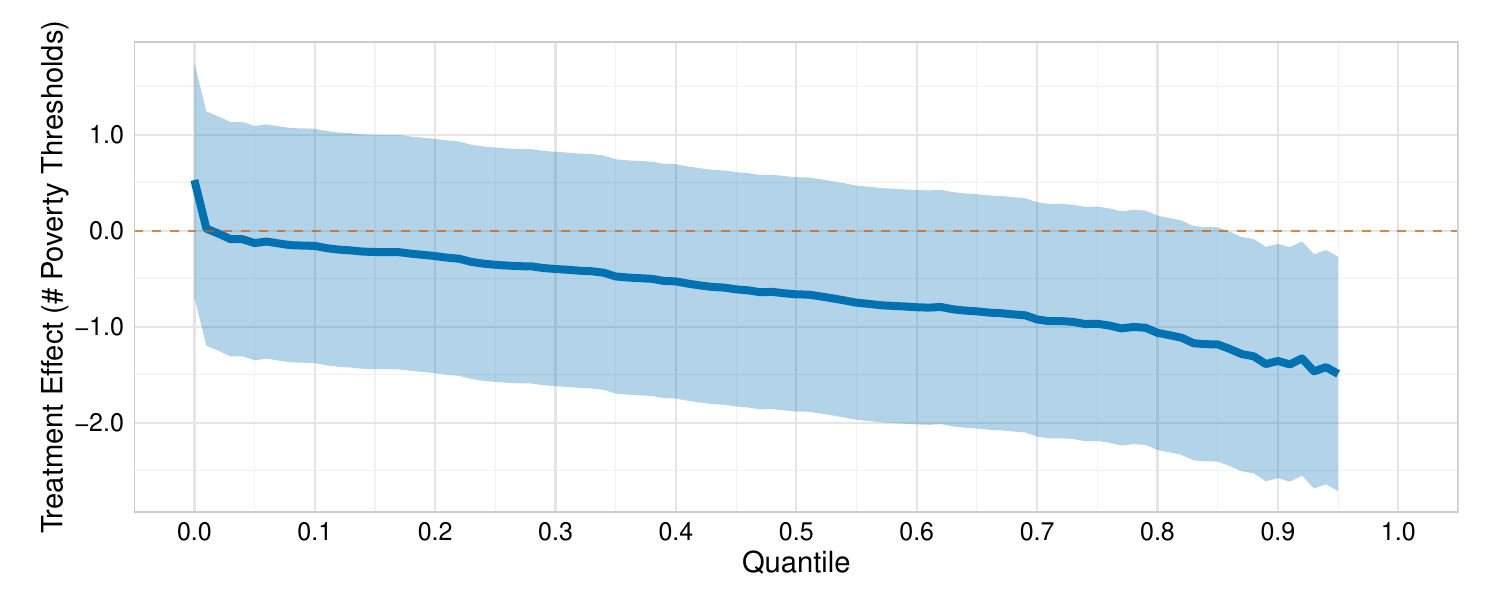}
\caption{Distributional Effects of Democratic Governor Control, 1984-2010}
\label{fig:cps_r3d_baseline}
\floatfoot{\textit{Note:} local average quantile treatment effects estimates and uniform 90\% confidence bands for R3D of effect of Democratic governor control on within-state income distribution. X-axis indicates quantile of the (average) income distribution while Y-axis indicates the difference in average state-level income distributions, in the final year of the governor's tenure, near the 50\% vote share threshold. Income is measured as real equivalized family income in multiples of the federal poverty threshold. Sample runs from 1984--2010, estimates are obtained using the second-order Fr\'echet estimator in Section \ref{sec:estimator} with first-order IMSE-optimal bandwidth and triangular kernel as in Section \ref{sec_app:bandwidth}, and uniform bands are constructed using Algorithm \ref{app:bootstrap} with 5,000 bootstrap repetitions. Treatment nullity p-value: 0.038, treatment homogeneity p-value: 0.061, IMSE-optimal bandwidth: 0.22.} 
\end{figure}

The main results are shown in Figure \ref{fig:cps_r3d_baseline}. The graph depicts the LAQTE estimates, with the Y-axis indicating the effect as a multiple of the federal income threshold for the quantile of the distribution indicated by the X-axis. The light blue band depicts the 90\% uniform confidence band. 

As shown, treatment effects are slightly positive at the lowest quantiles and become increasingly negative farther up the income distribution, with the top 10 percentiles seeing a decline in income of 1.5 times the federal poverty threshold. By contrast, the very bottom quantiles see their income increase by nearly half the poverty threshold. Only the effects for the top 10 percentiles (85th--95th) are uniformly significant at the 90\% level. The p-values for the uniform treatment nullity test and the treatment heterogeneity test are 0.0376 and 0.061, respectively, suggesting the observed negative relation between income quantile and effect size is significant.

 The estimated results are very similar for alternative specifications with the local polynomial estimator of Section \ref{sec:estimator_simple}, when using a uniform instead of a triangular kernel, or when using half the IMSE-optimal bandwidth in Figures \ref{fig:cps_r3d_simple_method},  \ref{fig:cps_r3d_uniform_kernel}, and \ref{fig:cps_r3d_1_2_bandwidth}. In contrast, when estimating the baseline specification with the income distribution of the same year as the election as outcome variable, none of the quantile treatment effects are significant, nor are the nullity and homogeneity tests. This suggests the results are not driven by reverse causality, where the pre-existing income distribution drives the election outcomes. This aligns with the small effects of local economic conditions on voting behavior estimated in the literature on retrospective voting \citep{healy2013retrospective}. Additionally, I check whether the results are not driven by families ``voting with their feet'' by moving across states. To that end, Figure \ref{fig:cps_r3d_migration} demonstrates that the results are near-identical when excluding families that moved across state borders in the previous year, barring some expected loss of precision.  

Finally, Table \ref{tab:standard_rd} reports estimates of the local average treatment effect using the standard RD estimator with robust confidence bands \citep{calonico2014robust}, using both the state-level weighted average family income and the raw family-level outcome data. The state-level treatment effect estimate is $-0.631$ and significant at the 90\% level, while the family-level estimate is $-0.525$ and is very precisely estimated. As predicted theoretically, the state-level estimates are in line with the average of the LAQTEs produced by the R3D estimators, which is -0.647, while the family-level estimates are 15\% less strong, because they do not account for the two-level sampling when using the disaggregated family-level data. Both standard RD estimates cloak the underlying heterogeneity, in particular the redistribution that is achieved at the cost of the estimated drop in overall income. I also report the quantile RD estimator of \citet{qu2019uniform} in Figure \ref{fig:r3d_cps_qte_rd}. Unfortunately, the confidence bands in the companion \texttt{R} package are not currently implemented. However, in line with the simulations above, the estimated effects exhibit substantial bias, effectively precluding the need for inference. Specifically, the estimated quantile treatment effects are more than twice as small as the R3D estimates, and the corresponding average effect is only -0.158, 4 times smaller than the standard RD estimates using the aggregated data.

Taken together, these results suggest a classical equality--efficiency trade--off under Democratic governorship, with some redistribution of income achieved at the cost of a loss of income for upper-income earners \citep{okun1975equality}. They also highlight the practical utility of the R3D estimator in uncovering distributional heterogeneity in treatment effects, compared to standard RD methods, while producing estimates that are consistent with those standard RD estimates in the aggregate.

\section{Conclusion}

This paper introduces the Regression Discontinuity Design with Distributions (R3D), a novel extension of the standard Regression Discontinuity Design (RDD) framework tailored to settings where the outcome of interest is a distribution rather than a single scalar value. This generalization is motivated by the common real-world setting where treatment is assigned at a higher level of aggregation than the outcome of interest, such as firm-level policies that affect employees, county-level policies that affect inhabitants, or school-level policies that affect students. Standard RD methods do not apply to such settings since they do not account for the two-level randomness involved in these settings, which introduces sampling at the level of distributions. To address this, I define the local average quantile treatment effect (LAQTE) as the primary estimand, which quantifies the difference in \emph{average} quantile functions, instead of observed ones, just above and below a treatment cutoff. This measure offers a natural and intuitive extension of the traditional RDD treatment effect to distribution-valued outcomes.

To estimate the LAQTE, I propose two complementary estimators: one based on local polynomial regression applied to random quantiles and another leveraging local Fr\'{e}chet regression in 2-Wasserstein space. The local polynomial approach adapts familiar RDD techniques to handle distribution-valued data pointwise, while the Fr\'{e}chet regression method treats the quantile function as a cohesive functional object, improving efficiency and finite-sample performance. Both estimators are developed for the sharp as well as the fuzzy R3D setting. I establish the asymptotic normality of both estimators and develop uniform, debiased confidence bands that can be estimated with a multiplier bootstrap. Additionally, I introduce a data-driven bandwidth selection procedure for functional outcomes. Simulations confirm the robustness of these theoretical properties, demonstrating good finite-sample performance and reliable coverage of the confidence bands.

The practical utility of the R3D framework is illustrated through an empirical application examining the effect of gubernatorial party control on within-state income distributions in the United States, using a close-election RDD. The findings reveal a classical equality--efficiency trade-off under Democratic governorship, with some redistribution achieved at the cost of an overall loss of income. In particular, incomes at the top of the distribution decline, while slight but not statistically significant improvements are observed at the lower end of the distribution. This evidence underscores the method’s ability to uncover nuanced distributional impacts that scalar-based approaches might overlook. Moreover, the implied average effect is in line with standard RD estimates at the aggregate state level, unlike quantile RD methods, which estimate effects that are up to 4 times smaller in magnitude. 

There are several avenues for future research. The R3D framework could be extended to allow for covariates \citep{jin2025identification, frolich2019including}, multiple running variables or cutoffs \citep{bertanha2020regression, gunsilius2023distributional, cheng2023estimation, cattaneo2016interpreting}, or multivariate outcome distributions \citep{chen2023sliced}. Further, applying these methods to empirical domains where the R3D setting occurs frequently, such as education, labor policy, or politics, promises to yield new insights into the distributional consequences of policy interventions.

In summary, the R3D framework offers a powerful and versatile new tool for causal inference with functional outcomes, making the estimation of distributional treatment effects practical in a novel but commonly occurring setting. By providing both theoretical foundations and practical estimation strategies, this article equips researchers with a new way to address pressing questions about how policies shape distributions. 

\clearpage

\bibliography{references}

\clearpage
\appendix

% (Optional) Reset counters so numbering starts fresh in the appendix.
\setcounter{table}{0}
\setcounter{figure}{0}
\setcounter{theorem}{0}
\setcounter{equation}{0}

% Redefine numbering to include the prefix "A-"
\renewcommand{\thetable}{A-\arabic{table}}
\renewcommand{\thefigure}{A-\arabic{figure}}
\renewcommand{\thesection}{A-\arabic{section}}
\renewcommand{\thetheorem}{A-\arabic{theorem}}
\renewcommand{\theproposition}{A-\arabic{proposition}}
\renewcommand{\thelemma}{A-\arabic{lemma}}
\renewcommand{\thecorollary}{A-\arabic{corollary}}
\renewcommand{\theequation}{A-\arabic{equation}}
\renewcommand{\thedefinition}{A-\arabic{definition}}

\section{Mathematical Notation and Definitions}

\begin{definition}[VC Type (Def. 3.6.10, \cite{gine2021mathematical})]
A class of measurable functions $\mathcal{F}$ is of $VC$ type with respect to a measurable envelope $F$ of $\mathcal{F}$ if there exist finite constants $A, v$ such that for all probability measures $Q$ on $(\Omega^x, \mathcal{F}^x)$
$$
N\left(\mathcal{F}, L^2(Q), \varepsilon\|F\|_{L^2(Q)}\right) \leq(A / \varepsilon)^v .
$$
\end{definition}

\section{Derivation of Local Polynomial Regression Weights} \label{sec_app:poly_weights}
The aim is to estimate
\[
  \hat{m}_{\pm,p}(q) =
  \Bigl.\Bigl(\text{polynomial fit at }x = c\Bigr)\Bigr|_{\text{order}=p}
\]
via the following one‐sided weighted least squares:
\[
  \hat{\boldsymbol{\alpha}}_{\pm,p}
  =
  \underset{\boldsymbol{\alpha}\,\in\,\mathbb{R}^{p+1}}{\arg\min}
  \;\sum_{i=1}^{n}
   \delta^+_i 
    K\!\Bigl(\tfrac{X_{i}-c}{h}\Bigr)\,\Bigl[
      Q_{Y_i}(q)-\boldsymbol{\alpha}^\top\,r_{p}\Bigl(\tfrac{X_{i}-c}{h}\Bigr)
    \Bigr]^2,
\]
where \(r_{p}(u) := (\,1,\;u,\;u^{2},\dots,u^{p}\,)^\top\).

Define:
\[
  \mathbf{X}_{\pm}
  =
  \begin{pmatrix}
    r_{p}\!\bigl(\tfrac{X_{1}-c}{h}\bigr)^{\!\top}\\[4pt]
    r_{p}\!\bigl(\tfrac{X_{2}-c}{h}\bigr)^{\!\top}\\
    \vdots\\
    r_{p}\!\bigl(\tfrac{X_{n}-c}{h}\bigr)^{\!\top}
  \end{pmatrix}
  ,\quad
  \mathbf{W}_{\pm} = \mathrm{diag}\Bigl\{ \delta_i^\pm\,K\!\bigl(\tfrac{X_{i}-c}{h}\bigr)
     : i=1,\dots,n
  \Bigr\},
\]
\[
  \mathbf{Q}
  =
  \bigl(Q_{Y_1}(q),\;Q_{Y_2}(q),\dots,Q_{Y_n}(q)\bigr)^\top.
\]
The solution to the above least‐squares problem is:
\[
  \hat{\boldsymbol{\alpha}}_{\pm,p}
  =
  \Bigl(
    \mathbf{X}_{\pm}^\top\,\mathbf{W}_{\pm}\,\mathbf{X}_{\pm}
  \Bigr)^{\!-1}
  \;\bigl(\mathbf{X}_{\pm}^\top\,\mathbf{W}_{\pm}\,\mathbf{Q}\bigr).
\]
Since the regression function at \(X=0\) is the \emph{intercept} component, let
\(e_{0}=(1,\,0,\,0,\dots,0)^\top\) as before.  Then
\[
  \hat{m}_{\pm,p}(q)
  =
  e_{0}^{\!\top}\,\hat{\boldsymbol{\alpha}}_{\pm,p}
  =
 e_{0}^{\!\top}
  \,\bigl(
    \mathbf{X}_{\pm}^\top\,\mathbf{W}_{\pm}\,\mathbf{X}_{\pm}
  \bigr)^{\!-1}
  \,\bigl(\mathbf{X}_{\pm}^\top\,\mathbf{W}_{\pm}\,\mathbf{Q}\bigr).
\]

Noting that everything in front of $\mathbf{Q}$ is independent of $Q_{Y_i}(q)$
and depends only on $\{X_i\}$, $K_h(\cdot)$, $h$, etc., it follows:
\[
  \hat{m}_{\pm,p}(q)
  =
  \sum_{i=1}^{n}
    \underbrace{
      e_{0}^{\!\top}
      \,\bigl(        \mathbf{X}_{\pm}^\top\,\mathbf{W}_{\pm}\,\mathbf{X}_{\pm}
      \bigr)^{\!-1}
      \,\bigl(
        \mathbf{X}_{\pm}^\top\,\mathbf{W}_{\pm}
      \bigr)_{:\,,\,i}
    }_{=:~s_{\pm,\,i n}^{(p)}(h)}
    \, Q_{Y_i}(q).
\]
Therefore, the one‐sided local‐polynomial estimator of order~\(p\) can be
written as a simple weighted average:
\begin{align*}
 &  \hat{m}_{\pm,p}(q)
  =
  \;\sum_{i=1}^n
    \Bigl[
      \,s_{\pm,\,i n}^{(p)}(h)
    \Bigr]\,
    Q_{Y_i}(q),
  \quad\text{where} \\
  & 
  s_{\pm,\,i n}^{(p)}(h)
  =
  \delta_i^\pm\;
  e_0^{\!\top}\!
\Bigl(\mathbf{X}_{\pm}^\top\,\mathbf{W}_{\pm}\,\mathbf{X}_{\pm}\Bigr)^{-1}
  r_{p}\!\Bigl(\tfrac{X_i - c}{h}\Bigr)\;K\!\Bigl(\tfrac{X_i - c}{h},\Bigr).
\end{align*}

\section{Overview of Local Fr\'echet Regression} \label{app:frechet}

This supplement provides a brief overview of local Fr\'echet regression as proposed in \citet{petersen2019frechet}. 

\subsection{Generalized Conditional Expectations}
The concept of the Fr\'echet mean arises as a natural generalization of the Euclidean mean. To see this, let $Z \in \mathbb{R}$, then the conditional expectation $E[Z \mid X=x]$ at $x$ can be defined as the unique minimizer $f$ of the mean squared error,
\[
E[Z \mid X=x] \coloneqq \argmin_{f \in \mathbb{R}} E[ d_E\left(Z,f  \right)^2 \mid X=x],
\]
where $d_E(x,y) \coloneqq \| x - y\|$ the standard Euclidean metric. The conditional Fr\'echet mean $m_{\oplus}(x)$ generalizes this to any metric space $(\Omega, d)$ equipped with a distance metric $d$ by replacing the squared Euclidean distance with the generalized squared distance $d(Y,\cdot), Y \in \Omega$ \citep{petersen2019frechet}, 
\begin{equation} \label{eq:conditional_frechet_mean}
m_{\oplus}(x)\coloneqq \underset{\omega \in \Omega}{\operatorname{argmin}} M_{\oplus}(\omega, x), \quad M_{\oplus}(\cdot, x) \coloneqq E\left[d^2(Y, \cdot) \mid X=x\right].
\end{equation}

The corresponding conditional Fr\'echet variance $V_{\oplus}(x)$ is defined analogously to the classical variance operator as the squared distance from the mean, 
\begin{equation*}
    V_{\oplus}(x) \coloneqq E \left[ d^2(Y, m_{\oplus}(x)) \mid X=x \right].
\end{equation*}
The ``unconditional'' Fr\'echet mean and variance are defined analogously. 

\subsection{The 2-Wasserstein Metric}
Consider using the 2-Wasserstein distance $d_{W_2}(Y_1, Y_2)$ to measure the distance between two distribution functions $Y_1, Y_2 \in \mathcal{Y}$. For one-dimensional distribution functions, this metric can be shown to equal \citep[Theorem 2.18]{villani2021topics},  
\begin{equation}
d_{W_2}^2\left(Y_1, Y_2\right)=\int_0^1\left(Q_{Y_1}(q)-Q_{Y_2}(q)\right)^2 \dd q,
\end{equation}
where remember that $Q_{Y_1}$ and $Q_{Y_2}$ are the quantile functions corresponding to $Y_1$ and $Y_2$, respectively. 

The reason for the asymptotic equivalence between Fr\'echet regression in 2-Wasserstein space $(\mathcal{Y}, d_{w_2})$ and local polynomial regression on quantiles is that the Fr\'echet mean of any random distribution $Y \in \mathcal{Y}$ equipped with $d_{W_2}$ is the unique cdf $F_{\oplus}$ with the quantile function \citep[Theorem 3.2.11]{panaretos2020invitation},
\[
Q_{F_\oplus}(q) = E Q_{Y}(q) = \int_{\mathcal{Y}} Q_Y(q) \dd P(Y).
\]
Informally, the ``average'' distribution computed by means of the Fr\'echet mean under the 2-Wasserstein distance is the only distribution that has a quantile function equal to the expected quantile function at each quantile $t$. In that sense, it is the ``correct'' metric for computing average quantile functions.

\subsection{Local Fr\'echet Regression}

Fr\'echet regression was introduced in \citet{petersen2019frechet} as a generalization of linear regression \citep{fan1996local} when the outcome $Y$ takes values in a general metric space $\Omega$ beyond just the Euclidean space $\mathbb{R}$.

In the definition of the conditional Fr\'echet mean introduced above, consider the case $Z \in \Omega=\mathbb{R}$ and write $m=m_\oplus$ for brevity. Then the (population) local linear estimate of $m(x)$ is $\tilde{l}(x) = \beta^*_0$, where,
$$
\begin{aligned}
\left(\beta_0^*, \beta_1^*\right)= \argmin_{\beta_0, \beta_1}\int K_h(x'-x)  \times\left[\int z \mathrm{~d} F_{Z \mid X}(x', z)-\left(\beta_0+\beta_1(x'-x)\right)\right]^2 \mathrm{~d} F_X(x'),
\end{aligned}
$$
with $K_h(\cdot) = h^{-1} K(\cdot / h)$ with $K$ a smoothing kernel and $h$ a bandwidth.
Defining $\mu_j=E\left[K_h(X-x)(X-x)^j\right]$ and $\sigma_0^2=\mu_0 \mu_2-\mu_1^2$, the solution $\beta^*_0$ can be written as,
\[
\tilde{l}(x) = \beta^*_0 = E[s(X, x, h) Z],
\]
with weight function,
\begin{equation} \label{eq:population_weights}
s(x', x, h)=\frac{1}{\sigma_0^2}\left\{K_h(x'-x)\left[\mu_2-\mu_1(x'-x)\right]\right\}
\end{equation}
which corresponds to the local Fr\'echet mean, 
\begin{equation}
\tilde{l}(x)=\underset{z \in \mathcal{R}}{\operatorname{argmin}} E\left[s(X, x, h)(Z-z)^2\right].
\end{equation}

Just as with the definition of the classical Fr\'echet mean, this can be generalized to $Y\in\Omega$ on a general metric space as,
\begin{equation*} 
\tilde{l}_{\oplus}(x)=\underset{\omega \in \Omega}{\operatorname{argmin}} \: \left\{ \tilde{L}_n(\omega) \coloneqq  E\left[s(X, x, h) d^2(Y, \omega)\right] \right\}
\end{equation*}
where the dependence on $n$ is through the bandwidth sequence $h=h_n$. 

Then, assume that $\left(X_i, Y_i\right) \sim F, i=1, \ldots, n$ are independent. The corresponding sample estimator is, 
\begin{equation} \label{eq:hat_l_main}
\hat{l}_{\oplus}(x)=\underset{\omega \in \Omega}\argmin \: \left\{ \hat{L}_n(\omega, x) \coloneqq  n^{-1} \sum_{i=1}^n s_{i n}(x, h) d^2\left(Y_i, \omega\right) \right\},
\end{equation}
with the empirical weights,
\begin{equation} \label{eq:weights}
s_{i n}(x, h)=\frac{1}{\hat{\sigma}_0^2} K_h\left(X_i-x\right)\left[\hat{\mu}_2-\hat{\mu}_1\left(X_i-x\right)\right],
\end{equation}
where 
\[
\hat{\mu}_j=n^{-1} \sum_{i=1}^n K_h\left(X_i-x\right)\left(X_i-x\right)^j, \quad \hat{\sigma}_0^2=\hat{\mu}_0 \hat{\mu}_2-\hat{\mu}_1^2.
\] 
These weights are identical to those for the classical local polynomial regression \citep{fan1996local}. The generalization lies in the use of the distance metric and the projection onto $\Omega$. 

Local Fr\'echet regression ``from the left and right'', as considered in the main text, simply requires adding a $\delta_i^\pm$ term to the appropriate equations. 

\subsection{Existence, Uniqueness, and Local Polynomial Equivalence}

The following result establishes the equivalence of the projected local polynomial regression estimator and the local Fr\'echet regression estimator from \citet{petersen2019frechet} in the metric space $(\mathcal{Y}, d_{W_2})$.

\begin{proposition} \label{prop:wasserstein}
The projected local polynomial regression estimator in \eqref{eq:local_linear} is equivalent to the quantile function of the local polynomial Fr\'echet regression estimator of order $p$ on the metric space $(\mathcal{Y}, d_{W_2})$,
\[
\argmin_{\omega \in \mathcal{Y}} \frac1n \sumn s_{\pm,in}^{(p)}(h) d^2_{W_2}(\omega, Y_i), 
\]
which exists and is unique.
Similarly, the projected conditional average quantile function $\Pi_Q(m_\pm)$ is equivalent to the quantile function of the conditional Fr\'echet mean on $(\mathcal{Y}, d_{W_2})$ in \eqref{eq:conditional_frechet_mean}, which exists and is unique. 
\begin{proof}
 Denote $\langle\cdot, \cdot\rangle_{L^2},\|\cdot\|_{L^2}$ and $d_{L^2}(\cdot, \cdot)$ the $L^2$ inner product, norm, and distance on $[0,1]$, respectively.

From the definition of $\hat{m}_{+,p}$, I have,
\begin{align*}
 & \frac1n \sum_{i=1}^n  s^{(p)}_{+,in}(h) d_{L^2}(Q_{Y_i}, \hat{m}_{+,p})^2 + d_{L^2}(Q_\omega, \hat{m}_{+,p})^2  \\ 
 & =   \langle \frac1n \sum_{i=1}^n   s^{(p)}_{+,in}(h), Q_{Y_i}^2 \rangle_{L^2} - 2   \langle \hat{m}_{+,p},  \frac1n \sum_{i=1}^n  s^{(p)}_{+,in}(h) Q_{Y_i} \rangle_{L^2} +   \frac1n \sum_{i=1}^n   s^{(p)}_{+,in}(h) \langle \hat{m}_{+,p} , \hat{m}_{+,p} \rangle_{L^2} \\
 & +  
\langle Q_\omega, Q_\omega \rangle_{L^2} -  2 \langle  \frac1n \sum_{i=1}^n   s^{(p)}_{+,in}(h) Q_{Y_i}, Q_\omega \rangle_{L^2} +   \langle \hat{m}_{+,p}, \hat{m}_{+,p} \rangle_{L^2} \\ 
  & =   \langle \frac1n \sum_{i=1}^n   s^{(p)}_{+,in}(h), Q_{Y_i}^2 \rangle_{L^2} -\langle Q_\omega, Q_\omega \rangle_{L^2} -  2 \langle \frac1n  \sum_{i=1}^n s^{(p)}_{+,in}(h) Q_{Y_i}, Q_\omega \rangle_{L^2} \\  
 & =  \frac1n \sum_{i=1}^n   s^{(p)}_{+,in}(h) d_{L^2}(Q_{Y_i}, Q_\omega)^2 \\
 & =  \frac1n \sum_{i=1}^n   s^{(p)}_{+,in}(h) d_{W_2}(Y_i, \omega)  = \eqref{eq:hat_l_main}
\end{align*}
where the second equality follows from  $ \frac1n \sum_{i=1}^n   s^{(p)}_{+,in}(h) = 1$ and the definition of $\hat{m}_{+,p}$ and the third equality follow from $ \frac1n \sum_{i=1}^n   s^{(p)}_{+,in}(h) = 1$. 
As a result, the local Fr\'echet regression estimator ``from the right'', $\hat{l}_{+,\oplus}(c)$ on $(\mathcal{Y}, d_{W_2})$ equals
\begin{align} \label{eq:frechet_as_projection}
\hat{l}_{+,\oplus}(c) & = Q^{-1} \left( \argmin_{h \in Q(\mathcal{Y})} d_{L^2}(h, \hat{m}_{+,p})^2  \right),
\end{align}
where $Q^{-1}$ is the quantile function’s generalized inverse, which maps it back to its corresponding CDF. An identical argument holds for the Fr\'echet estimator from the left, $\hat{l}_{-, \oplus}$. Thus, the quantile function of the local Fr\'echet regression estimator is the $L^2$ projection of the local polynomial estimator onto the space of quantile functions. Further, note that the solution to \eqref{eq:frechet_as_projection} always exists and is unique by the convexity of the objective function and of the space of quantile functions $Q(\mathcal{Y})$.

To see that the quantile function of the conditional Fr\'echet mean is equivalent to the projected conditional average quantile, rewrite the conditional Fr\'echet functional in \eqref{eq:conditional_frechet_mean} on $(
\mathcal{Y}, d_{W_2})$ as,
\begin{align*}
& M_{\oplus}(\Omega, x) = E[ d_{W_2}^2(Y_i, \omega) \mid X=x] \\
& = \int_{\mathcal{X}} \int_{0}^1 \left( Q_{Y_i}(q) - Q_{\omega}(q) \right)^2 \dd q \dd F_{Y|X=x} \\ 
& =   \int_{0}^1  \int_{\mathcal{X}} \left( Q_{Y_i}(q)^2 - 2 Q_{Y_i}(q) Q_{\omega}(q)  +  Q_{\omega}(q)^2 \right) \dd F_{Y|X=x} \dd q \\ 
& = C' + \int_{0}^1 \left( m(q) - Q_\omega(q) \right)^2 \dd q
\end{align*}
where $C' = \int_{0}^1  \int_{\mathcal{X}}  Q_{Y_i}(q)^2 \dd F_{Y|X=x} \dd q - \int_0^1 m(q)^2 \dd q $ is a constant that does not depend on $Q_\omega$, and the second equality follows from Fubini-Tonelli by the fact that all distributions in $\mathcal{Y}$ have finite variance. As a result, the conditional Fr\'echet mean $m_{\oplus}$ on $(\mathcal{Y}, d_{W_2})$ equals
\[
m_{\oplus}(x) = Q^{-1}\left( \argmin_{h \in Q(\mathcal{Y})} d_{L^2}(h, m) \right),
\]
the $L^2$ projection of $E[Q_Y(\cdot) \mid X=x]$ onto the space of quantile functions. But since $E[Q_Y(\cdot) \mid X=x]$ is a valid quantile function, the two functions are in fact equivalent. This is the well-known result that the Fr\'echet mean in 2-Wasserstein space has a quantile function equal to the average quantile function. Existence and uniqueness  follow by standard properties of conditional expectations.
 \end{proof}
\end{proposition}

\section{Implementation Details}

\subsection{First-Stage Estimators} \label{app:first_stage}

Let $t \leq p, \, t \in \mathbb{N}_+$ and denote $\delta^\pm_x \coloneqq 1\!\bigl\{x \,\gtreqlessslant\,0\bigr\}$. The uniformly consistent first-stage estimators $\hat{\mathcal{E}}_1(y, t, x,q), \hat{\mathcal{E}}_2(y, t, x,q)$ proposed by \citet[A.6]{chiang2019robust} are,
\begin{align*}
\hat{\mathcal{E}}_1(y, t, x,q) & = \left( Q_Y(q) -  \tilde{E}[Q_Y(q) \mid X=x] \right) \mathrm{1}\left( | x / h_{1}(q) | \leq 1 \right) \\
\intertext{and }
\hat{\mathcal{E}}_2(y, t, x,q) & = \left( T -  \tilde{E}[T \mid X=x] \right) \mathrm{1}\left( | x / h_{2}(q) | \leq 1 \right) 
\end{align*}
where
\[
 \tilde{E}[Q_Y(q) \mid X=x] \coloneqq r_t(x / h_1(q))' \hat{\boldsymbol{\alpha}}_{+,t}  \,\delta^+_x + r_t(x / h_2(q))' \hat{\boldsymbol{\alpha}}_{-,t} \delta^-_x
\]
and 
\[
 \tilde{E}[T\mid X=x] \coloneqq r_t(x / h_2(q))' \hat{\boldsymbol{\alpha}}_{+,T,t} \,\delta^+_x + r_t(x / h_2(q))' \hat{\boldsymbol{\alpha}}_{-,T,t}\, \delta^-_x
\]
with $ \hat{\boldsymbol{\alpha}}_{+,t}$ defined in \eqref{eq:local_linear} and similarly
\begin{align*}
  \hat{\boldsymbol{\alpha}}_{\pm,T,p}
 &  =  \underset{\boldsymbol{\alpha}\,\in\,\mathbb{R}^{p+1}}{\arg\min}
  \;\sum_{i=1}^{n}
    \delta_i^\pm \;
    K\!\Bigl(\tfrac{X_{i}}{h}\Bigr)\,\Bigl[
      T_i-\boldsymbol{\alpha}^\top\,r_{p}\Bigl(\tfrac{X_{i}}{h}\Bigr)
    \Bigr]^2\nonumber.
\end{align*}
For the corresponding Fr\'echet first-stage estimates, one simply projects $\tilde{E}[Q_Y(q) \mid X=x]$ onto the space of quantile functions before evaluating it at a given $q$ in the expression for $\hat{\mathcal{E}}_1(y, t, x,q)$.

Then, by Lemma 7 in \citet{chiang2019robust}, the first-stage local polynomial estimators are uniformly consistent for $\mathcal{E}_1(y, t, x,q) \mathrm{1}\left( | x / h_{1}(q) | \leq 1 \right)$ and $\mathcal{E}_2(y, t, x,q) \mathrm{1}\left( | x / h_{2}(q) | \leq 1 \right)$ on $[\underline{c}, \overline{c}] \times \mathcal{Y} \times \{0,1\}$. By the coerciveness of the projection onto quantile functions \citep{bauschke2017}, the local Fr\'echet version of these first-stage estimates is also uniformly consistent. The benefit of these estimators is that the same $p$-th order local polynomial estimators can be reused for both the first and second stage, reducing computational load. 

Furthermore, a standard consistent estimator for $f_X(0)$ is the kernel density estimator $\hat{f}_X(0) \coloneqq \frac{1}{n b} \sumn K\left( X_i / b \right)$ with $b=b_n \to 0$ and $nb \to \infty$. 

\subsection{Bandwidth Selection} \label{sec_app:bandwidth}

While Assumption \ref{asspt:bandwidth_chiang} prescribes asymptotic bandwidth rates, in practice, researchers need to choose a bandwidth in finite sample. Here, I derive MSE-optimal bandwidths for the local polynomial estimator (analogous to \citet[Supplement F]{chiang2019robust}), and integrated MSE-optimal (IMSE) bandwidths for the Fr\'echet regression estimator. Using the IMSE for the latter delivers a single bandwidth for the entire average quantile function, as assumed by the Fr\'echet estimator. To operationalize the one-step robust bias correction from \citet{calonico2014robust}, I need to compute the bandwidth that is optimal for the desired order of local polynomial estimation $s$, but then use that $s$-th order optimal bandwidth to estimate a $p$-th order local regression, with $p>s$. 

I remind the reader of the following notation,
$
  r_{s}(u) 
  = \bigl(1,\,u,\dots,\,u^{s}\bigr)
$, $\Gamma_{s}^{\pm} 
  =
  \int_{\mathbb{R}_{\pm}}
    K(u)\;r_{s}(u)\,r_{s}(u)^\prime
  \,du,$   $
  \Lambda_{s,s+1}
  =
  \int_{\mathbb{R}}
     u^{s+1}\,r_{s}(u)\,K(u)
  \,du $, 
  $
  \Lambda_{s,s+1}^{\pm}
  =
  \int_{\mathbb{R}_{\pm}}
     u^{s+1}\,r_{s}(u)\,K(u)
  \,du $, and 
$
\Psi_s^{ \pm}=\int_{\mathbb{R}_{ \pm}} r_s(u) r_s^{\prime}(u) K^2(u) d u
$.  Below, I drop the $\mathrm{R3D}$ superscript on the treatment effect estimators $\hat{\tau}^{\mathrm{R3D}}$ to ease notation. 

\subsubsection{Local Polynomial Estimator}

It is well known (cf. \citet{fan1992variable}, \citet{calonico2014robust}) that for a 
$p$‐th order local polynomial at a boundary, the leading bias is on the order of 
$\,h^{s+1}$.  Specifically:
\[
\mathrm{Bias}\bigl[\hat{m}_{\pm,s}(q)\bigr]
  =  h_1(q)^{s+1} B_\pm(q) := 
  \frac{h(q)^{s+1}}{(s+1)!}\;
  e_{0}^{\prime}\,
\bigl(\Gamma_{s}^{\pm}\bigr)^{-1}\,\Lambda_{s,s+1}^{\pm}
  \;\frac{\partial^{s+1} m_{\pm}(q)}{\partial x^{s+1}}
  + O\bigl(h^{s+2}\bigr),
\]
where $e_{0}=(1,0,\dots,0)^\prime\in\mathbb{R}^{s+1}$ is the row vector picking out the 
intercept term.  
Hence the bias of the difference \(\hat{\tau}_{s}(q)\) at each $q$ is, 
\[
  \mathrm{Bias}\bigl[\hat{\tau}_{s}(q)\bigr]
  \;\approx\;
  h_1(q)^{s+1} \left( B_+(q)  - B_-(q) \right)
\]
Similar derivations for the variance expressions imply that
\[
  \mathrm{Var}\bigl[\hat{m}_{\pm,s}(q)\bigr]
  \;\approx\;
  \frac{1}{n h} V_\pm(q) := 
\frac{1}{(n h) f_X(0)}  e_0^{\prime}\bigl(\Gamma_{s}^{\pm}\bigr)^{-1}\,
  \Bigl[\sigma_{1,1}(q,q\mid 0^\pm)\Bigr]\;
  \Psi_{s}^{\pm}\!\bigl((q,1),(q,1)\bigr)\,
\bigl(\Gamma_{s}^{\pm}\bigr)^{-1} e_0,
\]
where I remind the reader that $\sigma_{1,1}(q,q\mid 0^+)= \lim_{x \to 0^\pm }\mathrm{Var}\bigl(Q_{Y}(q)\mid X=x\bigr)$. Summing these expressions for both sides for $\hat{\tau}_{s}(q)$ 
(the difference) yields
\[
  \mathrm{Var}\bigl[\hat{\tau}_{s}(q)\bigr]
  \;\approx\;
  \frac{1}{nh}\;\Bigl[
    V_{+}(q) + V_{-}(q)
  \Bigr].
\]

 Based on the standard bias-variance expression $\mathrm{MSE} = \mathrm{Bias}^2 + \mathrm{Var}$, these expressions allow one to derive an MSE-optimal bandwidth $h_1^*(q)$ at each $q$ for the local polynomial estimator by optimizing with respect to $h_1$, which gives,
 \begin{equation} \label{eq:h_mse}
 h^*_1(q) = \left( \frac{1}{2(s+1)} \frac{V_+(q) - V_-(q)}{ (B_+(q) - B_-(q))^2} \right)^{1/(2s+3)} n^{-1/(2s+3)}.
 \end{equation}
The derivation for the denominator's bandwidth in the fuzzy RDD, $h_2^*$, follows identically by replacing $m_{\pm}(q)$ with $m_{\pm,T}$ and $\sigma_{1,1}(q,q | 0^\pm)$ with $\sigma_{2,2}(q,q | 0^\pm) = \lim_{x \to 0^\pm}\mathrm{Var}(T | X=x)$ in the formulas above and noting that the optimal bandwidth will be the same for all $q$.

\subsubsection{Fr\'echet Estimator}
Define
\[  \mathrm{IMSE}\bigl[\hat{\tau}_{s}\bigr]
  =
  \int_{a}^{b}
    \mathrm{MSE}\bigl[\hat{\tau}_{s}(q)\bigr]
  \dd q
  =
  \int_{a}^{b}\!\Bigl[
    \mathrm{Bias}\bigl[\hat{\tau}_{s}(q)\bigr]^{2}
    +\mathrm{Var}\bigl[\hat{\tau}_{s}(q)\bigr]
  \Bigr]\dd q.
\]
Putting the above expansions together, it follows that
\[
  \mathrm{IMSE}\bigl[\hat{\tau}_{s}\bigr]
  =
  h^{2(s+1)} A_{s}
  +
  \frac{1}{nh}B_{s}
  +
  o\left(h^{2(s+1)} + \frac1{nh}\right),
\]
where
\[
  A_{s}
  = \int_a^b (B_+(q)  - B_-(q))^2 \dd q
\]
and
\[
  B_{s}
  = \int_{a}^{b} (V_+(q) - V_-(q)) \dd q.
\]

Then remember the Fr\'echet conditional mean estimator $\hat{m}_{\pm,\oplus}$ is $\Pi_{\!\mathcal{Q}}[\hat{m}_{\pm,1}(\cdot)]$, 
the $L^2$‐projection of the local polynomial estimator onto quantile functions, and the corresponding treatment effect estimator $\hat{\tau}_{\oplus}(q)$.
Write the difference between the treatment effect estimators as
\[
  \Delta_{\oplus,s}(q)
  =
  \hat{\tau}_{\oplus,s}(q)-\hat{\tau}_{s}(q)
  =
  \bigl[\hat{m}_{+,\oplus, s}-\hat{m}_{+,s}\bigr](q)
  -
  \bigl[\hat{m}_{-,\oplus, s}-\hat{m}_{-,s}\bigr](q).
\]
I show in Lemma \ref{lemma:frechet_lp_squeeze} that
\[
  \bigl\|\hat{m}_{\pm,\oplus,s}-\hat{m}_{\pm,s}\bigr\|_{L^2([a,b])}
  =
  o_{p}\bigl((n h)^{-1/2}\bigr).
\]
Hence the difference $\Delta_{\oplus, s}(\cdot)$ is also $o_{p}\bigl((n h)^{-1/2}\bigr)$ in $L^2([a,b])$. 

Then write, 
\[
\mathrm{IMSE}\bigl[\hat{\tau}_{\oplus, s}\bigr]
  -
\mathrm{IMSE}\bigl[\hat{\tau}_{s}\bigr]
  =
  \int_{a}^{b}
  \Bigl[
\mathrm{MSE}\bigl(\hat{\tau}_{\oplus, s}(q)\bigr)
    - \mathrm{MSE}\bigl(\hat{\tau}_{s}(q)\bigr)
  \Bigr]\,dq.
\]
But pointwise in $q$,
\[
\mathrm{MSE}\bigl[\hat{\tau}_{\oplus, s}(q)\bigr]
  -
  \mathrm{MSE}\bigl[\hat{\tau}_{s}(q)\bigr]
  =
2\,E\bigl[(\hat{\tau}_{s}-\tau)\,\Delta_{\oplus, s}\bigr]
  +
E\bigl[\Delta_{\oplus, s}^{2}\bigr].
\]
By the Cauchy–Bunyakovsky–Schwarz inequality and the previously established convergence results, this difference is $o_p(1/nh)$. Hence
\[
  \mathrm{IMSE}\bigl[\hat{\tau}_{\oplus, s}\bigr]
  =
\mathrm{IMSE}\bigl[\hat{\tau}_{s}\bigr]
  +
 o_p((1/nh).
\]
As a result, the leading terms of the IMSEs of the local polynomial and the Fr\'echet estimator are the same, and thus,
\begin{equation} \label{eq:h_imse}
\mathrm{IMSE}\bigl[\hat{\tau}_{\oplus,s}\bigr]
  \;\approx\;
  A_{s}\,h^{2(p+1)}
  +
  \frac{1}{nh} B_{s}.
\end{equation}
Taking a derivative in $h$ and setting it to $0$ gives the IMSE-optimal bandwidth for the sharp Fr\'echet RD setting,
\[
  h_{\oplus, 1}^*
  =
  \biggl(\frac{B_{s}}{2(s+1)\,A_{s}}\biggr)^{\!1/(2s+3)}\,
  n^{-\tfrac{1}{2s+3}}.
\]
For the fuzzy Fr\'echet RD setting, one simply uses this rate for the numerator and $h_2^*$, derived above, for the denominator. 

Note that to obtain the estimates of the $A$ and $B$ terms, as explained below, I rely on standard local polynomial estimates rather than the projected Fr\'echet estimates. The reason is that the bias term involves the second derivative of the conditional expectation. Derivatives of quantile functions are not quantile functions themselves, and hence projecting them onto the space of quantile functions lacks meaning. This approach is justified by the above derivations, since the Fr\'echet and local polynomial estimators coincide asymptotically. 

\subsubsection{Practical Estimation}

To estimate the ``oracle'' bandwidths derived above in practice, I propose the following three-step procedure:

\noindent \textbf{Step 1: Preliminary Bandwidths.}

\begin{enumerate}
\item[(i)] 
Estimate the density of \(X\) at zero by a kernel‐density estimator using the rule of thumb of \citet{silverman2018density}:
\[
  \hat{f}_X(0)
  =
  \frac{1}{n c_n}\,
  \sum_{i=1}^n
    K\left(\frac{X_i}{c_n}\right)
  \quad\text{where}\quad
  c_n = 1.06\,\hat{\sigma}_X\,n^{-1/5},
\]
and \(\hat{\sigma}_X\) is the sample standard deviation of \(\{X_i\}_{i=1}^n\).

\item[(ii)]
Compute the \emph{pilot} bandwidths \(h_{k,n}^{0}\) for local polynomial fits of order \(p\) using the bias-variance formulas derived above,
\[
  h_{1,n}^0(q)
  =
  \Biggl(
     \tfrac{1}{\,2(p+1)\,}
     \,\frac{\,C_{1,0}(q)'\,}{\,C_{1,0}(q)^{2}\,}
  \Biggr)^{\!\!\frac{1}{\,2p+3\,}}
   n^{-\frac{1}{\,2p+3\,}},
\]
where \(C_{1,0}\) and \(C_{1,0}'\) are the bias and variance expressions derived above with first-stage estimates plugged in,
\[
\begin{aligned}
  C_{1,0}(q)
  &= 
  e_{0}^{\prime}
  \,\Bigl[
\bigl(\Gamma_{s}^{+}\bigr)^{-1}\,\Lambda_{s,\,s+1}^{+}\,\frac{\partial^{s+1} \overline{m}_{+}(q)}{\partial x^{s+1}}
    -
\bigl(\Gamma_{s}^{-}\bigr)^{-1}\,\Lambda_{s,\,s+1}^{-}\,\frac{\partial^{s+1} \overline{m}_{-}(q)}{\partial x^{s+1}}
  \Bigr]
  \,\Big/\,(s+1)!,
  \\[3pt]
  C_{1,0}^{\prime}(q)
  &= 
  \frac{\,1\,}{\,\hat{f}_X(0)\,}
  \;\,
  e_{0}^{\prime}
  \,\bigl[
\,\overline{\sigma}_{1,+}^{2}\,
\bigl(\Gamma_{s}^{+}\bigr)^{-1}\,\Psi_{s}^{+}\,\bigl(\Gamma_{s}^{+}\bigr)^{-1}
      +
\,\overline{\sigma}_{1,-}^{2}(q)\,
\bigl(\Gamma_{s}^{-}\bigr)^{-1}\,\Psi_{s}^{-}\,\bigl(\Gamma_{s}^{-}\bigr)^{-1}
    \bigr]\,
  e_{0}.
\end{aligned}
\]
where \(\frac{\partial^{s+1} \overline{m}_{\pm}(q)}{\partial x^{s+1}}\) and \(\overline{\sigma}_{1,\pm}^{2}(q)\) are preliminary guesses of the \((s+1)\)‐th derivative term and the variance, respectively. 
  In practice, one can obtain them by fitting a global polynomial of degree \(\,\ge s+1\) and computing the sample variance of the first term. As suggested in \citet[Supp. F]{chiang2019robust}, simply setting them to 1 can also deliver satisfactory performance. The pilot bandwidth for the denominator in the fuzzy RDD, $h_{2,n}^0(q)$ can be obtained entirely analogously by substituting a guess for \(\frac{\partial^{s+1} \overline{m}_{\pm, T}(q)}{\partial x^{s+1}}\) and its corresponding variance. 
\end{enumerate}

\medskip

\noindent
\textbf{Step 2: First‐Stage Local Polynomial Fits.}

Using the pilot bandwidths \(\{\,h_{1,n}^0(q),\,h_{2,n}^0(q)\}\) from Step~1, run local polynomial regressions of order \(s\) at each quantile $q$,
\[
\widecheck{\alpha}_{\pm,s}(q)
  =
  \underset{\alpha\in\mathbb{R}^{s+1}}{\arg\min}
  \;\sum_{i=1}^n
    \delta_{i}^{\pm}\;K\Bigl(\tfrac{X_i}{\,h_{k,n}^{0}}\Bigr)
      \Bigl[
         Q_{Y_i}(q)
         -
    \alpha^\top\,r_{s} \bigl(\tfrac{X_i}{\,h_{k,n}^{0}}\bigr)
      \Bigr]^2,
\]
which gives the first-stage estimates
\[ \bigl[\widecheck{m}_{\pm}(q),\dots,\dpar{^{s}\widecheck{m}_{\pm}(q))}{x^{s}}\bigr] = \widecheck{\alpha}_{\pm, s}^{\prime} \operatorname{diag}\left[1,1!/ h_{1, n}^0, \ldots, s!/\left(h_{1, n}^0\right)^s\right] \] and the corresponding first-stage $s$-th order expansion, 
$$\begin{aligned} \widecheck{E}[Q_Y(q) \mid X=x] = & 
\Bigl[
  \widecheck{m}_{+}(q)
  + \widecheck{m}_{+}^{(1)}(q)\,x
  + \cdots
  + \frac{\partial^{s}\widecheck{m}_{+}(q)}{\partial x^{s}}\,
    \frac{x^{s}}{s!}
\Bigr]
\,\delta_x^{+}
+ \\ 
& \Bigl[
  \widecheck{m}_{-}(q)
  + \widecheck{m}_{-}^{(1)}(q)\,x
  + \cdots
  + \frac{\partial^{s}\widecheck{m}_{-}(q)}{\partial x^{s}}\,
    \frac{x^{s}}{s!}
\Bigr]
\,\delta_x^{-}.
,\end{aligned}$$
as well as the corresponding variance estimates,
\[ \widecheck{\sigma}_{11}\left(q, q \mid 0^{ \pm}\right)=\left(\frac{\sum_{i=1}^n\left(Q_{Y_i}(q)-\widecheck{E}[Q_Y(q) \mid X=c] \right)^2 K\left(\frac{X_i}{h_{1, n}^0}\right) \delta_i^{ \pm}}{\sum_{i=1}^n K\left(\frac{X_i}{h^0_{1, n}}\right) \delta_i^{ \pm}}\right)^{1 / 2},
\]
and analogously for $\widecheck{E}[T \mid X=x]$, $ \widecheck{\sigma}_{22}\left(q, q \mid 0^{ \pm}\right)$. Then the uniform consistency of $\widecheck{E}[Q_Y(q) \mid X=x] \mathrm{1}\left\{|x| \leq h_{1, n}^0(q)\right\}$ and $\widecheck{E}[T \mid X=x] \mathrm{1}\left\{|x| \leq h_{2, n}^0\right\}$ is implied by Lemma 7 in \citet{chiang2019robust}, see the discussion in Appendix \ref{app:first_stage}. 

\medskip

\noindent
\textbf{Step 3: Final Bandwidth via MSE (or IMSE).}

Finally, plug these first‐stage expansions into the MSE- and IMSE-optimal bandwidth formulas derived in \eqref{eq:h_mse} and \eqref{eq:h_imse}. 

\begin{itemize}
\item \emph{Local polynomial estimator}, $h_{1}^*(q)$: MSE requires a separate \(\widehat{h}_{k,n}(q)\) for each quantile \(q\).  
\item \emph{Fr\'echet estimator}: $h_{\oplus,1}^*$: IMSE across \(q\in[a,b]\) can be obtained by averaging the bias$^2$ and variance from Step~2 over \(q\in[a,b]\) to get a single bandwidth for all $q$.   
\end{itemize}

Finally, one can optionally apply the rule-of-thumb bandwidth algorithm from \citet{calonico2018effect, calonico2020optimal} for optimal coverage error to these (I)MSE-optimal estimated bandwidths,
\begin{align*}
h^{\mathrm{ROT}}_1(q) = h_1^*(q) n^{-s/(2s+3)(s+3)}
\end{align*}
and similarly for $h^{\mathrm{ROT}}_2(q)$.

\subsection{Multiplier Bootstrap: Algorithm} \label{app:bootstrap}
% Simplified algorithm presentation

\noindent\textbf{Input:}
\begin{itemize}
\item A sample $\{(X_i, Y_i, T_i)\}_{i=1}^n$, where $Y_i \in \mathcal{Y}$ (distributional outcome), $T_i \in \{0,1\}$, and running variable $X_i \in \mathbb{R}$ with cutoff normalized to $0$.
\item A finite grid of $M$ quantiles $q_j$,  $\mathcal{T}^{*} \coloneqq (q_1, \ldots, q_M) \subset [a,b] \subset (0,1)$.
\item A chosen local polynomial order $p$.
\item A kernel function $K$ and bandwidth $h>0$. For simplicity, assume a single bandwidth here, but see \ref{sec_app:bandwidth} for more details on bandwidth selection. 
\item Number of bootstrap repetitions $B$ and significance level $\lambda\in (0,1)$.
\end{itemize}

\noindent\textbf{Remark.} In practice, $Y_i, Q_{Y_i}$ are computed using samples $\{ Z_{ij} \}_{j=1}^{n_i} \sim Y_i$ based on \eqref{eq:empirical_quantile}. If the entire population is observed, these estimates coincide with the true distribution and quantile function, otherwise the results in Section \ref{sec:empirical_cdf_stats} apply. 

\medskip

\noindent\textbf{Step 1: Estimate conditional means on a grid of quantiles.} \\
For each $q_j \in \mathcal{T}^{*}$:

\begin{enumerate}
    \item[(i)] 
     Form the local polynomial estimator as 
    \[
      \hat{m}_{\pm,p}(q_j) 
      =
      \sum_{i=1}^n 
      s_{\pm,i n}^{(p)}(h)
      \,Q_{Y_i}(q_j),
    \]
    where $Q_{Y_i}(q_j)$ is the $q_j$‐quantile of $Y_i$, and $s_{\pm,i n}^{(p)}(h)$ are the usual local polynomial weights for $X_i\gtreqlessslant 0$. 
    
    \item[(ii)] (Sharp RDD) Set
    \[
      \hat{\tau}_{p}^{\mathrm{R3D}}(q_j) = \hat{m}_{+,p}(q_j) - \hat{m}_{-,p}(q_j).
    \]
    \item[(iii)] (Fuzzy RDD only) Also compute $\hat{m}_{\pm,T,p} = \sum_{i=1}^n s_{\pm,i n}^{(p)}(h)\,T_i$, and form
    \[
      \hat{\tau}_{p}^{\mathrm{F3D}}(q_j) 
      =
      \frac{
        \hat{m}_{+,p}(q_j) -\hat{m}_{-,p}(q_j)
      }{
        \hat{m}_{+,T,p}-\hat{m}_{-,T,p}
      }.
    \]
\end{enumerate}

  \noindent \textit{Optional: Fr\'echet estimator}. Project $\hat{m}_{\pm,p}$ onto the space of monotone functions through the isotonic regression:
    \[
    \hat{m}_{\oplus,+,p} = \argmin_{u_1, \ldots, u_M \in \mathbb{R}^M} \sum_{j=1}^M \left( \hat{m}_{\pm,p}(q_j) - u_j \right)^2
    \]
    subject to the constraint $u_1 \leq \ldots \leq u_M$. 

\smallskip

\noindent Then, for each $q_j \in \mathcal{T}^*$, carry out the following steps. \\
\noindent\textbf{Step 2: Estimate residuals for first‐stage weighting.} \\
Obtain uniformly consistent first‐stage estimators of the residual functions. For instance, for $k\in\{1,2\}$ and each $i$,
\[
  \hat{\mathcal{E}}_k\bigl(Y_i,T_i,X_i,q_j\bigr)
  =
  \Bigl[\,g_k(Y_i,T_i,q_j)-\tilde{E}\bigl\{g_k(Y,T,q_j)\,\bigm|\,X_i\bigr\}\Bigr]
  \;\mathbf{1}\Bigl\{\bigl\lvert X_i/h_{k}(q_j)\bigr\rvert\le 1\Bigr\},
\]
where $g_1(Y_i,q)=Q_{Y_i}(q_j)$, $g_2(Y_i,T_i)=T_i$, and 
\(\tilde{E}\{ \cdots | X_i\}\) is a local‐polynomial fit of order $t\le p$ that reuses the $p$-th order estimates computed in Step 1 (see \ref{app:first_stage}).

\smallskip
\noindent\textbf{Step 3: Generate bootstrap draws.} \\
Draw $\left\{ \{ \xi_i^b \}_{i=1}^n\right\}_{b=1}^B$ i.i.d. from $N(0,1)$, independent of the data, for $b=1,\ldots,B$. 
For $k\in\{1,2\}$, compute
\[
  \hat{\nu}_{\xi,n}^{\pm,b}(q_j,k)
  =
  \sum_{i=1}^n 
  \,\xi_i^b
  \,\frac{
    e_0^\top\bigl[\Gamma_{\pm,p}\bigr]^{-1}
    \,\hat{\mathcal{E}}_k\bigl(Y_i,T_i,X_i,q\bigr)\,
    r_p\bigl(X_i/h_k(q_j)\bigr)\,
    K\bigl(X_i/h_k(q_j)\bigr)\,\delta_i^\pm
  }{
    \sqrt{nh_k(q_j)}\;\hat{f}_X(0)
  },
\]
where $\delta_i^\pm=1\{X_i\gtreqlessslant 0\}$, $r_p(\cdot)$ is the local‐polynomial basis, and $\Gamma_{\pm,p}=\int_{\mathbb{R}_\pm}K(u)\,r_p(u)\,r_p(u)^\top\,du$.  

\smallskip
\noindent\textbf{Step 4: Form the bootstrap processes.}
\begin{enumerate}
  \item[(i)] (Sharp RDD) For each $b$:
  \[
    \hat{\mathbb{G}}^{\mathrm{R3D},b}(q_j)
    = 
    c_1(q_j)^{-\tfrac12}\,
    \Bigl[\hat{\nu}_{\xi,n}^{+,b}(q_j,1)
           -\hat{\nu}_{\xi,n}^{-,b}(q_j,1)\Bigr].
  \]
  \item[(ii)] (Fuzzy RDD) For each $b$:
  \[
    \hat{\mathbb{G}}^{\mathrm{F3D},b}(q_j)
    =
    \frac{
      \bigl[\hat{m}_{+,T,p}-\hat{m}_{-,T,p}\bigr]\,
           \hat{\nu}_{\xi,n}^{\Delta,b}(q_j,1)
      -
      \bigl[\hat{m}_{+,p}(q_j)-\hat{m}_{-,p}(q_j)\bigr]\,
           \hat{\nu}_{\xi,n}^{\Delta,b}(q_j,2)
    }{
      \bigl[\hat{m}_{+,T,p}-\hat{m}_{-,T,p}\bigr]^2
    },
  \]
  where 
  $
     \hat{\nu}_{\xi,n}^{\Delta,b}(q_j,k)
     =\hat{\nu}_{\xi,n}^{+,b}(q_j,k)-\hat{\nu}_{\xi,n}^{-,b}(q_j,k).
  $
  \item[(iii)] (Optional local‐Fr\'echet version) In the above equations, replace 
   $\hat{m}_{\pm,p}$
   by the Fr\'echet estimator $\hat{m}_{\pm,\oplus,p}$ if needed.
\end{enumerate}

\smallskip
After carrying out step 2--4 for each $q_j \in \mathcal{T}^*$, do: \\ 
\noindent\textbf{Step 5: Compute the critical value and construct bands.}

For a given significance level $\lambda\in(0,1)$, define 
\[
  \hat{c}_n^B(a,b;\lambda)
  =
  \text{$(1-\lambda)$‐quantile of}\;
  \Bigl\{
    \max_{q\in\mathcal{T}^*}\bigl|\hat{\mathbb{G}}^{b}(q)\bigr|
    \;\colon\;
    b=1,\dots,B
  \Bigr\},
\]
where $\hat{\mathbb{G}}^b(q)$ stands for either $\hat{\mathbb{G}}^{\mathrm{R3D},b}(q)$ or $\hat{\mathbb{G}}^{\mathrm{F3D},b}(q)$ depending on the design.

Then, an asymptotically valid uniform $(1-\lambda)100\%$ confidence band for 
\(\tau^{\mathrm{R3D}}(q)\) (sharp) or \(\tau^{\mathrm{F3D}}(q)\) (fuzzy) on \(q\in [a,b]\) is given by:
\[
  \Bigl[\,
    \hat{\tau}_p(q)\;\pm\;
    \tfrac{1}{\sqrt{nh}}\;
    \hat{c}_n^B(a,b;\lambda)
  \Bigr],
  \quad
  \text{for }q\in\mathcal{T}^*.
\]

\subsection{Computational Details}

\begin{figure}[hb!]
\includegraphics[width=0.8\textwidth]{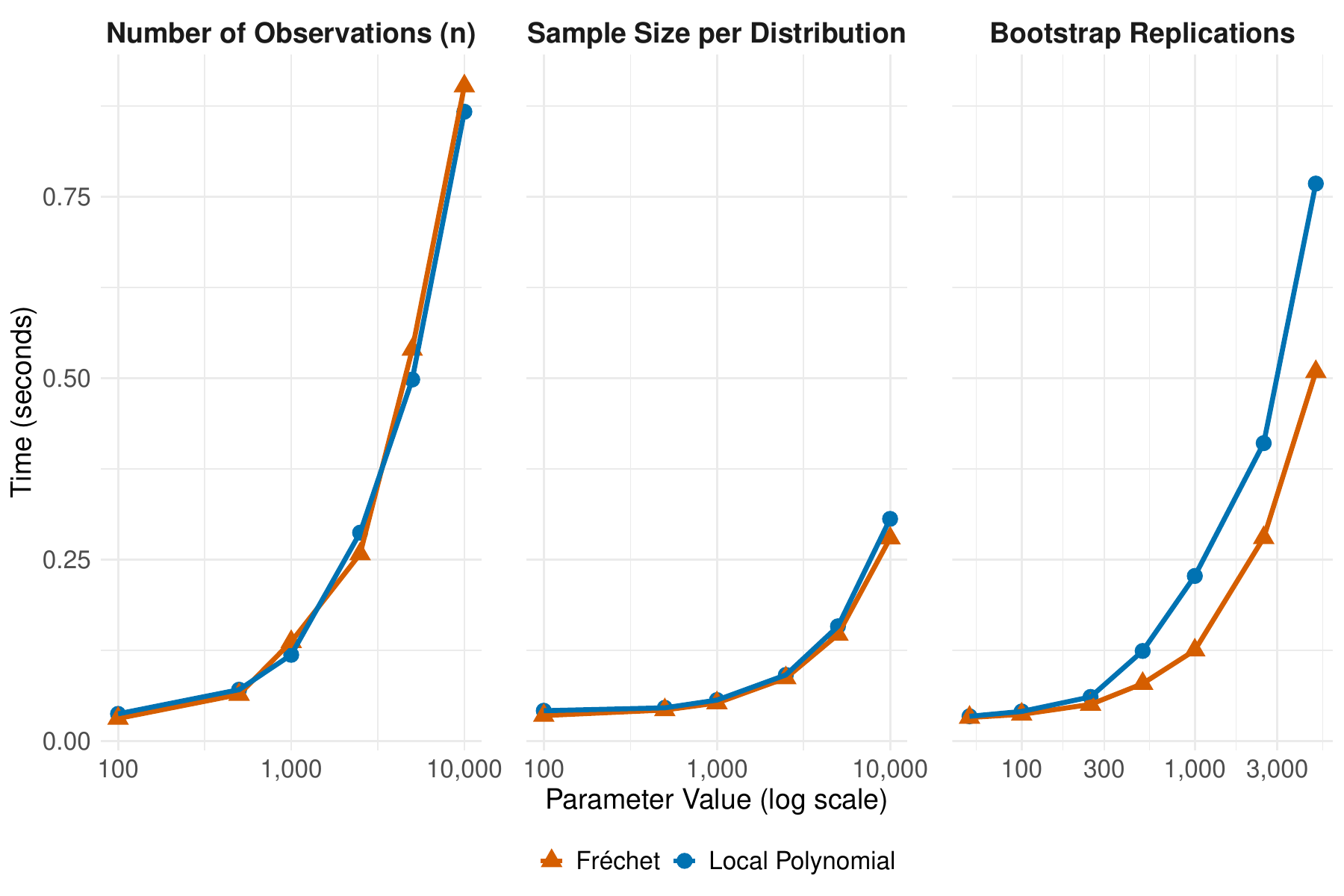}
\caption{Speed Benchmarks for R3D Package}
\label{fig:benchmarks}
\floatfoot{\textit{Note:} plots indicate seconds taken to estimate the model on an Apple M1 Pro computer with 16GB RAM, for various data and bootstrap sizes. The base model used for all computations, unless indicated otherwise, had $n=500$ with 500 samples per distribution, 100 bootstrap repetitions, and the quantile function evaluated at 20 quantiles.}
\end{figure}

An R implementation of the package can be found at \url{https://davidvandijcke.com/R3D}. The main polynomial weights estimation was implemented with a Fortran backend, leading to highly performant code, as illustrated in Figure \ref{fig:benchmarks}. For example, the model with 5 million total observations evaluated at 20 quantiles and 100 bootstrap repetitions solves in less than a second. The computational complexity scales linearly with the number of observations, the number of draws for the empirical distributions, and the number of bootstrap replications. The Fr\'echet estimator solves faster for increasing bootstrap repetitions than the local polynomial one, likely because the code can use optimized vector operations with one single bandwidth in the Fr\'echet case. The package also includes the option to parallelize the bootstrap for further speed improvements with large datasets.

\clearpage
\section{Proofs}

\subsection{Identification Results}

\paragraph{Proof of Lemma \ref{lemma:ident}.}
\begin{proof}
It holds that,
\begin{align*}
 \lim_{x \to 0^+} E[ Q_Y(q) \mid X=x ] 
 & =  \lim_{x \to 0^+} E[ Q_{Y^1}(q)  \mid X=x ] 
  =  E[ Q_{Y^1}(q)  \mid X=0 ] 
\end{align*}
and similarly for $x \to 0^-$. The first equality follows from the definition of $Y$ in terms of potential outcomes and the second from \ref{asspt:i_cont} and \ref{asspt:i_dens}. The result then follows from taking differences and using the linearity of the expectations operator.
\end{proof}

\paragraph{Proof of Lemma \ref{lemma:fuzzy_identification}.}
\begin{proof}
It holds that,
$$
\begin{aligned}
& \lim _{x \rightarrow 0^{+}} E[T \mid X=0]-\lim _{x \rightarrow 0^-} E[T \mid X=0] \\
& \quad=E\left[T^1 \mid X=0\right]-E\left[T^0 \mid X=0\right] \\
& \quad=E\left[T^1-T^0 \mid X=0\right] \\
& \quad=\operatorname{Pr}\left(T^1>T^0 \mid X=0\right)=\operatorname{Pr}\left(\mathrm{C} \mid X=0\right)
\end{aligned}
$$
where the first equality follows from the definition of $T^1$ and $T^0$, the continuity assumption \ref{asspt:i_treatment_continuity} and the zero-measure indefinites assumption in \ref{asspt:i_monotonicity}. The third equality follows from the law of total expectation and Assumption \ref{asspt:i_monotonicity}. Again, by Assumptions \ref{asspt:i_cont} and \ref{asspt:i_monotonicity},
$$
\lim _{x \rightarrow 0^{+}} E\left[Q_Y(q) \mid X=x\right]=E\left[Q_{Y^1}(q) \mathbf{1}\{\mathrm{C}\}+Q_{Y(\omega)}(q) \mathbf{1}\{\operatorname{not} \mathrm{C}\} \mid X=0\right] .
$$
and similarly for $x \to 0^-$. As a result,
\begin{align*}
& \lim _{x \rightarrow 0^{+}} E\left[Q_Y(q) \mid X=x\right]-\lim _{x \rightarrow 0^{-}} E\left[Q_Y(q) \mid X=x\right] \\
& = 
E\left[Q_{Y^1}(q)-Q_{Y^0}(q) \mid \text {C}, \, X=0\right] \times \operatorname{Pr}\left(\mathrm{C} \mid X=0\right).
\end{align*}
Combining these two derivations with Assumption \ref{asspt:i_fuzzyrd} gives the result. 
\end{proof}
\subsection{Asymptotic Results}

\begin{lemma}[Quantile functionals are VC type]
\label{lemma:quantile-vc-subgraph}
\begin{proof}
Let $\Theta = [a,b]$ a compact subset of $[0,1]$ or $[0,1]$ in the case where all cdfs in $\mathcal{Y}$ have compact support. For each $q \in \Theta$ and $Y \in \mathcal{Y}$, define 
\[
  f_{q}(Y) = Q_Y(q) = \inf\{\,x : Y(z) \ge q\}.
\]
  Then the family $\mathcal{F} = \{\,Y \to Q_Y(q) \colon q \in \Theta\}$ is a VC-subgraph class with index $V(\mathcal{F}) = 2$ (in the sense of \citet[\S 2.6.2]{vaart1996weak}). Indeed, the subgraph of $f_q$ is
\[
  G_{f_q}
  =\bigl\{\,(Y,z)\colon z \le Q_{Y}(q)\bigr\}
  =\bigl\{\,(Y,z)\colon Y(z)\,\ge\,q\bigr\},
\]
so membership in $G_{f_q}$ depends only on the scalar value $Y(z)$. A single point $(Y,z)$ can be shattered: include it by setting $q \leq Y(z)$, exclude it by setting $q > Y(z)$. However, 2 points cannot be shattered. To see this, let $\left(Y_1, z_1\right)$ and $\left(Y_2, z_2\right)$ have $Y_1\left(z_1\right) \leq Y_2\left(z_2\right)$. The subsets $\emptyset,\left\{\left(Y_2, z_2\right)\right\}$, and $\left\{\left(Y_1, z_1\right),\left(Y_2, z_2\right)\right\}$ can be realized by choosing $q>Y_2\left(z_2\right), q \in\left[Y_1\left(x_1\right), Y_2\left(z_2\right)\right]$, and $q \leq$ $Y_1\left(z_1\right)$, respectively. However, the subset $\left\{\left(Y_1, z_1\right)\right\}$ cannot be realized, due to the monotonicity of cumulative distribution functions. Hence, the largest shattered set has size 1, which implies $V(\mathcal{F}) = 2$.

Furthermore, let $F(Y) = \sup_{q \in \Theta} |Q_{Y}(q)|$. Then $F$ is a measurable envelope for the class of functions $f_q(Y)$ in $L^r(P)$ for any probability measure $P$ on $(\Omega^x, \mathcal{F}^x)$ and any $r \ge 1$, since each $Q_{Y}$ is bounded by the support of $Y$. By Theorem~2.6.7 of \cite{vaart1996weak}, there is a universal $K>0$ such that for all $0 < \varepsilon < 1$,
\[
  N\bigl(\varepsilon \|F\|_{Q,r}, \,\mathcal{F},\, L_r(Q)\bigr)
  \leq K\,V(\mathcal{F})\,(16e)^{V(\mathcal{F})}
         \,\Bigl(\tfrac{1}{\varepsilon}\Bigr)^{r\,[V(\mathcal{F})-1]}\!,
\]
and plugging in $V(\mathcal{F}) = 2$ gives a polynomial bound in $(1/\varepsilon)^r$.
\end{proof}
\end{lemma}

\begin{lemma}[Conditional Expected Quantile Functions are VC type]
\label{lemma:conditional-quantile-vc}
\begin{proof}
Let \(Y\) be a random distribution (with finite second moment), and for each \(q \in [0,1]\) define the real-valued function
\[
   g_q(x) = E\bigl[\,Q_Y(q)\,\mid\,X = x\bigr].
\]
Denote this family by
\[
  \mathcal{F}
  =\bigl\{\,x \to g_q(x) : q \in [0,1]\bigr\}.
\]
I claim \(\mathcal{F}\) is a VC-subgraph class of finite index. Indeed, by the results in Proposition \ref{prop:wasserstein}, each \(g_q(\cdot)\) can be identified with a one-dimensional quantile function: specifically, there is a conditional Fr\'echet mean \(m_{\oplus}(x) \in \mathcal{Y}\), as defined in \eqref{eq:conditional_frechet_mean}, such that
\[
  g_q(x)
  =
  Q_{\,m_{\oplus}(x)}(q),
\]
where \(Q_{\,m_{\oplus}(x)}\) is the quantile function of \(m_{\oplus}(x)\). In other words, for each \(q\), the subgraph of \(g_q\) can be written as
\[
  G_{g_q}
  =
  \bigl\{\,(x,z) : z \le Q_{m_{\oplus}(x)}(q)\bigr\}   =   \bigl\{\,(x,z) : m_{\oplus}(x)(z) \geq q \bigr\}.
\]
Since each $x \in \mathbb{R}$ defines a distinct, unique cdf $m_{\oplus}(x) \in \mathcal{Y}$ by Proposition \ref{prop:wasserstein}, the conclusion follows by an identical argument as in Lemma \ref{lemma:quantile-vc-subgraph}.
\end{proof}
\end{lemma}

\paragraph{Remark}

In the proofs that follow, I apply the results from \citet{chiang2019robust} to my R3D setting. For ease of comparison, note that their $\mu_{1}(x, \theta_1) = E[Q_Y(\theta_1) \mid X=x]$, $\mu_2(x, \theta_2) = 1$ in the sharp R3D setting, and $\mu_2(x, \theta_2) = E[T \mid X=x]$ in the fuzzy R3D setting. Further, the specific instances of their class of Wald estimands \citep[Eq. 4.1]{chiang2019robust} I consider are the sharp R3D \eqref{eq:estimator_R3D} and the F3D estimator \eqref{eq:estimator_F3D} so that in both cases, their functions $\Upsilon, \psi, \phi$ are all equal to the identity operator. The rest of their notation is closely followed for ease of comparison.

\paragraph{Proof of Theorem \ref{thm:asymptotic_conditional_means}}
\begin{proof} 
The result follows by an application of Theorem 1 in \citet{chiang2019robust}, which holds for any random object $Y$ as long as their assumptions are satisfied (despite the fact that the authors call the random element $(Y,T,X)$ a ``random vector''). To that end, I need to verify Assumptions 1 and 2 in that paper. I restate them in my notation for clarity.
{
\renewcommand{\theassumption}{1, \citealp{chiang2019robust}}
\begin{assumption} \label{asspt1_chiang}
Let $\underline{c} < 0 < \overline{c}$. (i) (a) This part is equivalent to Assumption \ref{asspt:sampling}-(i). (b) This part is equivalent to Assumption \ref{asspt:i_dens}. (ii) (a) The collections of real-valued functions $\{ x \to E[Q_Y(q) \mid X=x] : q \in [a,b] \}$, $ \{ Y \to Q_Y(q) : q \in [a,b] \}$ are of VC type with common integrable envelope $F_\mathcal{E}$ such that $\int_{\mathcal{Y} \times [\underline{c}, \overline{c}] }\left|F_{\mathcal{E}}(y, x)\right|^{2+\epsilon} d \mathrm{P}^x(y, x)<\infty$ for some $\epsilon>0$. (b) This part is equivalent to Assumption \ref{asspt:average_quantile_smoothness} (i). (c) For any $(q,k), (q',l) \in [a,b] \times \{1,2\}$, it holds that $\sigma_{kl}(q,q' | \cdot) \in C^{1}([\underline{c}, \overline{c}] \setminus \{0\} )$ with bounded derivatives in $x$ and $\sigma_{kl}(q,q' \mid 0^{\pm}) < \infty$. (d) For each $Y \in \mathcal{Y}$, $Q_Y(q)$ is left- or right-continuous in $q$. (iii) This part is equivalent to Assumption \ref{asspt:bandwidth_chiang}. (iv) (a) $K: [-1,1] \to \mathbb{R}^+$ is bounded and continuous, (b) $\{ K(\cdot / h) : h > 0 \}$ is of VC type, (c) $\Gamma_{\pm, p}$ is positive definite. 
\end{assumption}
}
\begin{itemize}
\item \textit{(ii) (a)} The fact that the function classes are of VC type is proved in Lemmas \ref{lemma:quantile-vc-subgraph} and \ref{lemma:conditional-quantile-vc}. A common integrable envelope can be constructed as follows. Define $F_{1}(y,x) = \sup_{q \in [a,b]} |Q_Y(q)|$ and $F_2(y,x) = \sup_{q \in [a,b]} |E[Q_Y(q) \mid X=x]|$ and define $F_{\varepsilon}(y,x) \coloneqq F_1(y,x) + F_2(y,x)$. Then clearly $\sup_{q \in [a,b]} |Q_Y(q)| \leq F_{\mathcal{E}}(y,x)$ and $\sup_{q \in [a,b]} |E[Q_Y(q) \mid X=x]| \leq F_{\mathcal{E}}(y,x)$. Moreover, by Assumption \ref{asspt:quantile_spread}, \begin{align*}
    & \int_{[\underline{c}, \overline{c}] \times \mathcal{Y}}\left(F_{\mathcal{E}}(x,y)\right)^{2+\varepsilon} \mathrm{d} P^x(y, x) \\
    & \leq 2^{1+\varepsilon} \int_{[\underline{c}, \bar{c}] \times \mathcal{Y}}\left(F_1(y, x)^{2+\varepsilon}+F_2(y, x)^{2+\varepsilon}\right) \mathrm{d} P^x(y, x) < \infty. 
\end{align*}
\item \textit{(ii) (c)} The covariance 
\begin{align*} 
& \sigma_{12}(q,q' \mid X=x)  = E[(Q_Y(q) - E[Q_Y(q) \mid X=x]) ( T - E[T \mid X=x]) \mid X=x]  \\
& = P(T=1 \mid X=x) E[Q_Y(q) \mid X=x] - P(T=1 \mid X=x) E[Q_Y(q) \mid X=x] \\
& + E[T \mid X=x] E[Q_Y(q) \mid X=x] - E[T \mid X=x] E[Q_Y(q) \mid X=x] = 0
\end{align*}
where the second equality follows from the law of total expectation. The variance term $\sigma_{22}(q,q' \mid X=x)=\var(T | X=x)$ is in $C^1([\underline{c}, \overline{c}] \setminus \{0 \} )$ by Assumption \ref{asspt:average_quantile_smoothness} (i). Finally, the cross-variance term $\sigma_{11}(q,q' \mid X=x)$ \begin{align*} 
& =E[(Q_Y(q) - E[Q_Y(q) \mid X=x]) (Q_Y(q') - E[Q_Y(q') \mid X=x]) \mid X=x].
\end{align*}
Expand the brackets and note that $E[Q_Y(q) Q_Y(q') \mid X=x]$ satisfies the assumption by Assumption \ref{asspt:average_quantile_smoothness} (ii) and the three other terms do so by Assumption \ref{asspt:average_quantile_smoothness} (i). 
\item \textit{(ii) (d)} follows by the left-continuity of quantile functions. 
\item \textit{(iv) (a)} Follows from Assumption \ref{asspt:kernel} where I can always normalize $K$ to have bounded support on $[-1,1]$ without loss of generality. 
\item \textit{(iv) (b)} To show that the function class $\{K(\cdot, / h): h>0\}$ is of VC type, consider the class of level sets $\{\{x: K(x / h)>t\}: h>0, t \in \mathbb{R}\}$.
For any $h>0$ and $t \in \mathbb{R}$, the set $\{x: K(x / h)>t\}$ is an interval in $\mathbb{R}$.
The class of intervals in $\mathbb{R}$ has a VC dimension of 2, which is finite. Hence, the function class $\{K(\cdot, / h): h>0\}$ is of VC type. 
\item \textit{(iv) (c)} Follows by the non-negativeness of $K$ in Assumption \ref{asspt:kernel}.
\end{itemize}
Under this set of assumptions, \citet{chiang2019robust} showed in their Lemma 1 that there exists a uniform Bahadur representation, 
$$
\begin{aligned}
& \sqrt{n h_{1}\left(q\right)}\left(\hat{m}_{\pm, p}\left(q\right)-m_{\pm}(q)-h_1(q)^{p+1}\left(q\right) \frac{e_0^{\prime}\left(\Gamma_{\pm, p}\right)^{-1} \Lambda_{p, p+1}^{ \pm}}{(p+1)!} \lim_{x \to 0^{\pm}}\dpar{m(q)^{p+1}}{x^{p+1}}\right) \\
= & \sum_{i=1}^n \frac{e_0^{\prime}\left(\Gamma_p^{ \pm}\right)^{-1} \mathcal{E}_1\left(Y_i, t_i, X_i, q\right) r_p\left(\frac{X_i}{h_{1}\left(q\right)}\right) K\left(\frac{X_i}{h_{1}\left(q\right)}\right) \delta_i^{ \pm}}{\sqrt{n h_{1}\left(q\right)} f_X(0)}+o_p^x(1)
\end{aligned}
$$
uniformly for all $q \in [a,b]$. An analogous expression obtains for $\hat{m}_{+,T,p}(q) - m_{+,T}(q)$. Note that this Bahadur representation is for the debiased estimator where the bias is of order $O(h^{p+1})$.
Then, by the proof of Theorem 1 in \citet{chiang2019robust}, $\nu^{+}_n(q, k) = \sum_{i=1}^n\left[f_{n i}(q, k)-E f_{n i}(q, k)\right]$ converges weakly to a tight zero-mean Gaussian process $\mathbb{G}_{H^+}$ with covariance function $H^+$ defined in the main theorem, where,
 \[ f_{n i}(q, k)=\frac{e_0^{\prime}\left(\Gamma_{+,p}\right)^{-1} r_p\left(\frac{X_i}{h_k\left(q\right)}\right)}{\sqrt{n h_{k}\left(q\right)} f_X(0)} \mathcal{E}_k\left(Y_i, T_i, X_i, q\right) K\left(\frac{X_i}{h_k\left(q\right)}\right) \delta_i^+\]
 and similarly for $\nu^-_n(q,k)$. Then Slutksy's theorem and Assumption \ref{asspt1_chiang} (iv) give the result. 
\end{proof}

\paragraph{Proof of Theorem \ref{thm:asymptotic_treatment_effects}}
\begin{proof}
The result for the sharp RD estimator in \eqref{eq:estimator_R3D} simply follows from Theorem \ref{thm:asymptotic_conditional_means} and the continuous mapping theorem, and similarly for the denominator of the fuzzy RDD estimator. Then, the ratio map
$$
\left(f_{+}, g_{+}\right) \to\left[f_{+}-f_{-}\right] /\left[g_{+}-g_{-}\right]
$$
is Hadamard differentiable tangentially to $\ell^{\infty}[a,b]$ on the subset where $g_{+}-g_{-} \neq 0$, which holds by \ref{asspt:i_monotonicity} \citep[Lemma 3]{chiang2019robust}. Then the functional delta method yields the result \citep[Lemma 3.9.3]{vaart1996weak}.
\end{proof}

\paragraph{Proof of Theorem \ref{thm:bootstrap}}
\begin{proof}
The result follows from Theorem \ref{thm:asymptotic_conditional_means} and Theorem 2 in \citet{chiang2019robust}. The latter applies because their Assumptions 1--4 are satisfied in my setting. Their Assumption 1 was shown to hold in the proof of Theorem \ref{thm:asymptotic_conditional_means}. Moreover, their Assumption 2 is satisfied since their operators $\psi, \phi, \Upsilon$ are trivially Hadamard differentiable in my setting, and by Assumptions \ref{asspt:i_monotonicity} and \ref{asspt:bandwidth_chiang}. Further, their Assumption 3 is equivalent to Assumption \ref{asspt:multiplier}. Finally, their Lemma 7 implies that the first-stage estimators $\hat{\mathcal{E}}_{k}(y, t, x, q)$ they propose are uniformly consistent for the population quantities $\mathcal{E}_{k}(y, t, x, q)$ on the kernel support $|X_i / h_k(q)| \leq 1$. Furthermore, I have assumed that $\hat{f}_X(0)$ is a consistent estimator of $f_X(0)$. As a result their Assumption 4 is also satisfied. Thus, their Theorem 2 follows. The final result obtains by combining it with Theorem \ref{thm:asymptotic_conditional_means} by plugging in $\hat{m}_{\pm,p}(\cdot), \hat{m}_{T,\pm,p}(\cdot)$ for $m_\pm(\cdot), m_{T,\pm}(\cdot)$.
\end{proof}

\paragraph{Proof of Theorem \ref{thm:convergence_frechet}}
\begin{proof}
  
From Theorem \ref{thm:asymptotic_conditional_means}, I already know that
\begin{equation}
\label{eq:unprojected-L2-conv}
  \sqrt{nh}\bigl[\hat{m}_{\pm,p}(q ) - m_\pm(q)\bigr]
    \leadsto \mathbb{G}_\pm(q)
  \quad\text{in }\ell^\infty([a,b]),
\end{equation}
where $\hat{m}_{\pm,p}(\cdot)$ are the unprojected local-polynomial estimators and $\mathbb{G}_{H^\pm}(\cdot)$ is a tight zero-mean Gaussian limit process. Moreover, the standard ordering of $L^p$ spaces implies that $\hat{m}_{\pm,p}-m_\pm$ also converges in $L^2$ norm on $[0,1]$. In
particular,
\[
  \|\hat{m}_{\pm,p}-m_\pm\|_{L^2}
  =
  O_p\bigl((nh)^{-\tfrac12}\bigr).
\]

\smallskip

\noindent
I then form the \emph{projected estimators}:
\[
  \hat{m}_{\pm, \oplus, p}(q) 
  :=
  \bigl[\Pi_{\!\mathcal{Q}}(\hat{m}_{\pm,p})\bigr](q),
  \qquad
  m_{\pm}(q)
  :=
  \bigl[\Pi_{\!\mathcal{Q}}(m_\pm)\bigr](q),
\]
where $\Pi_{\!\mathcal{Q}}$ is the \emph{metric projection} onto the closed,
convex cone $\mathcal{Q}\subset L^2$ of quantile functions. Since
$m_\pm(q)$ themselves are already valid quantile functions, they are unchanged
by the projection:
\[
  m_{\pm}^{\oplus} =
  \Pi_{\!\mathcal{Q}}(m_\pm) =
  m_\pm.
\]
Hence
\[
  \hat{m}_{\pm, \oplus, p} - m_\pm
  =
  \Pi_{\!\mathcal{Q}}(\hat{m}_{\pm,p})
  -\Pi_{\mathcal{Q}}(m_\pm).
\]
The goal is to show that
\[
\sqrt{nh}\,\bigl[\hat{m}_{\pm, \oplus, p} - m_\pm\bigr]
  \leadsto
  \text{the \emph{same} limit laws as in \eqref{eq:unprojected-L2-conv}.}
\]
Equivalently, I want to check that $\Pi_{\!\mathcal{Q}}$ has a suitable
Hadamard derivative at $m_\pm$ which acts as the identity map.

To that end, I distinguish two cases. 

\smallskip
\paragraph{Case 1: $m_\pm$ is strictly monotone.}\

\textit{Step 1.~$\Pi_{\mathcal{Q}}$ is 1-Lipschitz in $L^2$.}
Since $\mathcal{Q}$ is convex and closed in a Hilbert space, the map
\[
  \Pi_{\!\mathcal{Q}}\colon L^2 \to \mathcal{Q} \subset L^2
\]
is the usual orthogonal (metric) projection. By Proposition 4.16 in \citet{bauschke2017}, it is \emph{nonexpansive}, i.e.\
\[
  \|\Pi_{\mathcal{Q}}(g_1) -\Pi_{\mathcal{Q}}(g_2)\|_{L^2}
  \leq
  \|g_1 - g_2\|_{L^2},
  \quad\forall\,g_1,g_2\in L^2.
\]
Hence $\Pi_{\!\mathcal{Q}}$ is globally Lipschitz with constant~1.

\smallskip

\textit{Step 2.\;~Almost-everywhere Hadamard differentiability \citep{preiss2014gateaux}.}

In finite dimensions, Rademacher’s theorem implies that a Lipschitz function is
almost-everywhere differentiable (w.r.t.\ Lebesgue measure).  In the infinite-dimensional
space $L^2$, there is no Lebesgue measure.  Instead, \citet{preiss2014gateaux} showed that
any Lipschitz map $T\colon X\to Y$ between a separable Banach space $X$ and a Banach space $Y$ with the Radon-Nikodym property
admits Hadamard derivatives at all but a ``small'' set $N$, where $N$ belongs to
the $\sigma$-ideal of directionally porous sets.  (Equivalently, $X\setminus N$
is a full measure set in the sense that $N$ is
$\sigma$-directionally porous.)

Hence, since $\Pi_{\mathcal{Q}}\colon L^2\to L^2$ is Lipschitz and $L^2$ is
separable while $L^2$ also has the Radon–Nikodym property (being
a Hilbert space), there exists an
exceptional set $\mathcal{N}\subset L^2$ such that for each $x\in
L^2\setminus \mathcal{N}$, $\Pi_{\mathcal{Q}}$ is Hadamard differentiable
at $x$.  

Since $\hat{m}_{\pm,p}\to m_\pm$ in $L^2$ norm with probability~1,
eventually $(\hat{m}_{\pm,p}-m_\pm)$ remains in $L^2\setminus\mathcal{N}$.
Hence with probability~1, for large $n$, $\Pi_{\mathcal{Q}}$ is
Hadamard differentiable almost surely at $\hat{m}_{\pm,p}$, and in the relevant directions
$\hat{m}_{\pm,p}-m_\pm$.

\smallskip

\textit{Step 3.\;~The derivative at $m_\pm$ is the identity map.}
Since $m_\pm\in\mathcal{Q}$ (Proposition \ref{prop:wasserstein}), I have
\[
\Pi_{\mathcal{Q}}\bigl(m_\pm\bigr) = m_\pm.
\]
In a convex set, the projection acts as the identity on any $m_\pm$. Thus
$D\,\Pi_{\mathcal{Q}}[m_\pm]=\mathrm{Id}$ on the relevant tangent space in
$L^2$.  Concretely, if $m_\pm+\varepsilon u\in\mathcal{Q}$ then
$\Pi_{\mathcal{Q}}(m_\pm+\varepsilon u)-\Pi_{\mathcal{Q}}(m_\pm)$ is just
$\varepsilon u$, forcing the derivative to be the identity operator.

\smallskip

\textit{Step 4.\;~Hadamard derivative expansion.}
Then, write
\[
  \hat{m}_{\pm, \oplus, p} - m_\pm
   =   \Pi_{\mathcal{Q}}\bigl(\hat{m}_{\pm,p}\bigr)
    -\Pi_{\mathcal{Q}}\bigl(m_\pm\bigr)
  =
  \Pi_{\mathcal{Q}}\bigl(m_\pm + (\hat{m}_{\pm,p}-m_\pm)\bigr)
    - \Pi_{\mathcal{Q}}(m_\pm).
\]
Since $\hat{m}_{\pm,p}-m_\pm\in L^2\setminus\mathcal{N}$ eventually a.s., by the definition of the Hadamard derivative of
$\Pi_{\mathcal{Q}}$ at $m_\pm$ \citep[\S 20.2]{van2000asymptotic},
\[
  \Pi_{\mathcal{Q}}(m_\pm+u) 
    -\Pi_{\mathcal{Q}}(m_\pm)
  =
  D\Pi_{\mathcal{Q}}[m_\pm](u)
  + o(\|u\|_{L^2}),
\]
where $\|o(\|u\|_{L^2})\|_{L^2}/\|u\|_{L^2}\to 0$ as $\|u\|\to0$.  Substituting
$u=\hat{m}_{\pm,p}-m_\pm$ and recalling Step 4 implies
$D\Pi_{\mathcal{Q}}[m_\pm]=\mathrm{Id}$, I get
\[
  \hat{m}_{\pm,\oplus,p}-m_\pm
  =
  (\hat{m}_{\pm,p}-m_\pm) 
  + o\bigl(\|\hat{m}_{\pm,p}-m_\pm\|_{L^2}\bigr).
\]
Then, since $\|\hat{m}_{\pm,p}-m_\pm\|_{L^2}=O_p\bigl((nh)^{-1/2}\bigr)$, 
\[
\sqrt{nh}\bigl[\hat{m}_{\pm, \oplus,p}-m_\pm\bigr]
  =
\sqrt{nh}\bigl[\hat{m}_{\pm,p}-m_\pm\bigr]
    +o_p(1),
\]
so $\hat{m}_{\pm, \oplus, p}$ inherits the same uniform limit law as
\eqref{eq:unprojected-L2-conv}.

\smallskip

\paragraph{Case 2: $m_{\pm}$ is weakly monotone.}\

Suppose $m_{\pm}$ is only weakly increasing and hence may lie on the boundary
of the cone $\mathcal{Q}$. To move it inside the interior of $\mathcal{Q}$, fix
a small ``hill'' function $H(\cdot)$ on $[0,1]$ (e.g.\ strictly positive on
$(0,1)$ and zero near the endpoints). For each $n$, pick a scalar
$\eta_n>0$ that shrinks to zero as $n\to\infty$ at rate
$\eta_n=o\bigl((n h)^{-1/2}\bigr)$. Define the \emph{$\eta$-perturbed}
functions
\[
  m_{\pm}^\eta(\cdot)= m_{\pm}(\cdot)+\eta_n\,H(\cdot),
  \quad
  \hat{m}_{\pm,p}^\eta(\cdot)=
    \hat{m}_{\pm,p}(\cdot)+\eta_n\,H(\cdot).
\]
Each $m_{\pm}^\eta$ is now \emph{strictly} increasing on $(0,1)$, so
$m_{\pm}^\eta$ is an interior point of $\mathcal{Q}$. In similar vein, consider the
projected estimator
\[
  \hat{m}_{\pm,\oplus,p}^\eta
 = \Pi_{\mathcal{Q}}\Bigl[\hat{m}_{\pm,p}^\eta\Bigr],
\]
which falls under the ``strictly monotone'' scenario of Case 1.
In particular, the same Hadamard-differentiability and functional delta method
argument implies that
\[
  \sqrt{n h}\bigl[\hat{m}_{\pm,\oplus,p}^\eta - m_{\pm}^\eta\bigr]
  \to
  \mathbb{G}_{\pm},
\]
the same limiting law as before. Then, by 1-Lipschitz continuity of the
projection,
\[
  \bigl\|
    \hat{m}_{\pm,\oplus,p}^\eta - \hat{m}_{\pm,\oplus,p}
  \bigr\|_{L^2}
  \leq
  \bigl\|
    \hat{m}_{\pm,p}^\eta - \hat{m}_{\pm,p}
  \bigr\|_{L^2}
 =\eta_n\,\|H\|_{L^2},
  \quad\text{and}\quad
  \bigl\|m_{\pm}^\eta - m_{\pm}\bigr\|_{L^2}
 =\eta_n \|H\|_{L^2}.
\]
Hence
\[
  \sqrt{n h}\,\Bigl(
    \bigl[\hat{m}_{\pm,\oplus,p}^\eta - m_{\pm}^\eta\bigr]
    \;-\;
    \bigl[\hat{m}_{\pm,\oplus,p} - m_{\pm}\bigr]
  \Bigr)
 =
  \sqrt{n h} \,O\bigl(\eta_n\bigr),
\]
which vanishes since $\eta_n=o\bigl((nh)^{-1/2}\bigr)$.
 Consequently,
\[
  \sqrt{n h}\,\bigl[\hat{m}_{\pm,\oplus,p} - m_{\pm}\bigr]
  \;\;\Longrightarrow\;\;
  \mathbb{G}_{\pm}
\]
also holds in the weakly-monotone case.  This completes the proof.
\end{proof}

\paragraph{Proof of Corollary \ref{corollary:frechet_treatment}.} 
\begin{proof}
The result follows by an identical argument as the proof of Theorem \ref{thm:asymptotic_treatment_effects}.
\end{proof}

\paragraph{Proof of Proposition \ref{prop:empirical_quantiles}.}
\begin{proof}
Remember the definition of the local polynomial estimator with empirical quantile functions, 
\[
\bar{m}_{\pm, p}(q) = \frac{1}{n} \sum_{i=1}^n s_{ \pm, i n}^{(p)}(h) \widehat{Q}_{Y_i}(q).
\]
I have that,
\begin{align*}
& \sqrt{nh} \left( \bar{m}_{\pm, p}(q) - m_{\pm}(q) \right) \\
& = \sqrt{nh} \left(  \bar{m}_{\pm, p}(q) - \hat{m}_{\pm, p}(q)\right) +  \sqrt{nh} \left( \hat{m}_{\pm, p}(q)  - m_{\pm}(q) \right)  \\ 
& = \sqrt{nh} \left(\frac{1}{n} \sum_{i=1}^n s_{ \pm, i n}^{(p)}(h) \left( \widehat{Q}_{Y_i}(q) - Q_{Y_i}(q) \right) \right) + \sqrt{nh} \left(\left( \hat{m}_{\pm, p}(q)  - m_{\pm}(q) \right) \right) \\
& = o_p(1) + \sqrt{nh} \left( \hat{m}_{\pm, p}(q)  - m_{\pm}(q) \right) ,
\end{align*}
uniformly over $q \in [a,b]$. The last equality follows from Assumption \ref{asspt:E_convergence} and Corollary 21.5 in \citet{van2000asymptotic} which guarantees the uniform convergence of each empirical quantile function $\widehat{Q}_{Y_i}$ at rate $n_i^{-1/2}$ for $i=1,\ldots,n$. Then Assumption \ref{asspt:E_n} translates this to a convergence rate of at least $O_p(n^{-1/2})$ for all $i=1,\ldots,n$, which is faster than the rate $\sqrt{nh}$ for any $h \to 0$ as required by Assumption \ref{asspt:bandwidth_chiang}.

Then, define the Fr\'echet estimator with empirical distribution functions as,
\[
\underset{\omega \in \mathcal{Y}}{\operatorname{argmin}} \frac{1}{n} \sum_{i=1}^n s_{ \pm, i n}^{(p)}(h) d_{W_2}^2\left(\omega, \hat{Y}_i\right).
\]
An identical argument to the one in Proposition \ref{prop:wasserstein} shows that the quantile function of this estimator is,
\[
\bar{m}_{\pm, \oplus, p}=\underset{h \in Q(\mathcal{Y})}{\operatorname{argmin}} \, d_{L^2}\left(h, \bar{m}_{\pm, p}\right)^2,
\]
that is, the projection of the local polynomial estimator with empirical quantile functions. Then, since I have established that the latter converges uniformly to the same limiting process as the standard local polynomial estimator, the same exact argument as in the proof of Theorem \ref{thm:convergence_frechet} implies that $\bar{m}_{\pm, \oplus, p}(\cdot) - m_{\pm, p}(\cdot)$ has the same limiting law as $\bar{m}_{\pm, p}(\cdot) - m_{\pm}(\cdot)$, with the only difference that $m_\pm$ is always strictly increasing under Assumption \ref{asspt:E_convergence} so that we only need to consider Case 1 from that proof. Finally, the results for the treatment effects with empirical quantile functions, $\bar{\tau}^{\mathrm{R3D}}_{p}(\cdot)$, $\bar{\tau}^{\mathrm{F3D}}_{p}(\cdot)$, $\bar{\tau}^{\mathrm{R3D}}_{\oplus, p}(\cdot)$, $\bar{\tau}^{\mathrm{F3D}}_{\oplus, p}(\cdot)$, then follow from identical arguments as in the proof of Theorem \ref{thm:asymptotic_treatment_effects}.
\end{proof}

The following lemma establishes the intuitive result that the difference between the Fr\'echet and local polynomial estimators converges faster than each of them converges to the population moment. In Section \ref{sec_app:bandwidth}, I use it to derive the IMSE-optimal bandwidth for the Fr\'echet estimator based on the standard MSE-optimal bandwidth for the local polynomial one,
\begin{lemma} \label{lemma:frechet_lp_squeeze}
Under the Assumptions of Theorem \ref{thm:asymptotic_conditional_means},
\[
  \|\hat{m}_{\pm,\oplus,p}-\hat{m}_{\pm,p}\|_{L^2}
  =
  o_p\bigl((nh)^{-1/2}\bigr).
\]

\begin{proof}
I write $L^2:=L^2([a,b])$ for brevity, and let $\|\cdot\|_{L^2}$ denote the usual $L^2$ norm in $q\in[a,b]\subset(0,1)$.  

 As argued in the proof of Theorem \ref{thm:convergence_frechet}, $m_{\pm}(\cdot)$ is a point in the convex set $\mathcal{Q}$ of quantile functions. Assume first that it is an interior point (strictly increasing). Therefore, the metric projection $\Pi_{\mathcal{Q}}\colon L^2 \to \mathcal{Q}$ is Hadamard differentiable at $m_{\pm}$ with derivative equal to the identity operator.
Concretely, this means there is a remainder function $r_n(h)$ with $\|r_n(\epsilon)\|_{L^2}/\|\epsilon\|_{L^2}\to0$ whenever $\|\epsilon\|_{L^2}\to0$, such that 
\[
\Pi_{\mathcal{Q}}\bigl(m_{\pm}+\epsilon\bigr)
  =
  m_{\pm}
  +
  \epsilon
  +
  r_n(\epsilon),
  \quad
  \text{where }\|r_n(\epsilon)\|_{L^2}=o\bigl(\|\epsilon\|_{L^2}\bigr).
\]
Furthermore, $\Pi_{\mathcal{Q}}(m_{\pm})=m_{\pm}$.
Setting $\epsilon=\epsilon_n:=\hat{m}_{\pm,p}-m_{\pm}$ in the above expansion, and noting that 
$\|\epsilon_n\|_{L^2}=O_p((nh)^{-1/2})\to0$ by Theorem \ref{thm:asymptotic_conditional_means}, I obtain
\[
  \hat{m}_{\pm,\oplus,p}
  =
\Pi_{\mathcal{Q}}\bigl(m_{\pm}+\epsilon_n\bigr)
  =
  m_{\pm} + \epsilon_n + r_n(\epsilon_n).
\]
Hence
\[
  \hat{m}_{\pm,\oplus,p}
  -
  \hat{m}_{\pm,p}
  =
  \bigl[m_{\pm} + \epsilon_n + r_n(\epsilon_n)\bigr]
  -
  \bigl[m_{\pm} + \epsilon_n\bigr]
  =
  r_n(\epsilon_n).
\]
By construction, $\|r_n(\epsilon_n)\|_{L^2} = o(\|\epsilon_n\|_{L^2})$ and from above, 
$\|\epsilon_n\|_{L^2} = O_p((nh)^{-1/2})$. Combining gives,
\[
  \|\hat{m}_{\pm,\oplus,p} - \hat{m}_{\pm,p}\|_{L^2}
  =
  \|r_n(\epsilon_n)\|_{L^2}
  =
  o \bigl(\|\epsilon_n\|_{L^2}\bigr)
  =
  o_p \bigl((nh)^{-1/2}\bigr).
\]
The argument for the case where $m_\pm$ is a boundary point of $\mathcal{Q}$ follows analogously as in the proof of Theorem \ref{thm:convergence_frechet} by perturbing the quantile functions to make them strictly monotonic and then letting the perturbation go to $0$ at rate $o_p((nh)^{-1/2})$.

\end{proof}
\end{lemma}

\section{Additional Results}

\subsection{Tables}

\begin{table}[h!]

\begin{tabular}{lcc}
\hline\hline
\multicolumn{1}{c}{Level:} & \multicolumn{1}{c}{State} & \multicolumn{1}{c}{Family} \\
\hline
Treatment Effect & -0.631 & -0.525 \\
 & (0.419) & (0.062) \\
\\
\hline 
Robust 95\% CI & [-1.537, 0.105] & [-0.681, -0.438] \\
P-value & 0.087 & 0 \\
Bandwidth & 0.083 & 0.036 \\
Effective Observations & 203 & 138692 \\
\hline\hline
\end{tabular}
\caption{Canonical RD Estimates}
\label{tab:standard_rd}
\floatfoot{\textit{Note}: table presents canonical RD estimates using both state-level average family income (weighted by the family-level probability weights) and family-level income as outcome variable, computed using the \texttt{rdrobust} command in R \citep{calonico2015rdrobust}. MSE-optimal bandwidth was selected using the method in \citet{calonico2020optimal} and robust confidence intervals were calculated as in \citet{calonico2014robust}, clustered at the state level for the state-level data and the state-year level for the family-level data.}
\end{table}

\subsection{Figures}

\begin{figure}[htbp!]
\includegraphics[width=0.85\textwidth]{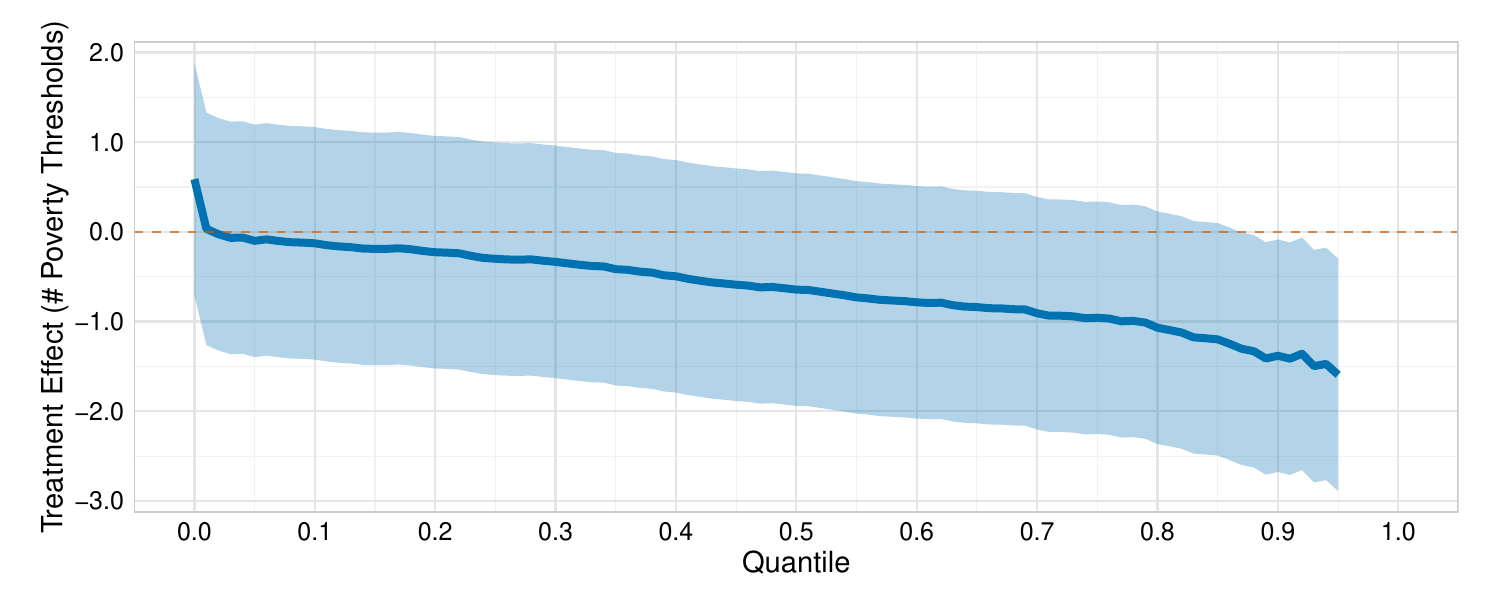}
\caption{Distributional Effects of Democratic Governor Control, 1984-2010: Local Polynomial}
\label{fig:cps_r3d_simple_method}
\floatfoot{\textit{Note:}  local average quantile treatment effects estimates and uniform 90\% confidence bands for R3D of effect of Democratic governor control on within-state income distribution. X-axis indicates quantile of the (average) income distribution while Y-axis indicates the difference in average state-level income distributions, in the final year of the governor's tenure, near the 50\% vote share threshold. Income is measured as real equivalized family income in multiples of the federal poverty threshold. Sample runs from 1984--2010, estimates are obtained using the second-order local polynomial estimator in Section \ref{sec:estimator_simple} with first-order IMSE-optimal bandwidth and triangular kernel as in Section \ref{sec_app:bandwidth}, and uniform bands are constructed using Algorithm \ref{app:bootstrap} with 5,000 bootstrap repetitions. Treatment nullity p-value: 0.04, treatment homogeneity p-value: 0.067. Average MSE-optinal bandwidths: 0.262.   } 
\end{figure}

\begin{figure}[h!]
\includegraphics[width=0.85\textwidth]{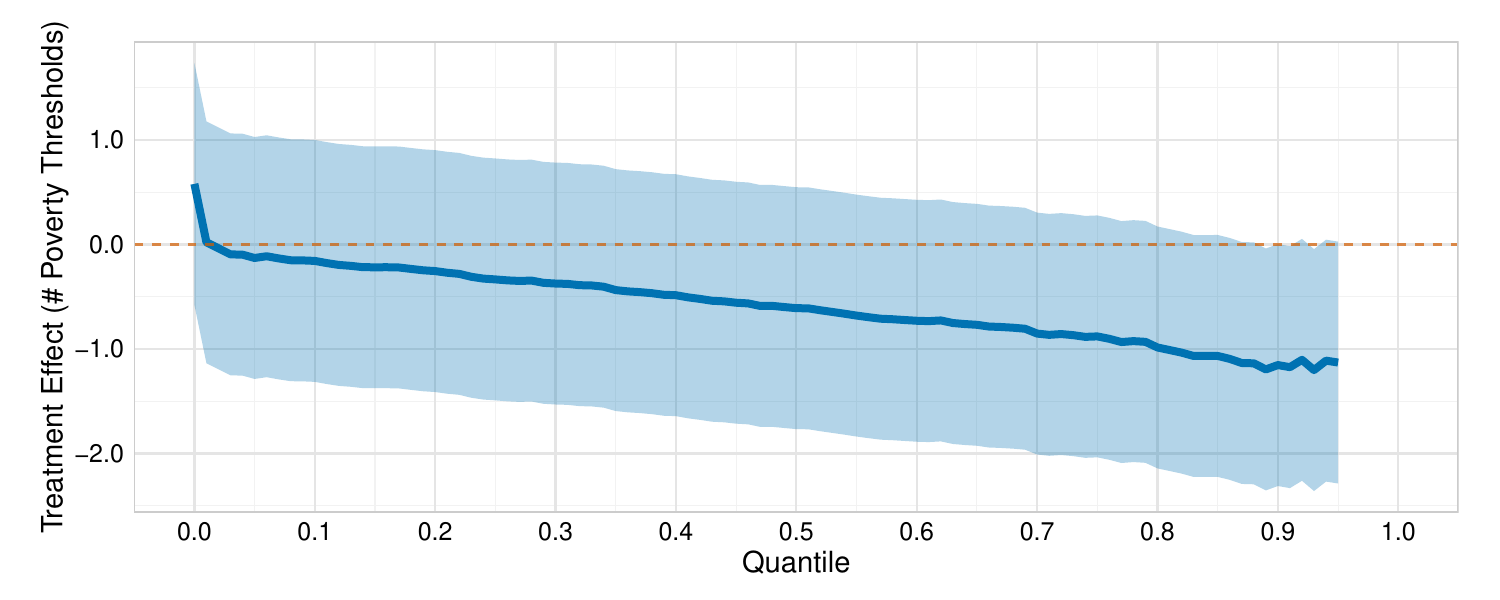}
\caption{Distributional Effects of Democratic Governor Control, 1984-2010: Uniform Kernel}
\label{fig:cps_r3d_uniform_kernel}
\floatfoot{\textit{Note:}  local average quantile treatment effects estimates and uniform 90\% confidence bands for R3D of effect of Democratic governor control on within-state income distribution. X-axis indicates quantile of the (average) income distribution while Y-axis indicates the difference in average state-level income distributions, in the final year of the governor's tenure, near the 50\% vote share threshold. Income is measured as real equivalized family income in multiples of the federal poverty threshold. Sample runs from 1984--2010, estimates are obtained using the second-order Fr\'echet estimator in Section \ref{sec:estimator} with first-order IMSE-optimal bandwidth and uniform kernel as in Section \ref{sec_app:bandwidth}, and uniform bands are constructed using Algorithm \ref{app:bootstrap} with 5,000 bootstrap repetitions. Treatment nullity p-value: 0.071, treatment homogeneity p-value: 0.140, IMSE-optimal bandwidth: 0.223. } 
\end{figure}

\begin{figure}[htbp!]
\includegraphics[width=0.85\textwidth]{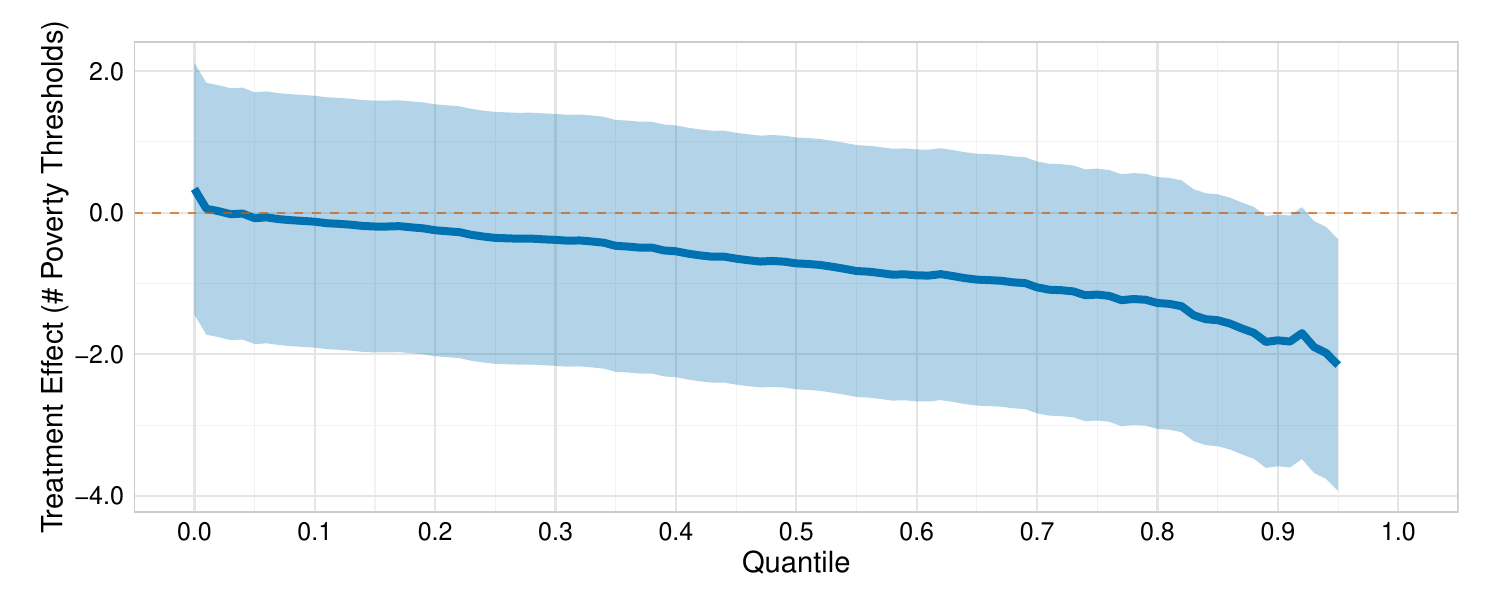}
\caption{Distributional Effects of Democratic Governor Control, 1984-2010: $1/2$ Bandwidth}
\label{fig:cps_r3d_1_2_bandwidth}
\floatfoot{\textit{Note:}  local average quantile treatment effects estimates and uniform 90\% confidence bands for R3D of effect of Democratic governor control on within-state income distribution. X-axis indicates quantile of the (average) income distribution while Y-axis indicates the difference in average state-level income distributions, in the final year of the governor's tenure, near the 50\% vote share threshold. Income is measured as real equivalized family income in multiples of the federal poverty threshold. Sample runs from 1984--2010, estimates are obtained using the second-order local polynomial estimator in Section \ref{sec:estimator_simple} with $1/2 \times$the first-order IMSE-optimal bandwidth (0.16) and triangular kernel as in Section \ref{sec_app:bandwidth}, and uniform bands are constructed using Algorithm \ref{app:bootstrap} with 5,000 bootstrap repetitions. Treatment nullity p-value: 0.045, treatment homogeneity p-value: 0.047, IMSE-optimal bandwidth: 0.11.  } 
\end{figure}

\begin{figure}[h!]
\includegraphics[width=0.85\textwidth]{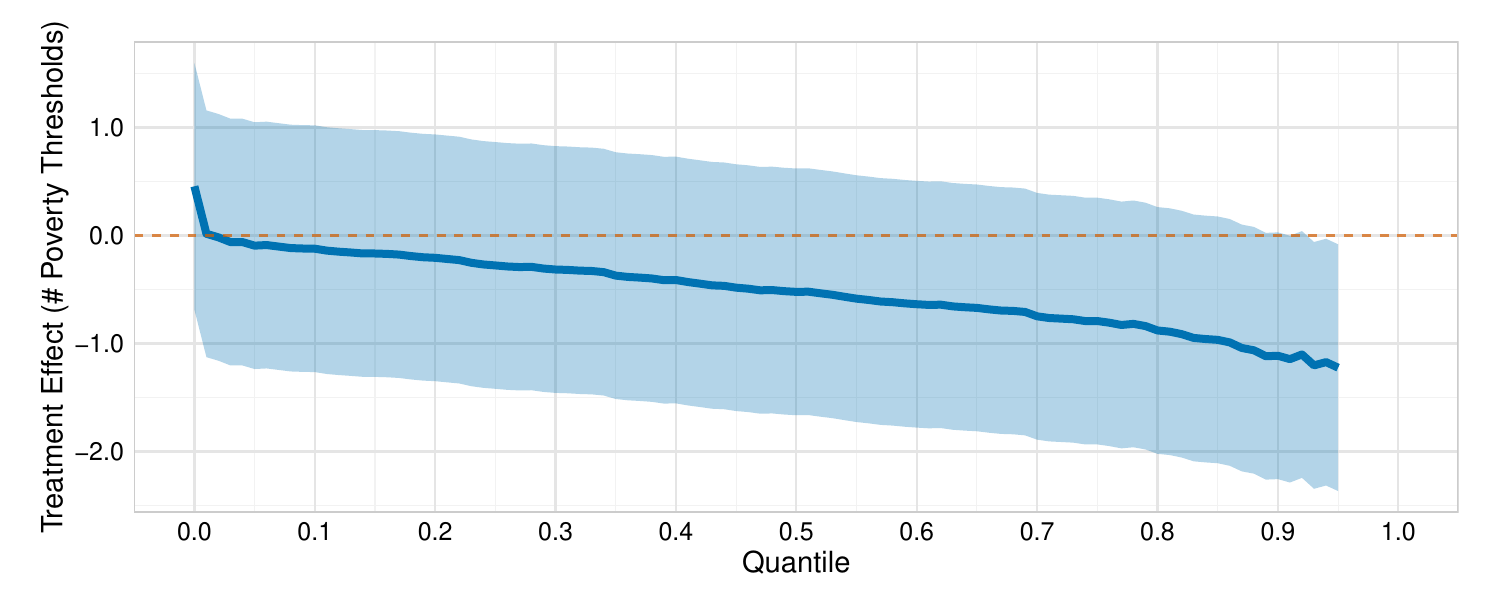}
\caption{Distributional Effects of Democratic Governor Control, 1984-2018}
\label{fig:cps_2018}
\floatfoot{\textit{Note:}  local average quantile treatment effects estimates and uniform 90\% confidence bands for R3D of effect of Democratic governor control on within-state income distribution. X-axis indicates quantile of the (average) income distribution while Y-axis indicates the difference in average state-level income distributions, in the final year of the governor's tenure, near the 50\% vote share threshold. Income is measured as real equivalized family income in multiples of the federal poverty threshold. Sample runs from 1984--2018, estimates are obtained using the second-order Fr\'echet estimator in Section \ref{sec:estimator} with first-order IMSE-optimal bandwidth and triangular kernel as in Section \ref{sec_app:bandwidth}, and uniform bands are constructed using Algorithm \ref{app:bootstrap} with 5,000 bootstrap repetitions. Treatment nullity p-value: 0.067, treatment homogeneity p-value: 0.087, IMSE-optimal bandwidth: 0.255. } 
\end{figure}

\begin{figure}[htbp!]
\includegraphics[width=0.85\textwidth]{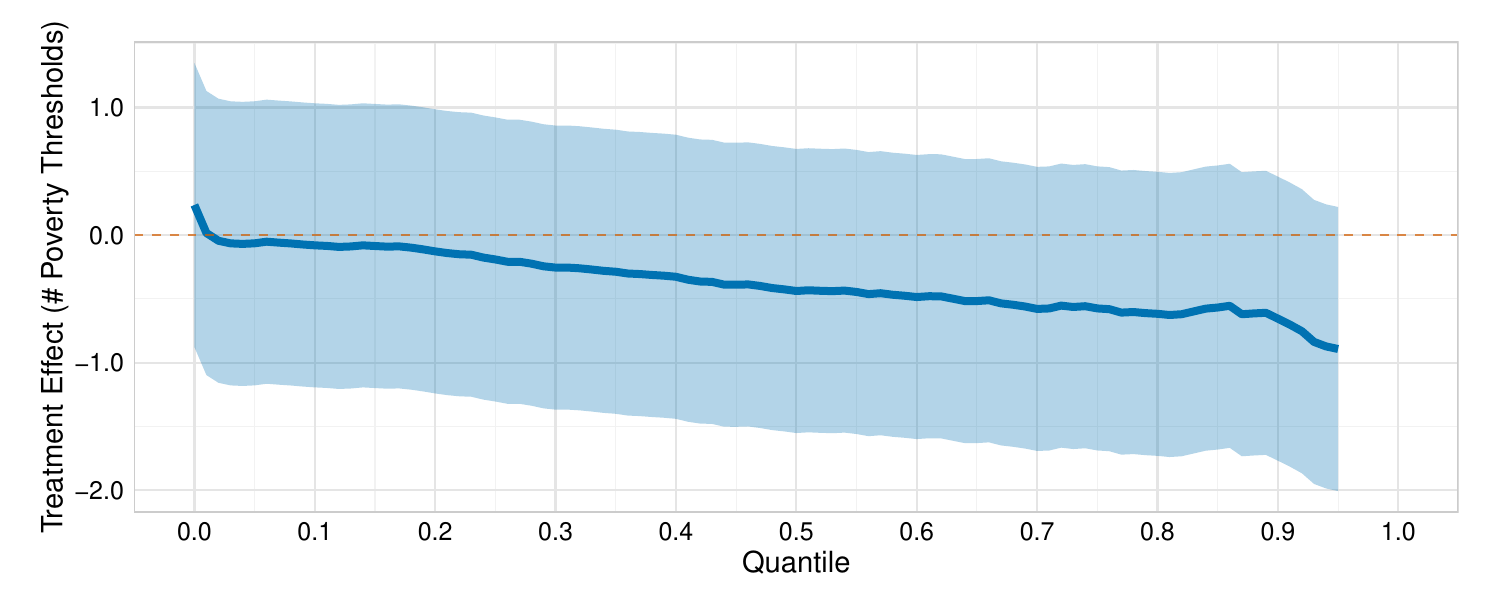}
\caption{Distributional Effects of Democratic Governor Control, Robustness: Election-Year Incomes}
\label{fig:cps_r3d_sameyear}
\floatfoot{\textit{Note:}  local average quantile treatment effects estimates and uniform 90\% confidence bands for R3D of effect of Democratic governor control on within-state income distribution. X-axis indicates quantile of the (average) income distribution while Y-axis indicates the difference in average state-level income distributions, in the election year, near the 50\% vote share threshold. Income is measured as real equivalized family income in multiples of the federal poverty threshold. Sample runs from 1984--2010, estimates are obtained using the second-order local polynomial estimator in Section \ref{sec:estimator_simple} with first-order IMSE-optimal bandwidth and triangular kernel as in Section \ref{sec_app:bandwidth}, and uniform bands are constructed using Algorithm \ref{app:bootstrap} with 5,000 bootstrap repetitions. Treatment nullity p-value: 0.183, treatment homogeneity p-value: 0.224, IMSE-optimal bandwidth: 0.241. } 
\end{figure}

\begin{figure}[htbp!]
\includegraphics[width=0.85\textwidth]{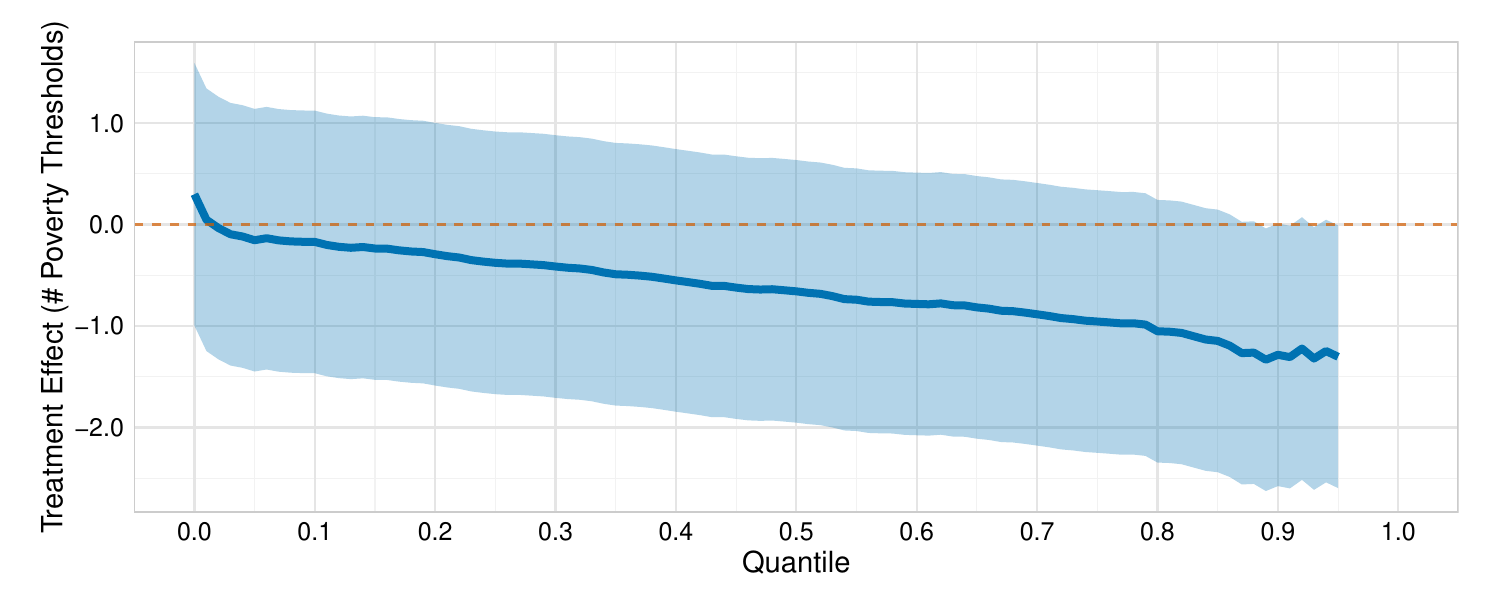}
\caption{Distributional Effects of Democratic Governor Control, Robustness: No Cross-State Migration}
\label{fig:cps_r3d_migration}
\floatfoot{\textit{Note:}  local average quantile treatment effects estimates and uniform 90\% confidence bands for R3D of effect of Democratic governor control on within-state income distribution. X-axis indicates quantile of the (average) income distribution while Y-axis indicates the difference in average state-level income distributions, in the final year of the governor's tenure, near the 50\% vote share threshold. Only families that did not migrate across state borders in the previous year are included. Income is measured as real equivalized family income in multiples of the federal poverty threshold. Sample runs from 1984--2010, estimates are obtained using the second-order local polynomial estimator in Section \ref{sec:estimator_simple} with first-order IMSE-optimal bandwidth and triangular kernel as in Section \ref{sec_app:bandwidth}, and uniform bands are constructed using Algorithm \ref{app:bootstrap} with 5,000 bootstrap repetitions. Treatment nullity p-value: 0.072, treatment homogeneity p-value: 0.137, IMSE-optimal bandwidth: 0.229. } 
\end{figure}

\begin{figure}[h!]
\includegraphics[width=0.85\textwidth]{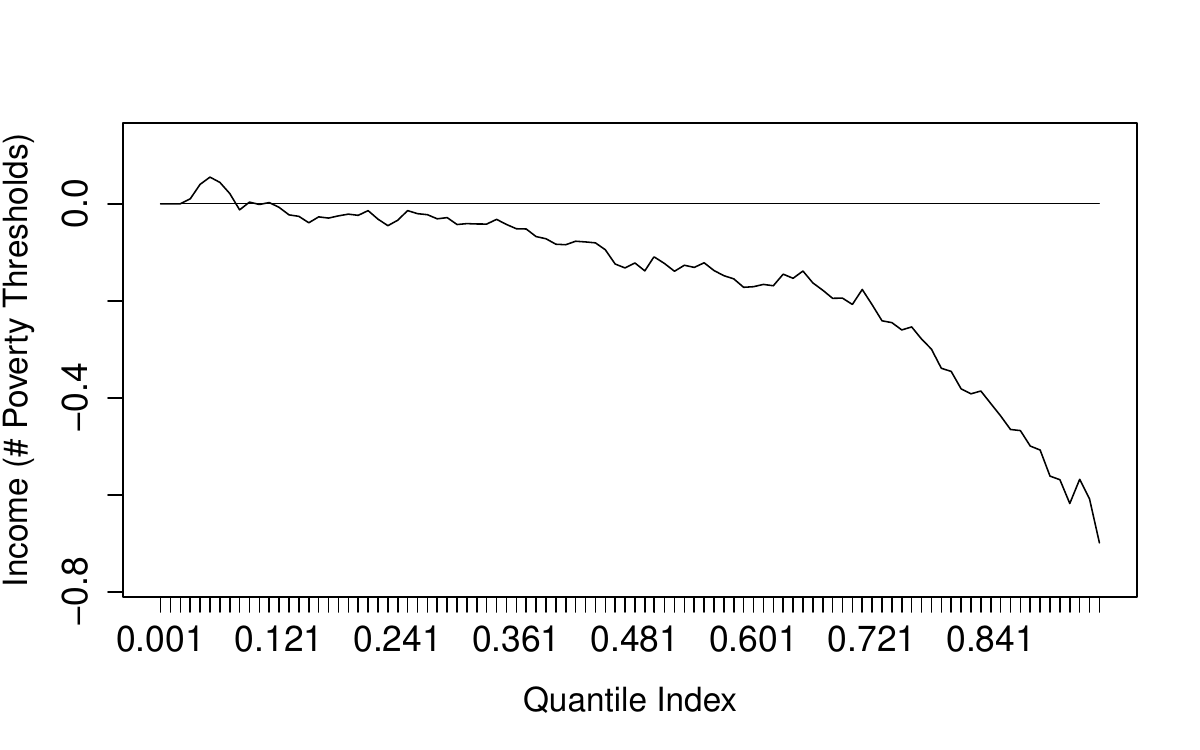}
\caption{Distributional Effects of Democratic Governor Control: Quantile RD Estimates}
\label{fig:r3d_cps_qte_rd}
\floatfoot{\textit{Note}: plot shows quantile RD estimates of effect of Democratic governor control on within-state income distribution. X-axis indicates quantile of the (average) income distribution while Y-axis indicates the difference in state-level income distributions, in the final year of the governor's tenure, near the 50\% vote share threshold. Income is measured as real equivalized family income in multiples of the federal poverty threshold. Sample runs from 1984--2010, estimates are obtained using the quantile RD estimator of \citet{qu2019uniform} with bias correction \citep{qu2024inference}, with the same bandwidth as Figure \ref{fig:cps_r3d_baseline} and triangular kernel \ref{sec_app:bandwidth}.}
\end{figure}

\clearpage
\section{Software Appendix}

All results in this paper were produced in \texttt{R} using \texttt{RStudio}. A complete reference list of packages used is provided below.

\end{document}